\documentclass{aa}
\usepackage{graphicx}
\usepackage{lscape}
\usepackage{subfigure}
\usepackage{natbib}
\usepackage[varg]{txfonts}
\usepackage{longtable}
\usepackage{booktabs}
\usepackage[normalem]{ulem} 
\usepackage{epstopdf}
\usepackage{xcolor}
\usepackage{soul}

\usepackage{scalerel}
\usepackage{tikz}
\usetikzlibrary{svg.path}

\definecolor{orcidlogocol}{HTML}{A6CE39}
\tikzset{
  orcidlogo/.pic={
    \fill[orcidlogocol] svg{M256,128c0,70.7-57.3,128-128,128C57.3,256,0,198.7,0,128C0,57.3,57.3,0,128,0C198.7,0,256,57.3,256,128z};
    \fill[white] svg{M86.3,186.2H70.9V79.1h15.4v48.4V186.2z}
                 svg{M108.9,79.1h41.6c39.6,0,57,28.3,57,53.6c0,27.5-21.5,53.6-56.8,53.6h-41.8V79.1z M124.3,172.4h24.5c34.9,0,42.9-26.5,42.9-39.7c0-21.5-13.7-39.7-43.7-39.7h-23.7V172.4z}
                 svg{M88.7,56.8c0,5.5-4.5,10.1-10.1,10.1c-5.6,0-10.1-4.6-10.1-10.1c0-5.6,4.5-10.1,10.1-10.1C84.2,46.7,88.7,51.3,88.7,56.8z};
  }
}

\newcommand\orcidicon[1]{\href{https://orcid.org/#1}{\mbox{\scalerel*{
\begin{tikzpicture}[yscale=-1,transform shape]
\pic{orcidlogo};
\end{tikzpicture}
}{|}}}}

\usepackage{hyperref} 
\hypersetup{
    colorlinks=true,
    citecolor=blue,
    linkcolor=blue,
    urlcolor=blue,
    }
\makeatletter
\renewcommand*\aa@pageof{, page \thepage{} of \pageref*{LastPage}}
\makeatother

\defcitealias{decicco19}{D19}
\defcitealias{decicco15}{D15}
\newcommand{\dd}{\citetalias{decicco19}}
\newcommand{\ddc}{\citetalias{decicco15}}

\bibpunct{(}{)}{;}{a}{}{,} 
\usepackage{amstext} 
\definecolor{cadmiumred}{rgb}{0.89, 0.0, 0.13}
\newcommand{\red}[1]{\textcolor{cadmiumred}{\bf{#1}}}
\newcommand{\blue}[1]{\textcolor{blue}{\bf{#1}}}
\newcommand{\violet}[1]{\textcolor{violet}{\bf{#1}}}
\definecolor{ste}{rgb}{0., 0.26, 0.15}

\newcommand{\orcid}[1]{\href{https://orcid.org/#1}{\includegraphics[width=8pt]{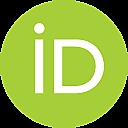}}}

\begin{document} 
   \title{A random forest-based selection of optically variable AGN in the VST-COSMOS field\thanks{Observations were provided by the ESO programs 088.D-4013, 092.D-0370, and 094.D-0417 (PI G. Pignata).}}

   \author{D. De Cicco\inst{1,2\orcidicon{0000-0001-7208-5101}}, 
   F. E. Bauer\inst{1,2,3\orcidicon{0000-0002-8686-8737}}, 
   M. Paolillo\inst{4,5,6\orcidicon{0000-0003-4210-7693}}, 
   S. Cavuoti\inst{6,5\orcidicon{0000-0002-3787-4196}}, 
   P. S\'{a}nchez-S\'{a}ez\inst{2,1\orcidicon{0000-0003-0820-4692}}, 
   W. N. Brandt\inst{7,8,9\orcidicon{0000-0002-0167-2453}}, 
   G. Pignata\inst{10,2\orcidicon{0000-0001-6003-8877}}, 
   M. Vaccari\inst{11,12\orcidicon{0000-0002-6748-0577}}, 
   M. Radovich\inst{13\orcidicon{0000-0002-3585-866X}}}

   \titlerunning{Optically variable AGN in the VST-COSMOS field}
   \authorrunning{D. De Cicco et al.}

\institute{Instituto de Astrof\'{i}sica, Pontificia Universidad Cat\'{o}lica de Chile, Av. Vicu\~{n}a    Mackenna 4860, 7820436 Macul, Santiago, Chile    
\\e-mail: demetradecicco@gmail.com, ddecicco@astro.puc.cl
\and
 Millennium Institute of Astrophysics (MAS), Nuncio Monse\~nor Sotero Sanz 100, Providencia, Santiago, Chile     
\and
 Space Science Institute, 4750 Walnut Street, Suite 2015, Boulder, CO 80301, USA 
 \and
 Department of Physics, University of Napoli ``Federico II'', via Cinthia 9, 80126 Napoli, Italy 
 \and
 INFN - Sezione di Napoli, via Cinthia 9, 80126 Napoli, Italy 
 \and
 INAF - Osservatorio Astronomico di Capodimonte, via Moiariello 16, 80131 Napoli, Italy 
 \and
 Department of Astronomy and Astrophysics, 525 Davey Laboratory, The Pennsylvania State University, University Park, PA 16802, USA 
 \and
  Institute for Gravitation and the Cosmos, The Pennsylvania State University, University Park, PA 16802, USA	
 \and
 Department of Physics, 104 Davey Laboratory, The Pennsylvania State University, University Park, PA 16802, USA 
\and
 Departamento de Ciencias Fisicas, Universidad Andres Bello, Avda. Republica 252, Santiago, Chile 
 \and
 Department of Physics and Astronomy, University of the Western Cape, Private Bag X17, 7535 Bellville, Cape Town, South Africa 
 \and
 INAF - Istituto di Radioastronomia, via Gobetti 101, 40129 Bologna, Italy 
        \and
 INAF - Osservatorio Astronomico di Padova, vicolo dell'Osservatorio 5, I-35122 Padova, Italy 
}

   \date{}
  \abstract
   {The survey of the COSMOS field by the VLT Survey Telescope is an appealing testing ground for variability studies of active galactic nuclei (AGN). With 54 $r$-band visits over 3.3 yr and a single-visit depth of 24.6 \emph{r}-band mag, the dataset is also particularly interesting in the context of performance forecasting for the Vera C. Rubin Observatory Legacy Survey of Space and Time (LSST).}
   {This work is the fifth in a series dedicated to the development of an automated, robust, and efficient methodology to identify optically variable AGN, aimed at deploying it on future LSST data.}
   {We test the performance of a random forest (RF) algorithm in selecting optically variable AGN candidates, investigating how the use of different AGN labeled sets (LSs) and features sets affects this performance. We define a heterogeneous AGN LS and choose a set of variability features and optical and near-infrared colors based on what can be extracted from LSST data.}
   {We find that an AGN LS that includes only Type I sources allows for the selection of a highly pure (91\%) sample of AGN candidates, obtaining a completeness with respect to spectroscopically confirmed AGN of $69\%$ (vs. 59\% in our previous work).\\ 
   The addition of colors to variability features mildly improves the performance of the RF classifier, while colors alone prove less effective than variability in selecting AGN as they return contaminated samples of candidates and fail to identify most host-dominated AGN.\\
   We observe that a bright ($r \lesssim 21$ mag) AGN LS is able to retrieve candidate samples not affected by the magnitude cut, which is of great importance as faint AGN LSs for LSST-related studies will be hard to find and likely imbalanced.\\
   We estimate a sky density of $6.2\times10^{6}$ AGN for the LSST main survey down to our current magnitude limit.}
   {}

   \maketitle

\section{Introduction}
\label{section:intro}
Time-domain astronomy is entering a new era marked by the availability of a new generation of telescopes designed to perform wide, deep, and high-cadence surveys of the sky. In the context of optical and near-infrared (NIR) variability studies focused on active galactic nuclei (AGN), the revolution will be pioneered by the Legacy Survey of Space and Time (LSST; see, e.g., \citealt{lsst,ivezich19}) which will be conducted with the Simonyi Survey Telescope at the Vera C. Rubin Observatory. Studies of AGN with LSST should: lead to a dramatic improvement in the determination of the constraints on the AGN demography and luminosity function, and hence on the accretion history of supermassive black holes (SMBHs), over cosmic time; provide novel constraints on the physics and structure of AGN and their accretion disks, as well as the mechanisms by which gas feeds them; and, given the reach to high redshifts, allow for the investigation of the formation and coevolution of SMBHs, their host galaxies, and their dark matter halos over most of cosmic time. 

The main LSST survey (Wide-Fast-Deep survey, WFD) will focus on an ${\approx}$18,000 sq. deg. area, which is expected to be surveyed ${\approx}1,000$ times in 10 yr, using about 90\% of the observing time. About 2--4\% of the remaining time will be devoted to ultra-deep surveys of well-known areas, collectively referred to as deep drilling fields (DDFs; e.g., \citealt{ddf,scolnic}), where extensive multiwavelength information is available from previous surveys. The first 10 yr observing program includes a proposal for high-cadence (up to ${\sim}14,000$ visits) multiwavelength observations of a ${\approx}9.6$ sq. deg. area per DDF, down to \emph{ugri} ${\sim}28.5$, \emph{z} ${\sim}28$, and \emph{y} ${\sim}27.5$ mag coadded depths. Given all of that, the DDFs will be excellent laboratories for AGN science. The current selection of DDFs includes the Cosmic Evolution Survey (COSMOS; \citealt{scoville07b})  field, one of the best studied extragalactic survey regions in the sky. The search for optically variable AGN in COSMOS is the subject of the present work, which is the fifth in a series aimed at constraining the performance of variability selection in the DDFs, making use of \emph{r}-band data from the SUpernova Diversity And Rate Evolution (SUDARE; \citealt{botticella13}) survey by the VLT Survey Telescope (VST; \citealt{VST}).

\citealt{decicco19} (hereafter, \dd) presented an AGN variability study in COSMOS over a 1 sq. deg. area; the dataset consists of 54 visits covering an observing baseline of ${\approx}$3.3 years. The adopted approach recovered a high-purity ($> 86$\%) sample of AGN, comprising 59\% of the X-ray detected AGN with spectroscopic redshifts. The number of confirmed AGN and the completeness are ${\sim}3$ and ${\sim}4$ times larger, respectively, than the corresponding values obtained in \citealt{decicco15} (hereafter, \ddc), which adopted a similar methodology but studied only the first five months of observations available at that time, obtaining results consistent with those obtained in the \emph{Chandra} Deep Field-South in \citet{falocco} and \citet{poulain}. \citetalias{decicco19} also demonstrated that a high-cadence sampling should be a primary characteristic of AGN monitoring campaigns as it has a considerable impact on the detection efficiency (e.g., adopting only ten visits instead of 54 over the 3.3 yr baseline, we would expect to recover only ${\sim}20$\% of the AGN that were identified and confirmed using 54 visits; see Fig. 10 in \dd).

Wide-field surveys such as LSST will provide information about millions of sources per night. The availability of reliable classification methods, as well as effective means to analyze the physical properties of the returned catalogs of sources, will hence be crucial. In response to the technological and scientific challenges posed by this new era of big data science, there has been an exponential increase in the development of fast and automated tools for data visualization, analysis, and understanding over the past decade, as classical techniques have often proven inadequate. In this context, there has been a natural evolution toward the use of machine-learning (ML) algorithms, and these are now widely used in the astronomical community. Some ML algorithms rely on supervised training, via ``labeled'' sets (LSs) of data, that is to say, samples of objects whose classification is known, and characterized by means of a set of features selected on the basis of the properties of interest. The training process heavily depends on the chosen features and on the so-called balance of the selected LS, which should ideally sample in the most complete and unbiased way possible the source population to be studied through broad and homogeneous coverage of the parameter space (in practice, this is never possible). The investigation of the performance of the algorithm when different features are used and the availability of adequate LSs are therefore essential in view of the revolutionary amount of data the astronomical community is about to be delivered.

In this work we test the performance of a random forest (RF; \citealt{Breiman2001}) classifier to identify optically variable AGN in the same dataset used in \dd. \citet{sanchezsaez} demonstrated that an RF algorithm making use of variability plus optical color features is able to identify variable AGN with 81.5\% efficiency. The present project applies a similar method to the VST-COSMOS dataset. Although smaller in size (${\sim}1$ dex), this dataset has the great advantage of probing ${\sim}1$ dex fainter sources (down to \emph{r(AB)} = 23.5 mag) compared to \citealt{sanchezsaez} (\emph{r} ${\sim}21$ mag) and most other works on AGN variability to date \citep[e.g.,][]{Schmidt, graham, simm}. It therefore allows for the exploration of areas of the parameter space that have been poorly investigated so far. Indeed, the VST-COSMOS dataset is to date one of the few that take advantage both of considerable depth and high observing cadence, two fundamental requirements for variability surveys that are often mutually exclusive when data from ground-based observatories are used.

With this project we mean to answer a series of questions in view of future AGN studies based on LSST data. We aim to characterize the performance of our classifier determining how different AGN LSs and different sets of features affect AGN selection in terms of purity, completeness, and size of the candidate samples. In particular, we aim to identify the most relevant features to be used for AGN classification, and to characterize the AGN samples obtained when using variability features alone, or coupled with color features, thus sketching out lower-limit expectations for LSST performance. Our analysis is centered on the use of data that LSST will provide, that is to say light curves and optical and NIR colors.

The paper is organized as follows: Section \ref{section:vst} provides a description of our dataset, of the variability and color features used in the analysis, and of the selected LS of sources. Section \ref{section:RF} introduces RF classifiers and characterizes their performance; Section \ref{section:tests} investigates how the inclusion of different AGN types in the LS, and the adoption of different features, can potentially affect the recovery of the labeled and unlabeled sets. Section \ref{section:results} is dedicated to the classification of the unlabeled set and focuses on the sample of AGN candidates obtained by the RF classifier elected as most suitable for this analysis. We characterize this sample, describe how we validate it making use of different diagnostics, and compare our findings with results from \dd. We also include a description of the results obtained from two additional classifiers of interest in two dedicated subsections. Section \ref{section:MIR_col} presents the results obtained including a mid-infrared (MIR) color as an additional feature, on the basis of the data that will be available for LSST-related studies of AGN. We summarize our findings and draw our conclusions in Section \ref{section:discussion}.

\section{VST-COSMOS dataset}
\label{section:vst}
This work makes use of the same dataset used in \citetalias{decicco19}, consisting of $r$-band observations from the three observing seasons (hereafter, seasons) of the COSMOS field by the VST. The telescope allows imaging of a field of view (FoV) of $1^\circ\times1^\circ$ in a single pointing (the focal plane scale is $0.214$\arcsec/pixel). The three seasons cover a baseline of 3.3 yr, from December 2011 to March 2015, and consist of 54 visits in total, with two gaps; detailed information about the dataset can be found in Table 1 from \citetalias{decicco19}. As discussed in \citetalias{decicco19}, they excluded 11 out of the 65 visits that constitute the full dataset. We refer the reader to \citetalias{decicco15} for details about the process of exposure reduction and combination, performed making use of the VST-Tube pipeline \citep{grado}, and for source extraction and sample assembly. VST-Tube magnitudes are in the AB system. 

Observations in the $r$-band are characterized by a three-day observing cadence, depending on observational constraints. The visit depth is $r \lesssim 24.6$ mag for point sources, at a ${\sim}5\sigma$ confidence level. This makes our dataset particularly interesting in the context of studies aimed at LSST performance forecasting, as the depth of LSST single visits is expected to be approximately the same as the VST images. 

The sample consists of 22,927 sources detected in at least 50\% of the visits in the dataset (i.e., there are at least 27 points in their light curves) and with a $1\arcsec$-radius aperture magnitude of \emph{r} $\leq 23.5$ mag. We start from the same sample for our analysis but, while \citetalias{decicco19} focused on a sample of variability-selected sources, here we use the entire sample; as a consequence, we need to exclude from it problematic sources, as was done in \citetalias{decicco19} for the variable sample. These problematic sources are mainly objects blended with a neighbor in at least some of the visits (those with poor seeing), making it substantially harder to measure fluxes correctly. 

To identify blended sources, we rely on the COSMOS Advanced Camera for Surveys (ACS) catalog \citep{koekemoer, scoville} from the \emph{Hubble Space Telescope} (\emph{HST}), which was constructed from 575 ACS pointings and includes counterparts for 22,747 out of our 22,927 sources. We match the catalog with itself within a $1.5\arcsec$ radius, in order to identify 1,171 sources with close neighbors and, in 90\% of the cases where we find neighbors, there is just one within the defined circular area. We adopt empirical criteria similar to \citetalias{decicco15} and \citetalias{decicco19}, but somewhat less conservative, whereby we require the maximum centroid-to-centroid distance between two neighbors to be $1.5\arcsec$ if their magnitude difference is $<1.5$ mag. The remaining 10\% of the cases with neighbors consist of sources with two to five neighbors. Following the same criterion adopted for source pairs, and after a visual check, we reject all of the sources with more than one neighbor.
The described cleaning process leads to the exclusion of 1,074 sources (4.7\% of the initial sample of 22,927 objects), returning a sample of 21,852 sources.

In Sects. \ref{section:var_feat}, \ref{section:stellarity}, and \ref{section:col_feat}, we introduce the set of features that we use to build our RF algorithm. Since we are interested in including a set of optical and NIR colors as features, we require that VST-COSMOS objects have $5\sigma$ detections available in $uBrizy$ imaging\footnote{$g$-band magnitudes are not available in COSMOS2015, hence we use the Subaru $B$ band, which is similar though not identical in wavelength coverage, as a replacement.} provided by the COSMOS2015 catalog \citep{laigle}, and also that they have a match in the already mentioned COSMOS ACS catalog, as this provides a morphological indicator that we use as an additional feature. These requirements reduce the available number of sources to 20,670 (hereafter, the main sample). In Sect. \ref{section:MIR_col} we also test the inclusion of a MIR color as an additional feature.

\begin{figure*}[ht!]
 \centering
\subfigure
            {\includegraphics[width=9cm]{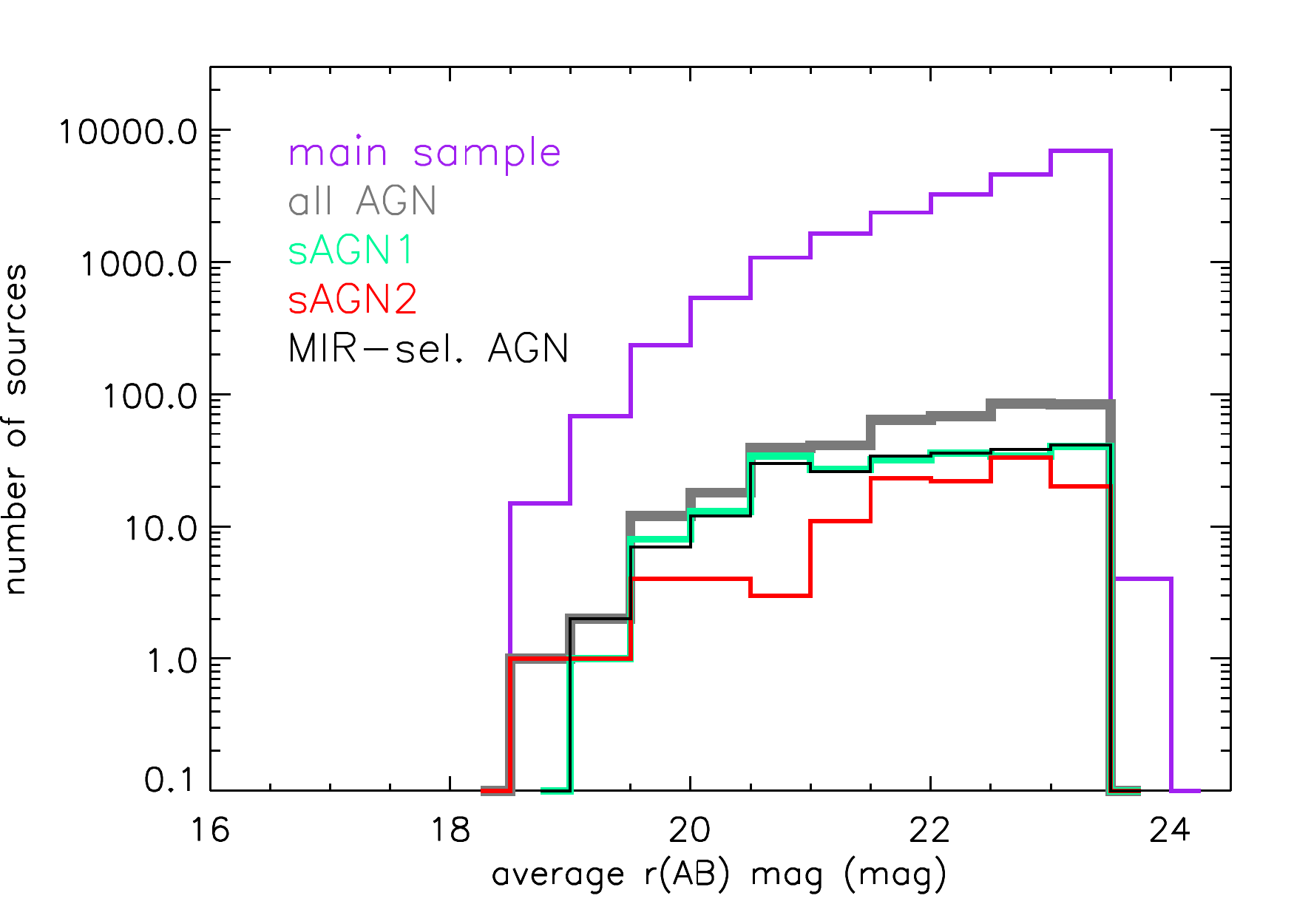}}
\subfigure
            {\includegraphics[width=9cm]{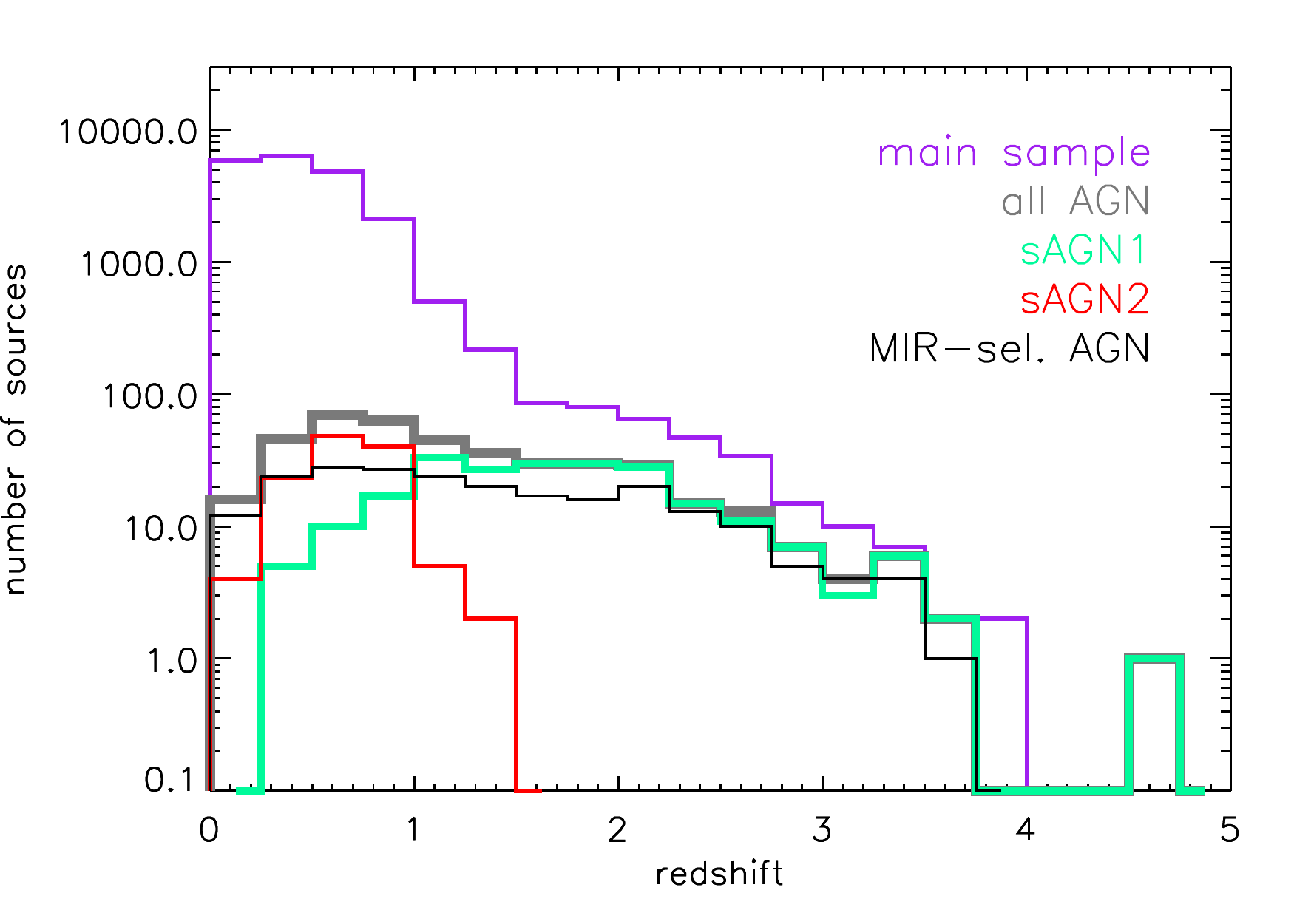}}
\subfigure
            {\includegraphics[width=9cm]{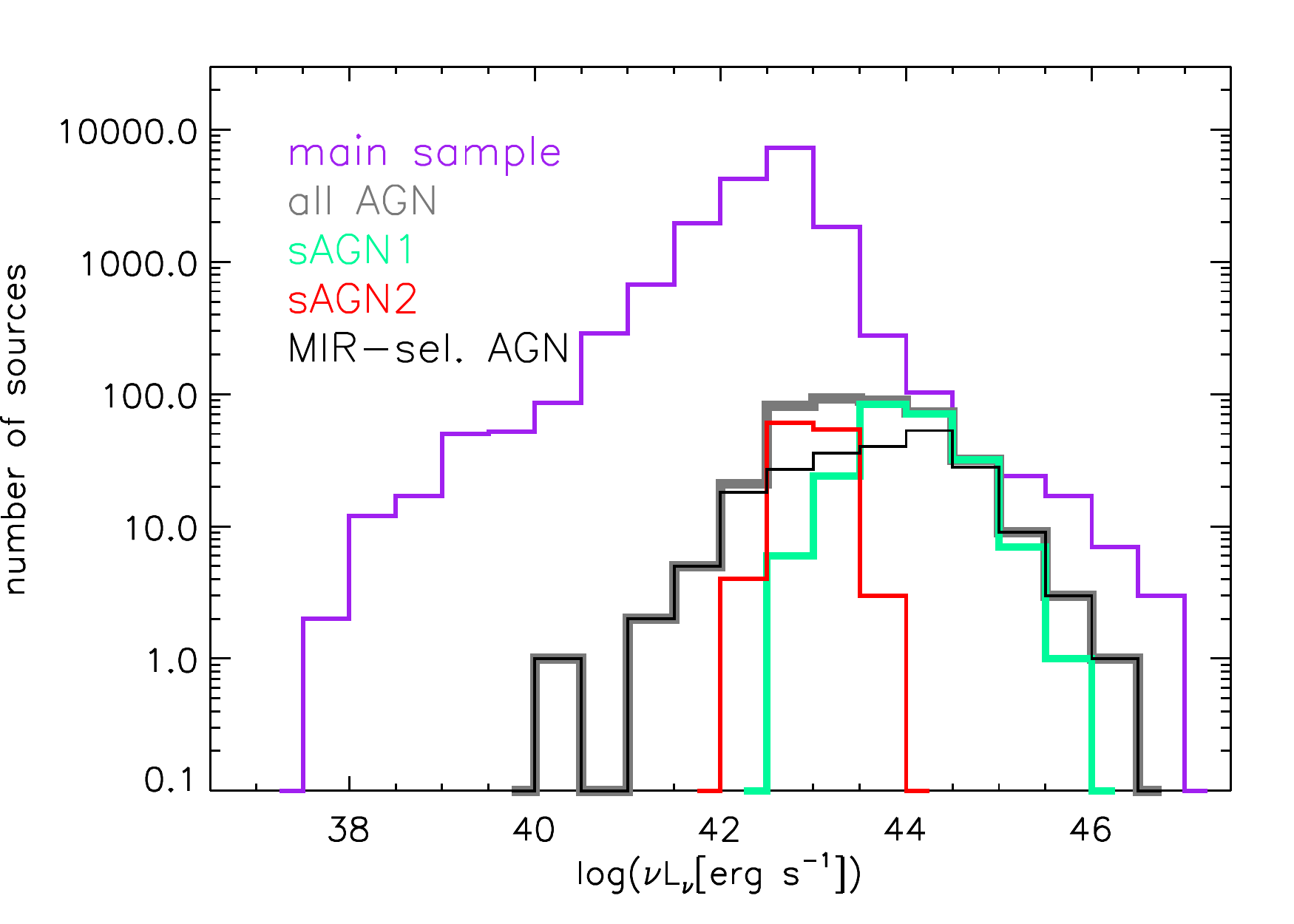}}\\
\subfigure
            {\includegraphics[width=9cm]{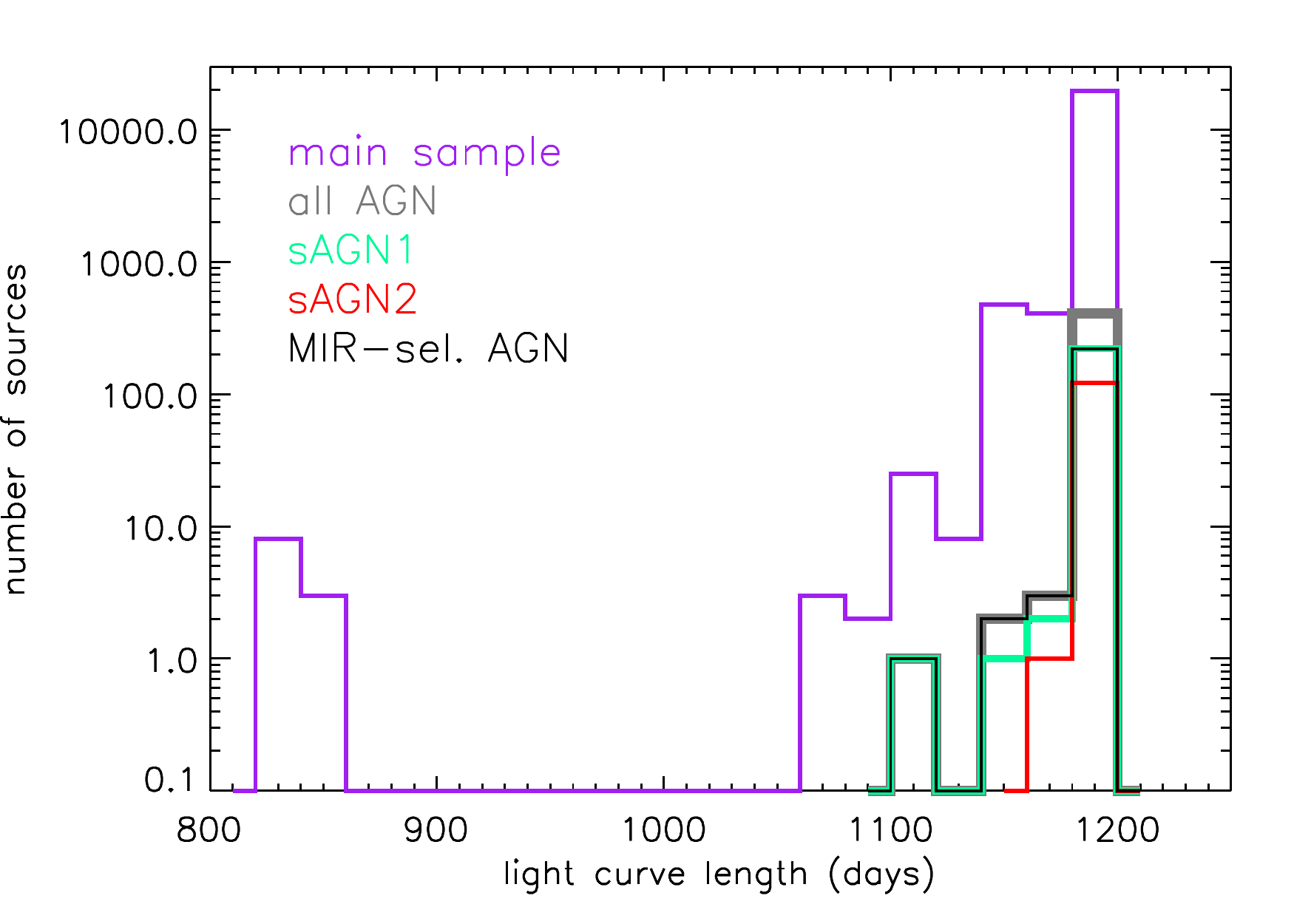}}
\subfigure
            {\includegraphics[width=9cm]{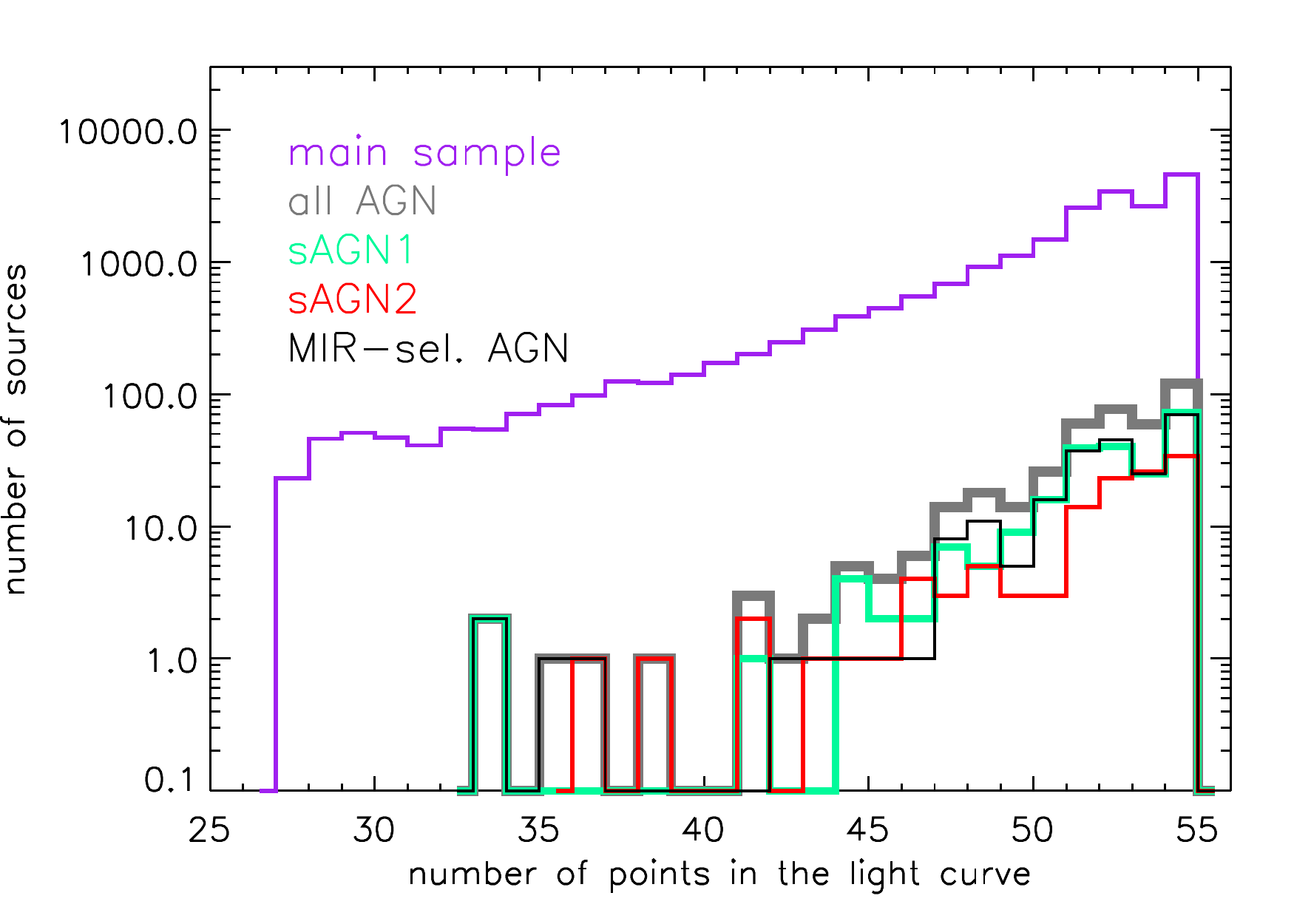}}
   \caption{\footnotesize{Distribution of magnitude (\emph{upper left panel}), redshift (\emph{upper right}), $r$-band luminosity (\emph{center}), length of the light curves (\emph{lower left}), and number of points in the light curves (\emph{lower right}) for the sources in the main sample (purple), the whole AGN LS (gray), and for the various types of AGN included in the LS, which will be introduced in Sect. \ref{section:labeled_set}: Type I AGN (sAGN1, green), Type II AGN (sAGN2, red), and MIR AGN (black). A redshift estimate, either spectroscopic or photometric, is available for all but 2.5\% of the sources in the main sample.}}\label{fig:hists_with_colors}
   \end{figure*}

In Fig. \ref{fig:hists_with_colors}, we provide a set of histograms characterizing the sources in the main sample. We show the average magnitude, redshift, and $r$-band luminosity distributions, as well as the distributions of the length and the number of points of each light curve. The light curves of the sources in the main sample have different lengths, but 88\% of them cover the full length of the baseline, that is, 1,187 days (${\approx}$3.3 yr). Only 11 sources have a light curve shorter than 850 days. 
The figure also shows that a large fraction of the sources are detected in almost all the visits. Indeed, 22\% of the sources have 54 points (i.e., the maximum number available based on our dataset) in their light curves, while 95\% of the main sample have at least 40 points.
We note that 91\% of the sources have \emph{r} $> 21$ mag, and hence are beyond the typical depth of most variability studies to date, as mentioned in Sect. \ref{section:intro}. Our dataset therefore gives us the opportunity to investigate how well the variability-based identification of AGN performs when dealing with fainter objects.

\subsection{Available redshift estimates}
Redshift estimates are available for all but four objects in the sample, and come from different COSMOS catalogs, here listed according to the order we adopt when assigning a redshift to our sources. We note that we accept only reliable redshift estimates, according to the reliability criteria defined for each catalog (see the corresponding reference papers for details).
\begin{itemize}
\item[--] The DEIMOS 10K Spectroscopic Survey Catalog of the COSMOS Field \citep{hasinger}: contains reliable spectroscopic redshifts for 8,415 of 10,718 objects (79\%) selected through heterogeneous criteria in COSMOS, using the DEep Image Multi-Object Spectrograph, DEIMOS). We match redshifts for 1,607 of the sources in the main sample.\\
\item[--] The catalog of optical and NIR counterparts of the X-ray sources in the \emph{Chandra}-COSMOS Legacy Catalog \citep{marchesi, civano16}: contains reliable spectroscopic redshifts for 2,151 out of 4,016 X-ray selected sources (54\%) and photometric redshifts for 3870 sources (96\% of the sample). We match unique spectroscopic redshifts for 423 additional sources in the main sample.\\
\item[--] The zCOSMOS Bright Spec Catalog \citep{Lilly09}: it was derived from the spectroscopic survey of 20,689 $I$-band selected galaxies (redshift $z < 1.2$, DR3-bright) observed with the VIsible Multi-Object Spectrograph (VIMOS), and provides redshift estimates for ${\approx}$93\% of them. We match unique redshifts for 7,529 additional sources in the main sample.\\
\item[--] The COSMOS2015 catalog \citep{laigle}: contains photometry and physical parameters for ${\approx}$1,200,000 sources in several filters, and includes photometric redshift estimates for ${\approx}$98\% of them. We find photometric redshifts for 8,364 sources from the main catalog with no previous available match.\\ 
\item[--] The already mentioned \emph{Chandra}-COSMOS Legacy Catalog, where we find photometric redshifts for 35 additional sources.\\
\item[--] The COSMOS Photometric Redshift Catalog \citep{Ilbert1}: contains photometric redshift estimates for 385,065 sources with $i < 25$ mag, computed making use of 30 broad, intermediate, and narrow bands, from the ultraviolet (UV) to the NIR. Here we find photometric redshifts for an additional 2,195 sources from the main sample with no previous available match.
\end{itemize}
In summary, among 20,670 sources in the main sample, 9,559 (46.2\%) have spectroscopic redshifts, 10,594 sources (51.3\%) have photometric redshifts, and 517 (2.5\%) do not have reliable redshift estimates: they do not have counterparts in most of the catalogs used here, since they fall in masked areas; for 33 of them we find redshift estimates in the zCOSMOS Bright Spec Catalog, but they are flagged as unreliable, so we do not take them into account. In order to evaluate the quality of photometric redshifts, where available we compare each spectroscopic measurement with each photometric measurement for the same source, and find a very good agreement, the average difference being $|z_{spec}-z_{phot}|\leq0.02\pm0.07$. 

Redshift estimates allow for the computation of the observed $r$-band luminosity $\nu L_\nu$ of our sample of sources. This is in the range $10^{37}-10^{46}$ erg s$^{-1}$ (Fig. \ref{fig:hists_with_colors}).

\subsection{Variability features}
\label{section:var_feat}
As an initial step to classify the main sample, we select a set of features that will be used to build our RF classifier.  
We adopt a set of 29 extraction features used to characterize the variability (or lack thereof) observed in the light curves of the sources. We list these in the first two blocks of Table \ref{tab:features} and briefly describe some of them below. About half the features are identical to the ones used in \citet{sanchezsaez}, while the full set arises from the development of the ALeRCE\footnote{\url{http://alerce.science/}} (Automatic Learning for the Rapid Classification of  Events; \citealt{forster}) broker light curve classifier \citep{sanchezsaez20} and, in particular, some features, such as the $IAR_\phi$, are there used for classification for the first time. 

We first focus on the group of features reported in the upper block of the table. One way to characterize the time dependence of the variability of a sample of sources is through their structure function (SF). Several definitions for the SF have been proposed in the literature \citep[e.g.,][and references therein]{Simonetti, diClemente,Schmidt, graham}. In essence, the SF measures the ensemble root mean square (rms) of the magnitude variation as a function of the time lag between different observations, computed as an average over a given time interval $\Delta t$, for all the pairs of observations characterized by a time lag $t_2 - t_1$ falling in the time interval $\Delta t$. A very general definition is therefore: 
\begin{equation}
SF=\langle[\mbox{mag($t_2$)}-\mbox{mag($t_1$)}]^2\rangle\mbox{   ,}
\label{eqn:mag_diff}
\end{equation} 
where mag($t_1$) and mag($t_2$) are two observations of the same source at different times, with $t_2 - t_1 \leq \Delta t$. Some specific definitions also take into account the noise contribution, the influence of outliers, and/or assume a Gaussian distribution both for intrinsic variability and noise. Several works \citep[e.g.,][]{VandenBerk} have proposed a power-law model for the SF of quasars: 
\begin{equation}
SF(\tau) = A_{SF}\left(\frac{\tau}{1\mbox{yr}}\right)^{\gamma_{SF}}\mbox{ .}
\label{eqn:sf_power_law}
\end{equation} 
This is characterized by two parameters: the variability amplitude $A_{SF}$ and the exponent $\gamma_{SF}$, which represents the logarithmic gradient of the mean magnitude variation. These two parameters are included in the features we compute for our sample of sources and, specifically, $A_{SF}$ is computed over a 1 yr timescale.

Quasar light curves are also often modeled as damped random walk (DRW) processes \citep[e.g.,][]{kelly,macleod10}. The model is characterized by two parameters: the relaxation time, that is to say the time $DRW\_\tau$ for the magnitude variations in a source light curve to become uncorrelated, and the variability amplitude $DRW\_\sigma$, computed from the light curve over times $t <$ $DRW\_\tau$. We obtain these two parameters by Gaussian process regression with a Ornstein-Uhlenbeck kernel, following \citet{graham17}, and include both parameters in our feature list.

We also include the $P_{var}$ parameter, which measures the detection confidence of variability \citep{mclaughlin}. Specifically, we compute from the source light curve the $\chi^2$ with respect to the mean magnitude value and then compute the probability $Q$ to obtain by chance a higher value of the $\chi^2$ for a non-variable source. $P_{var}$ is defined as $1 - Q$, and is therefore a measure of the intrinsic variability of a source.

An additional measure of the intrinsic variability amplitude of a source is given by the excess variance $\sigma_{rms}$ \citep[e.g.,][and references therein]{allevato}. This is defined as
\begin{equation}
\sigma_{rms} = \frac{\sigma_{LC}^2 - \overline\sigma_{mag}^2}{\overline{mag}^2}\mbox{ ,}
  \label{eqn:ev}
\end{equation}
where $\sigma_{LC}^2$ is the total variance of the light curve, $\overline\sigma_{mag}^2$ is the mean square photometric error associated with the magnitude measurements, and $\overline{mag}^2$ is the squared average magnitude. This is therefore an index comparing the total variance of a light curve and the variance expected based on the photometric errors measured from the light curve.

The features reported in the second block of the table are from the Feature Analysis for Time Series (FATS) Python library\footnote{\url{http://isadoranun.github.io/tsfeat/FeaturesDocumentation.html}} \citep{nun}, dedicated to feature extraction from astronomical light curves and time series in general. They include some well-known basic features, such as the standard deviation of a light curve, as well as features with more complex definitions. Following \citet{sanchezsaez20}, we chose not to include the mean magnitude in the feature list, as possible biases in the magnitude distribution of the LS can affect it and lead to biased classifications of fainter or brighter sources.
Table \ref{tab:features} reports a short description and reference paper for each feature. 

\begin{table*}[tb]
\caption{List of the features used in this work. The first two blocks of the table report variability features; \texttt{class\_star} is the only morphology feature used. The bottom part of the table reports the color features used, where \texttt{ch21} is the only MIR color used, while the others are optical or NIR colors.}
\label{tab:features}      
\centering
 \renewcommand\arraystretch{1.2}
 \footnotesize
 \resizebox{\textwidth}{!}{
 \begin{tabular}{l l l}
\hline Feature & Description & Reference\\
\hline
\ \texttt{$A_{SF}$} & rms magnitude difference of the SF, computed over a 1 yr timescale & \citet{Schmidt}\\
\ \texttt{$\gamma_{SF}$} & Logarithmic gradient of the mean change in magnitude & \citet{Schmidt}\\
\ \texttt{GP\_DRW\_$\tau$} & Relaxation time $\tau$ (i.e., time necessary for the time series to become uncorrelated), & \citet{graham17}\\
\  & from a DRW model for the light curve & \\
\ \texttt{GP\_DRW\_$\sigma$} & Variability of the time series at short timescales ($t << \tau$), & \citet{graham17}\\
\ & from a DRW model for the light curve & \\
\ \texttt{ExcessVar} & Measure of the intrinsic variability amplitude & \citet{allevato}\\
\ \texttt{$P_{var}$} & Probability that the source is intrinsically variable & \citet{mclaughlin}\\
\ \texttt{$IAR_\phi$} & Level of autocorrelation using a  discrete-time representation of a DRW model & \citet{eyheramendy18}\\
\hline
\ \texttt{Amplitude} & Half of the difference between the median of the maximum 5\% and of the minimum & \citet{richards11}\\
\ & 5\% magnitudes & \\
\ \texttt{AndersonDarling} & Test of whether a sample of data comes from a population with a specific distribution & \citet{nun}\\
\ \texttt{Autocor\_length} & Lag value where the autocorrelation function becomes smaller than $\eta^e$ & \citet{kim11}\\
\ \texttt{Beyond1Std} & Percentage of points with photometric mag that lie beyond 1$\sigma$ from the mean & \citet{richards11}\\
\ \texttt{$\eta^e$} & Ratio of the mean of the squares of successive mag differences to the variance & \citet{kim14}\\
\ & of the light curve & \\
\ \texttt{Gskew} & Median-based measure of the skew & -\\
\ \texttt{LinearTrend} & Slope of a linear fit to the light curve & \citet{richards11}\\
\ \texttt{MaxSlope} & Maximum absolute magnitude slope between two consecutive observations & \citet{richards11}\\
\ \texttt{Meanvariance} & Ratio of the standard deviation to the mean magnitude & \citet{nun}\\
\ \texttt{MedianAbsDev} & Median discrepancy of the data from the median data & \citet{richards11}\\
\ \texttt{MedianBRP} & Fraction of photometric points within amplitude/10 of the median mag & \citet{richards11}\\
\ \texttt{MHAOV Period} & Period obtained using the \texttt{P4J} Python package (\url{https://github.com/phuijse/P4J}) & \citet{Huijse18}\\
\ \texttt{PairSlopeTrend} & Fraction of increasing first differences minus the fraction of decreasing first differences & \citet{richards11}\\
\ & over the last 30 time-sorted mag measures & \\
\ \texttt{PercentAmplitude} & Largest percentage difference between either max or min mag and median mag & \citet{richards11}\\
\ \texttt{Q31} & Difference between the third and the first quartile of the light curve & \citet{kim14}\\
\ \texttt{Period\_fit} & False-alarm probability of the largest periodogram value obtained with LS & \citet{kim11}\\
\ \texttt{$\Psi_{CS}$} & Range of a cumulative sum applied to the phase-folded light curve & \citet{kim11}\\
\ \texttt{$\Psi_\eta$} & $\eta^e$ index calculated from the folded light curve
 & \citet{kim14}\\
\ \texttt{R$_{cs}$} & Range of a cumulative sum & \citet{kim11}\\
\ \texttt{Skew} & Skewness measure & \citet{richards11}\\
\ \texttt{Std} & Standard deviation of the light curve & \citet{nun}\\
\ \texttt{StetsonK} & Robust kurtosis measure & \citet{kim11}\\
\hline
\ \texttt{class\_star} & \emph{HST} stellarity index & \citet{koekemoer},\\
\  &  & \citet{scoville}\\
\hline
\ \texttt{u-B} & CFHT $u$ magnitude -- Subaru $B$ magnitude & \citet{laigle}\\
\ \texttt{B-r} & Subaru SuprimeCam $B$ mag -- Subaru SuprimeCam $r$+ mag & \citet{laigle}\\
\ \texttt{r-i} & Subaru SuprimeCam $r+$ mag -- Subaru SuprimeCam $i+$ mag & \citet{laigle}\\
\ \texttt{i-z} & Subaru SuprimeCam $i+$ mag -- Subaru SuprimeCam $z$++ mag & \citet{laigle}\\
\ \texttt{z-y} & Subaru SuprimeCam $z$++ mag -- Subaru Hyper-SuprimeCam $y$ mag & \citet{laigle}\\
\hline
\ \texttt{ch21} & \emph{Spitzer} 4.5 $\mu$m (\emph{channel2}) mag -- 3.6 $\mu$m (\emph{channel1}) mag & \citet{laigle}\\
\hline
\end{tabular}
}
\end{table*}

\subsection{Morphology feature}
\label{section:stellarity}
At present, the only morphology feature we use is the stellarity index. This is a neural network-based star/galaxy classifier computed via \emph{SExtractor} \citep{bertin} and ranging from 0 to 1 (extended to unresolved sources, respectively). Specifically, here we use the stellarity index provided by the COSMOS ACS catalog (\texttt{class\_star} parameter), which is based on F814W imaging. In general, since the resolution of ground-based imaging is lower than \emph{HST} resolution, we cannot obtain a quality as high as for the \emph{HST}-based stellarity; nonetheless for LSST we can expect better resolution (as good as ${\sim}0.3-0.4\arcsec$ for some subsets) than what we currently obtain from ground-based telescopes. In addition, the \emph{Euclid} \citep[e.g.,][]{euclid} and \emph{Roman} \citep[e.g.,][and references therein]{roman} space missions, whose launches are expected in 2022 and 2025, respectively, will eventually provide partial coverage at high resolution ($0.1-0.2\arcsec$). Joint \emph{Euclid} plus LSST analyses at the pixel level would provide LSST with a high-resolution model for deblending and differential chromatic refraction corrections.

\subsection{Color features}
\label{section:col_feat}
Color-color diagrams obtained using optical and NIR colors allow for the identification of specific loci occupied by pure AGN, pure galaxies, and stars, thus favoring the separation of these three classes of sources \citep[e.g.,][]{richards01}. 
In the following, we make use of magnitudes from the COSMOS2015 catalog to build a set of color features (bottom part of Table \ref{tab:features}): we choose to focus the analysis on colors that will be available in the LSST dataset. The selected colors are therefore \texttt{u-B}, \texttt{B-r}, \texttt{r-i}, \texttt{i-z}, and \texttt{z-y}, and we use them together with the other features from Table \ref{tab:features}. 

Magnitudes are available from the already mentioned COSMOS2015 catalog, where a single static value per band is reported for each detected source. Magnitudes come from different catalogs obtained with different instruments, but a point spread function homogenization is made among all bands in the catalog (see \citealt{laigle}). Consistent with the VST $r$-band magnitudes used in this work, we select $2\arcsec$-diameter apertures for each band. We note again that the sources with all the required magnitudes and a measurement of the stellarity index available number 20,670.

\subsection{Labeled set}
\label{section:labeled_set}
Our principal aim is the identification of AGN, and hence we do not consider subclassifications among non-AGN sources in the present work. We do however want to understand how different training sets affect the classifier. Thus, we establish a parent LS of sources here, which we further subdivide in various subsets to help interpret the results. Our parent LS consists of three classes of sources: AGN, stars, and inactive galaxies. The AGN class consists of 414 objects, selected on the basis of different properties: 
\begin{itemize}
    \item[--]225 Type I AGN: these correspond to all the VST-COSMOS sources with a counterpart in the \emph{Chandra}-COSMOS Legacy Catalog and there classified through optical spectroscopy as Type I AGN (hereafter, sAGN1).\\
    \item[--]122 Type II AGN:  these correspond to all the VST-COMOS sources with a counterpart in the \emph{Chandra}-COSMOS Legacy Catalog and there classified through optical spectroscopy as Type II AGN (hereafter, sAGN2; for details about how Type I and II AGN are identified in the \emph{Chandra}-COSMOS Legacy Catalog, see \citealt{marchesi} or Sect. 4.2 of \dd.)\\
    \item[--]67 additional VST-COSMOS sources lacking spectroscopic classification but classified as AGN according to the MIR selection criterion of \citet{donley}, which define an AGN locus in the plane comparing $8.0\mbox{ }\mu m/4.5\mbox{ }\mu m$ and $5.8\mbox{ }\mu m/3.6\mbox{ }\mu m$ flux ratios (hereafter, Donley non-sAGN). This MIR selection criterion is typically not biased against the identification of heavily obscured Type II AGN, as the MIR emission of AGN originates from hot dust and is much less affected by intervening obscuration. The criterion also proves effective in identifying AGN missed by X-ray surveys, hence representing a useful complement to X-ray selection. We note that, in total, there are 226 sources in the Donley selection region: apart from the above-mentioned 67 objects, there are 135 Type I AGN and 24 Type II AGN that are already included in the LS as part of the subsamples of sources with spectroscopic classification. Hereafter, we will use the label ``MIR AGN'' when referring to the full sample of 226 sources in the Donley selection region. 
\end{itemize}
Optical variability is typically biased toward Type I AGN, as it corresponds to flux variations in the AGN inner regions \citep[e.g.,][and references therein]{padovani}; nevertheless the identification of Type II AGN is still possible. We therefore include Type II AGN, as well as MIR AGN, in our initial LS, and investigate the impact of their inclusion upon the performance of the classifier. 

In Fig. \ref{fig:hists_with_colors}, we show the distributions of average magnitude, redshift, luminosity, light curve length, and number of points in the light curve for the whole AGN LS and the three subsamples of sAGN1, sAGN2, and MIR AGN, in addition to the same distributions shown for the sources in the main sample. It is apparent that each subsample spans most of the magnitude range of the main sample, with good coverage of the faint end. As for redshift, sAGN2 have $z < 1.45$, while sAGN1 and MIR AGN cover the whole range of the main sample. The luminosities of sAGN1 are in the range ${\approx}4\times10^{42} - 4\times10^{46}$ erg s$^{-1}$, while sAGN2 are characterized by lower luminosities, in the range ${\approx}10^{42} - 5\times10^{44}$ erg s$^{-1}$, and MIR AGN cover a broader range, their luminosities being ${\approx}2\times10^{39} - 4\times10^{46}$ erg s$^{-1}$. 

Starting from this parent LS of AGN, we use the following sets of sources for the tests that will be described in Sect. \ref{section:ls_tests}: 
\begin{itemize}
    \item[--] Only Type I AGN (sAGN1): this is a pure sample of unobscured AGN and, as stated above, we expect this selection to be the most suitable for this analysis.\\
    \item[--] Only Type II AGN (sAGN2): this is a pure sample of obscured AGN, and we choose it to analyze the performance of the classifier when using only this class of sources, generally not easy to identify through optical variability.\\
    \item[--] sAGN1 plus sAGN2: this is our most reliable selection of AGN, and we aim to assess to what extent the addition of Type II to Type I AGN improves or worsens the performance of the classifier.\\
    \item[--] sAGN1 plus sAGN2, with \emph{r} $\leq 21$ mag: when LSST data will be available, we will not have spectroscopic LSs of AGN covering as faint magnitude ranges as LSST data. Therefore, we mean to investigate whether a bright sample of AGN used as LS is effective in retrieving a sample of AGN candidates including fainter sources. Furthermore, we mean to assess whether the inclusion of faint sources in our LS affects our classification, increasing the fraction of misclassified sources.\\
    \item[--] Only MIR AGN: this is a heterogeneous selection of sources, typically including many obscured AGN, hence we do not expect optical variability to favor their identification and aim to verify our guess.\\
    \item[--] Full LS of AGN: this will allow us to analyze the performance of the classifier when using a more complete and heterogeneous LS of AGN compared to the previous ones.
\end{itemize}

With regard to non-AGN, we select a set of 1,000 stars from the already mentioned COSMOS ACS catalog. This catalog provides a star/galaxy classifier, the \texttt{mu\_class} parameter, which also allows for rejection of spurious sources, such as cosmic rays \citep{Leauthaud}. \texttt{mu\_class} is defined on the basis of a diagram comparing the \texttt{mag\_auto} and \texttt{mu\_max} parameters (i.e., Kron elliptical aperture magnitude and peak surface brightness above the background level, respectively), where point sources define a sharp locus. \texttt{mu\_class} equals 1 for galaxies, 2 for stars, and 3 for artifacts. In order to be confident that our sample of stars is reliable, we require the sources classified as stars in the COSMOS ACS catalog not be classified as AGN or galaxies in the \emph{Chandra}-COSMOS Legacy Catalog, and we resort to the color-color diagram proposed by \citet{nakos} to further clean the sample. The diagram compares  the \emph{r-z} and \emph{z-K} colors of a sample of sources, and it allows for the discrimination of stars from galaxies, as the first form a sharp sequence while galaxies tend to lie in a more scattered area of the diagram. We used this diagram in our previous works based on VST-COSMOS data (\citetalias{decicco15} and \citetalias{decicco19}). As in \citetalias{decicco19}, we obtain magnitudes in the various bands from the COSMOS2015 catalog. We require our sources to have $r-z$ $\leq 1.5$ mag in order to prevent contamination from galaxies. These criteria return a sample of 1,709 stars, and we randomly select 1,000 of them to be included in our LS. We verified that our selection of stars included objects classified as variable on the basis of the analysis presented in \citetalias{decicco19}: in this way, our LS is not unduly biased against variable stars.

We also select a set of 1,000 ``inactive'' galaxies (i.e., galaxies showing no nuclear activity) from our main sample based on the best-fit templates from \citet{bc03} reported in the COSMOS2015 catalog. We cross-match the selected samples of stars and inactive galaxies with all the available COSMOS catalogs that we are aware of, to be sure that no conflicting classification exists for these sources. The parent LS for all object classes therefore consists of 2,414 sources. 

\section{RF classifiers}
\label{section:RF}
Our dataset consists of the parent LS introduced in Sect. \ref{section:labeled_set} plus a set of sources with no classification, which we aim to classify. Each source in the dataset can be characterized by a number of properties that can be measured or computed from its light curve and ``static'' (or average), non-contemporaneous measurements. The LS is used to establish connections between the source features and their classification. In this way, a classifier algorithm can be trained, and is then able to classify unlabeled sources based on what it learned from the labeled ones. This is known as supervised learning.

The RF algorithm belongs to the supervised learning domain, and is based on the use of decision trees \citep[e.g.,][]{Morgan63, Messenger72}. A decision tree is a predictor named after its tree-like structure. Decision trees are built by means of recursive binary splitting of the source sample based on the characteristics of their features, according to specific criteria that define a cut-point for each split. Each split corresponds to a decision node. Trees are grown top-down, meaning that the data are recursively split into a growing number of nonoverlapping regions. The split goes on until a stopping criterion is fulfilled. The regions thus delimited are the leaves of the tree, and correspond to different classes, defined by the properties of the source features in each region. The sources will therefore be attributed to a specific class depending on the region to which they belong.

An RF classifier combines together an ensemble of trees, working together in order to obtain a single prediction with improved accuracy. Usually a number of bootstrapped sets of sources (training sets) are extracted from the LS and are used in the training process of the algorithm. Decision trees are built from each training set. The various trees are decorrelated, that is, at each split only a limited number of features are taken into account: this reduces the chance of involving always the same features (typically, the most relevant ones) in the splitting process. Both bootstrapping and decorrelation are used in order to reduce the variance of the classification, which is a measure of how much the obtained result depends on the specific training sets used. Both are also responsible for the ``random'' nature of the results obtained from an RF classifier (vs. results from a single decision tree algorithm, which is deterministic). 
Part of the LS is not used in the training process, but saved for the validation of the classifier. Finally, the classifier is fed with the unlabeled sources and returns a classification for them. The final prediction obtained from the ensemble of trees is the average of the predictions from the individual trees.

The code we use to apply an RF algorithm to our sample of sources takes its cue from the one used in \citet{sanchezsaez}. It is based on the use of the Python RF classifier library\footnote{\url{https://scikit-learn.org/stable/modules/generated/sklearn.ensemble.RandomForestClassifier.html}} included in the \emph{scikit-learn} library, which provides a number of tools for machine learning-based data analysis \citep{scikit-learn}.

\subsection{Performance of the classifier}
\label{section:performance}
One major concern is what types of AGN should be included in the LS, and which properties they should be selected on. As discussed in Sect. \ref{section:labeled_set}, one of the aims of this work is to assess how different AGN subsamples (sAGN1/sAGN2/MIR-selected) from the parent LS affect the performance of the classifier.

It is common practice to use a fraction (typically 70\%) of the LS for the training of the classifier, and the remaining fraction (30\%) for the validation. Here the heterogeneity of the AGN in the parent LS, which are selected on the basis of different characteristics, makes such a choice heavily dependent on the fraction of each AGN subsample falling into the training set, resulting in significantly different classifications for the sources in the validation set. In addition, the AGN LS selected for this study comprise an important (dominant) fraction of the total AGN in the region of interest (i.e., the LS is not independent from the total sample), and thus must be considered in purity and completeness estimates. In order to address these issues, we resort to the leave-one-out cross-validation \citep[LOOCV,][]{loocv}: this approach consists of using each source -one at a time- as a single-unit validation set, while all the rest of the LS constitutes the training set. A prediction is therefore made for the excluded source, based on the training set; this is done for each source in the LS so that, in the end, a prediction is available for each source. The full LS is therefore used as both training and validation set. 

The performance of a binary classifier is usually characterized through a number of standard metrics obtained from the results over the validation set, and computed from the so-called confusion matrix (CM). In order to define them, we first introduce the following terms, which correspond to the four frames of the CM obtained for a binary classifier:
\begin{itemize}
\item[--] true positives (TPs): known AGN correctly classified as AGN;
\item[--] true negatives (TNs): known non-AGN correctly classified as non-AGN;
\item[--] false positives (FPs): known non-AGN erroneously classified as AGN;
\item[--] false negatives (FNs): known AGN erroneously classified as non-AGN.
\end{itemize}
The metrics we use here are accuracy ($A$), precision ($P$, also known as purity), recall ($R$, also known as completeness), and $F1$, and are defined as in the following:
\begin{equation*}
A = \frac{\mbox{TPs}+\mbox{TNs}}{\mbox{Tot. Sample}}\mbox{ }
\end{equation*}
provides an overall evaluation of the performance of the classifier, telling how often the classification is correct, both for AGN and non-AGN, hence it is computed with respect to the total sample;
\begin{equation*}
P = \frac{\mbox{TPs}}{\mbox{TPs}+\mbox{FPs}}\mbox{ }
\end{equation*}
tells how often the classification as AGN is correct, hence it is computed with respect to all the sources classified as AGN;
\begin{equation*}
R = \frac{\mbox{TPs}}{\mbox{TPs}+\mbox{FNs}}\mbox{ }
\end{equation*}
tells how often known AGN are classified correctly, hence it is computed with respect to all the known AGN in the sample;
\begin{equation*}
F1 = 2\times\frac{P\times R}{P+R}\mbox{ }
\end{equation*}
is the harmonic mean of $P$ and $R$ and provides a different estimate of the accuracy, taking into account FPs and FNs.

The use of an RF algorithm allows for the assessment of the importance of the various features used in the classification process. In our classifier, the quality of a split is measured through the Gini index \citep{Gini}, which determines the condition of a split, with the aim of increasing the purity of the sample after the split. Essentially, this is obtained through the minimization of the Gini index when comparing its value before and after the split (the lower the value, the purer the node). If the position of the node in the tree is high, the split will affect the purity of a larger fraction of the dataset, so the improvement in the purity is weighted accordingly. 
The feature importance is evaluated taking into account the improvement in the purity due to that feature over the whole tree; this is done for each tree, then an average is performed over the total number of trees. Identifying the most important variables helps understand and potentially improve the model (e.g., we could choose to remove the less important variables in order to shorten the training time without significant loss in the performance).

\section{Finding the most appropriate feature set and labeled set}
\label{section:tests}
As mentioned in Sect. \ref{section:vst}, in this work we investigate the use of variability and color features plus a morphology feature, and test the dependence of the performance of an RF classifier on different classes of AGN adopted as LS.

\subsection{Tests about color relevance}
\label{section:color_test}
In order to assess the relevance of variability and color features in the context of the present analysis, we test the performance of different RF classifiers making use of different sets of features: for each test we always include the variability features and the stellarity index. In addition, we include two to five colors, as detailed below:
\begin{itemize}
\item[--] RF1: only variability plus stellarity index; 
\item[--] RF2: RF1 features plus the two colors \texttt{i-z} and \texttt{z-y}; 
\item[--] RF3: RF2 features plus the color \texttt{r-i}; 
\item[--] RF4: RF3 features plus the color \texttt{B-r}; 
\item[--] RF5: RF4 features plus the color \texttt{u-B}.
\end{itemize}
\noindent For completeness, we test the performance of an additional classifier where the only features used are the five colors mentioned above and the stellarity index (RF$_{col}$ classifier). 

\begin{figure}[tbh]
 \centering
            {\includegraphics[width=\columnwidth]{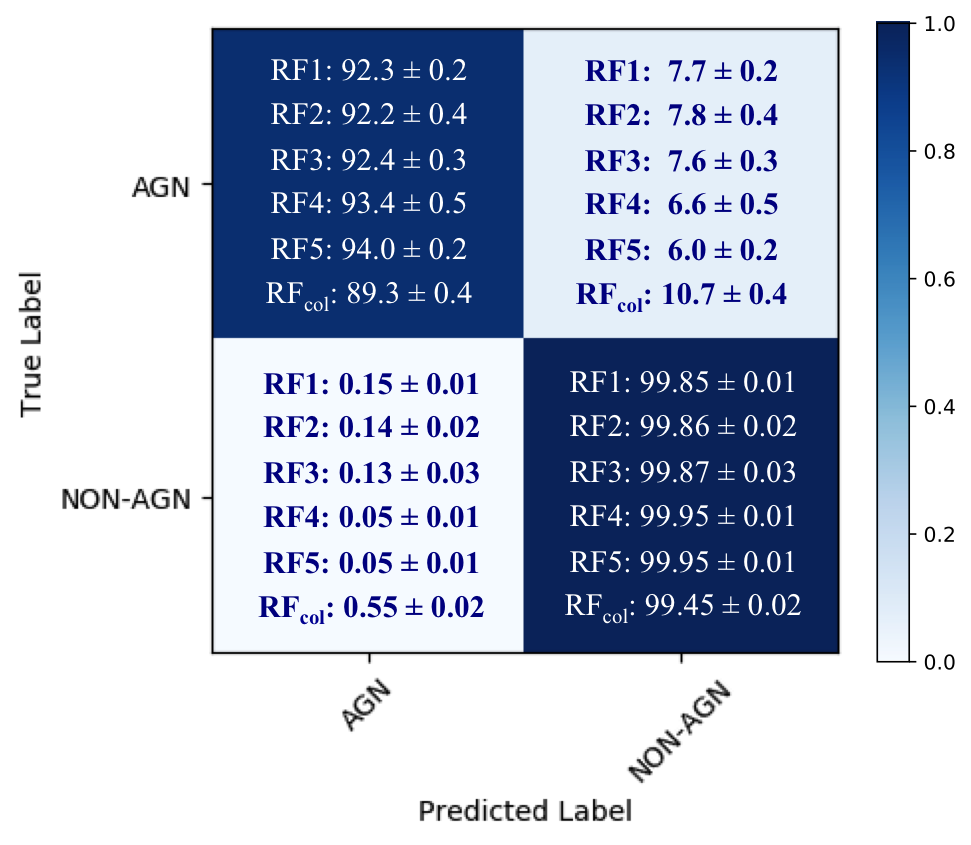}}
   \caption{Confusion matrices showing percent values from the classification of the validation set for each of the five RF classifiers where: only variability features and stellarity are used to identify AGN (RF1); the five colors \texttt{z-y}, \texttt{i-z} (RF2), \texttt{r-i} (RF3), \texttt{B-r} (RF4), and \texttt{u-B} (RF5) are progressively included in the feature set; and, finally, for the classifier RF$_{col}$, where the only features used are the five colors and the stellarity index. Each percent error is the standard deviation from the average value obtained from a set of ten simulations per classifier, where each classifier builds each time 500 trees in order to determine the final classification for each source.}\label{fig:cm_12345_col}
   \end{figure}

The LS we choose to use for these tests includes 1,000 stars, 1,000 inactive galaxies, and only the 225 sAGN1 as AGN, as it is well-known that the optical variability-based search for AGN favors Type I AGN, as shown for example in \citetalias{decicco19}. We show the confusion matrix for each of these classifiers in Fig. \ref{fig:cm_12345_col}. We note that, regardless of the number of colors taken into account, the first five results are very similar to each other and, in particular, these classifiers are always able to identify TNs with very high efficiency. As for TPs, the number of correctly classified AGN goes from 208 (${\approx}92\%$, RF1, RF2, and RF3 classifiers) to 212 (${\approx}94\%$, RF5 classifier) out of 225 sources. The latter classifier therefore contains $<2\%$ more AGN with respect to the other classifiers. Though these numbers are not very different from each other, they show that the introduction of colors in the feature list allows for the recovery of more sources, and a 2\% difference can turn into larger absolute numbers of AGN when larger samples of sources are involved. In the case of the RF$_{col}$ classifier the fraction of misclassified AGN is slightly worse, being almost 11\%, while the fraction of misclassified non-AGN is $< 1\%$, consistent with the other classifiers.

In Table \ref{tab:scores}, we report the values obtained for the above mentioned scores $A$, $P$, $R$, and $F1$, obtained for all the RF classifiers introduced so far. The table also includes the number of TPs, TNs, FPs, and FNs obtained from each test, in order to allow for comparison among the various numbers of sources rather than just fractions.
It is apparent that, for the first five classifiers, the accuracy of the classification is generally high, with the highest value obtained for the RF5 classifier. Precision is very high as well, the RF1 and RF3 classifiers being the ones returning a slightly lower value; in general, we find a very low number of FPs, which the definition of $P$ depends on. The RF5 classifier also returns the highest values for $R$ and $F1$. We can therefore state that, overall, the best scores correspond to the RF5 classifier, where all five optical and NIR colors are added to variability and morphology features. We therefore choose to work with the full set of features used for this classifier in what follows. The values obtained for the RF$_{col}$ classifier are always slightly lower than the values obtained for the other classifiers but, based on the obtained scores and CM, its performance does not seem very different from the others: $A$ decreases by $<0.1$, while the other scores show a maximum decrease of ${\approx}0.05\%$. Therefore, the addition of variability features to colors does not seem to bring substantial changes. We will discuss this focusing on the results obtained for the unlabeled set for the RF$_{col}$ classifier in Sect. \ref{section:col_only}.

\begin{table*}[thb]
\renewcommand\arraystretch{1.2}
\caption{TPs, FNs, FPs, TNs, Accuracy, Precision, Recall, and $F1$ computed for the five RF classifiers tested so far, including morphology and variability features (RF1) and optical and NIR colors (two to five from RF2 to RF5, respectively), and for the RF$_{col}$ classifier, where only morphology and color features are used. Each error is the standard deviation from the average value obtained from a set of ten simulations per classifier, where each classifier builds each time 500 trees to return the final classification for each source.}\label{tab:scores}
\centering \resizebox{\textwidth}{!}{  
\begin{tabular}{l c c c c c c c c}
\hline  & TPs & FNs & FPs & TNs & Accuracy & Precision & Recall & F1\\
 & (/225) & (/225) & (/2000) & (/2000) & & & & \\
\hline
RF1 & $208\pm0$ & $17\pm0$ & $3\pm0$ & $1997\pm0$ & $0.9909\pm0.0002$ & $0.98576\pm0.00003$ & $0.923\pm0.002$ & $0.953\pm0.001$\\
RF2 & $208\pm1$ & $17\pm1$ & $3\pm0$ & $1997\pm0$ & $0.9908\pm0.0005$ & $0.996\pm0.002$ & $0.922\pm0.004$ & $0.953\pm0.002$\\
RF3 & $208\pm1$ & $17\pm1$ & $3\pm1$ & $1997\pm1$ & $0.9912\pm0.0005$ & $0.988\pm0.002$ & $0.924\pm0.003$ & $0.955\pm0.002$\\
RF4 & $211\pm1$ & $14\pm1$ & $1\pm0$ & $1999\pm0$ & $0.9929\pm0.0005$ & $0.99526\pm0.00003$ & $0.934\pm0.005$ & $0.964\pm0.003$\\
RF5 & $212\pm1$ & $13\pm1$ & $1\pm0$ & $1999\pm0$ & $0.9935\pm0.0002$ & $0.99529\pm0.00001$ & $0.940\pm0.002$ & $0.967\pm0.001$\\
\hline
RF$_{col}$ & $201\pm1$ & $24\pm1$ & $11\pm0$ & $1989\pm0$ & $0.9842\pm0.0004$ & $0.948\pm0.002$ & $0.893\pm0.004$ & $0.919\pm0.002$\\
\hline
\end{tabular}}
\end{table*}

\subsection{Tests with different AGN LSs}
\label{section:ls_tests}
\begin{figure*}[tbh]
 \centering
            {\includegraphics[width=\textwidth]{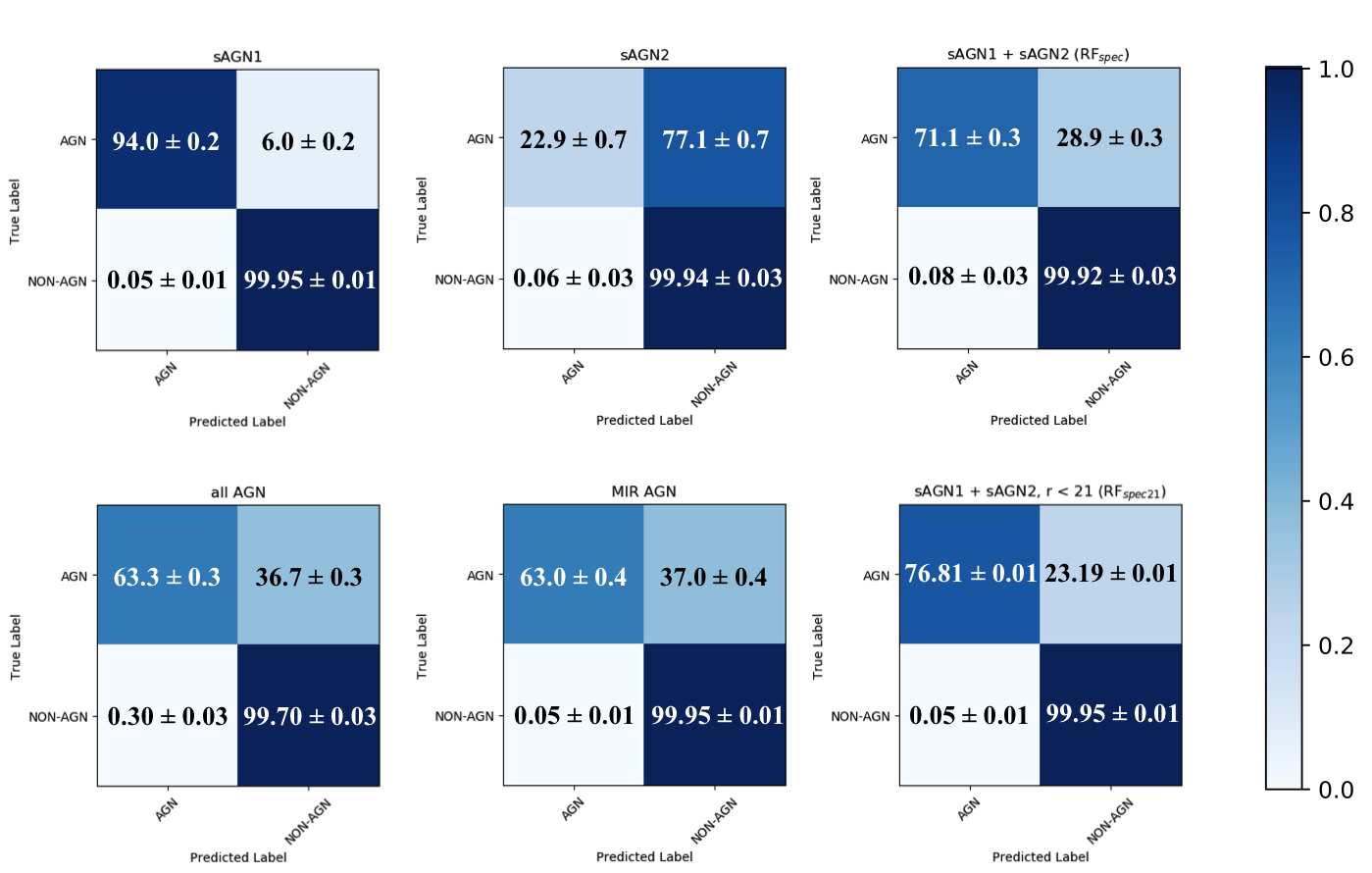}}
   \caption{Percent values for the confusion matrices obtained from the classification of different subsamples of AGN in the LS for each test: only sAGN1 (\emph{upper left}); only sAGN2 (\emph{upper center}); sAGN1 plus sAGN2 (\emph{upper right}); sAGN1, sAGN2, and Donley non-sAGN (\emph{lower left}); MIR AGN (\emph{lower center}); sAGN1 plus sAGN2, with \emph{r} $\leq 21$ mag (\emph{lower right}). True Label is the one we used to include the sources in the LS, as detailed in Sect. \ref{section:labeled_set}. Predicted Label is the outcome of the classification. Each percent error is the standard deviation from the average value obtained from a set of ten simulations per classifier, where each classifier builds each time 500 trees in order to determine the final classification for each source.}\label{fig:cm_ls_tests}
   \end{figure*}

Our aim is to identify the best LS, in the context of AGN selection based on variability and on optical and NIR colors.
We therefore make a number of tests, redefining each time the LS to include different subsets of AGN while always retaining the same set of stars and inactive galaxies. We note that throughout this section the LS changes for each test, while the set of features that we use is always the same and consists of all of the variability features plus all five color features and the stellarity index, all listed in Table \ref{tab:features}. We show the obtained results in Fig. \ref{fig:cm_ls_tests}. It is apparent that:
\begin{itemize}
    \item[--]Our classifier is always excellent in the identification of TNs (in the worst case we find 0.3\% FPs), independent of the type of AGN included in the LS.\\ 
    \item[--]As expected, we get the best results when we include only sAGN1 in the LS (upper left CM), obtaining the highest fraction of correctly classified AGN (94.0\% TPs).\\
    \item[--]As expected, including only sAGN2 in the LS (upper central CM) leads to the lowest fraction of correctly classified AGN (22.9\% TPs), presumably due to strong overlap with galaxy properties (colors and weaker variability).\\
    \item[--]Including both sAGN1 and sAGN2 leads to a good compromise, with 71.1\% of the known AGN correctly classified (i.e., 71.1\% recall $R$; we label this classifier RF$_{spec}$, upper right CM in the figure). As for the FNs, 12.0\% are sAGN1 and 88.0\% are sAGN2. We can also compute the fractions of misclassified sAGN1 and sAGN2 with respect to the total number of sAGN1 and sAGN2, and these are 5.3\% and 72.1\%, respectively. Notably, this is almost consistent with the combination of the sAGN1 and sAGN2 results alone.\\
    \item[--]The performance of the classifier with an AGN LS consisting of sAGN1 and sAGN2 is slightly improved when cutting the LS to $r \le 21$ mag, as the fraction of TPs is higher (76.8\% vs. 71.1\% in the classifier with no magnitude cut; we label this classifier RF$_{spec21}$, lower right CM in the figure). We will discuss the results obtained for the unlabeled set from the RF$_{spec21}$ classifier in Sect. \ref{section:spec21}.\\
    \item[--]Including all AGN types in the LS (lower left CM of Fig. \ref{fig:cm_ls_tests}), we obtain a relatively low fraction of TPs (63.3\%). The complementary fraction of 36.7\% FNs includes 4.9\% of all the sAGN1, 69.7\% of sAGN2, and 83.6\% of the Donley non-sAGN in the LS. We note that the 4.9\% and 69.7\% values are better than the fractions of FNs obtained when using sAGN1 and sAGN2 alone, suggesting that here we are able to recover a few more sources. This suggests that what is bringing the fractions down is the inclusion of MIR AGN, which have few or no discernable features in the optical bands that can pinpoint obscured AGN embedded within galaxies, and hence allow normal galaxies to enter as FNs.\\
    \item[--]Since Donley non-sAGN in our LS are mostly misclassified when mixed with the other AGN subsamples, we investigate further to assess whether AGN in the Donley region alone are good candidates for a LS in the context of our analysis. We therefore select all the 226 MIR AGN, regardless of their spectroscopic classification (which is in any case available for part of this sample), and do not include in our LS any other AGN coming from different selection methods (lower central CM of Fig. \ref{fig:cm_ls_tests}). While there are just 0.05\% FPs, we find that 37.0\% of the AGN are FNs. In particular, 10.7\% of these misclassified sources turn out to be sAGN1, 19.1\% are sAGN2, and 70.2\% are Donley non-sAGN, and these correspond to 4.0\%, 13.1\%, and 88.1\% of the total sAGN1, sAGN2, and Donley non-sAGN included in this set of 226 MIR AGN. Thus, as expected, an AGN LS based on the MIR selection criterion of \citet{donley} is not ideal for an optical variability-based study. 
\end{itemize}
These results suggest that a hierarchical classifier may be the best option. 

\subsection{Feature analysis for misclassified AGN}
\label{section:misclass}
The results illustrated in this section, together with the discussion of the various confusion matrices from Sect. \ref{section:ls_tests}, highlight how the misclassified AGN are mostly Type II and MIR AGN, regardless of the AGN subsamples included in the LS. 

In Table \ref{tab:top5feat} we report the top five features from the ranking obtained for each classifier built through variability features. We notice that generally colors are among the most important features when Type II or MIR AGN are included in the LS, while they are not when only Type I AGN are used as an LS, or when they prevail in size on the other classes (RF$_{spec}$, RF$_{spec21}$). In the case of the RF5 classifier, none of the five colors is in the top five features and the most important color, \texttt{u-B}, is ninth in the ranking, \texttt{B-r} is fourteenth, and the other colors place themselves in the lower half of the list (the full ranking for the RF5 classifier is shown in Fig. \ref{fig:f_i_comparison}). In the case of the classifier using only sAGN2 in the AGN LS, we notice that four out of the five colors used are among the top five features. As shown in Fig. \ref{fig:cm_ls_tests}, this classifier is also the one with the lowest fraction of TPs, which suggests that the colors used are not very efficient in separating Type II AGN from non-AGN.
We also notice that \texttt{ExcessVar}s \texttt{GP\_DRW\_$\sigma$}, \texttt{GP\_DRW\_$\tau$}, and $P_{var}$ are commonly found in the top five of the different classifiers.

\begin{table*}[htb]
\renewcommand\arraystretch{1.2}
\caption{Top five features in the ranking for each of the tested classifiers making use of variability features.}\label{tab:top5feat}
\centering \resizebox{\textwidth}{!}{
\footnotesize
\begin{tabular}{l c c c c c c c c c c}
\hline  & sAGN1 (RF5) & sAGN2 & all AGN types & MIR AGN & RF$_{spec}$ & RF$_{spec21}$ & RF1 & RF2 & RF3 & RF4\\
\hline 
I & \texttt{GP\_DRW\_$\sigma$} & \texttt{HST\_class\_star} & \texttt{u-B} & \texttt{u-B} & \texttt{ExcessVar} & \texttt{P$_{var}$} & \texttt{ExcessVar} & \texttt{ExcessVar} & \texttt{GP\_DRW\_$\sigma$} & \texttt{ExcessVar}\\
II & \texttt{ExcessVar} & \texttt{B-r} & \texttt{HST\_class\_star} & \texttt{ExcessVar} & \texttt{GP\_DRW\_$\sigma$} & \texttt{$\eta^e$} & \texttt{GP\_DRW\_$\sigma$} & \texttt{GP\_DRW\_$\sigma$} & \texttt{ExcessVar} & \texttt{GP\_DRW\_$\sigma$}\\
III & \texttt{GP\_DRW\_$\tau$} & \texttt{u-B} & \texttt{ExcessVar} & \texttt{GP\_DRW\_$\sigma$} & \texttt{GP\_DRW\_$\tau$} & \texttt{R$_{cs}$} & \texttt{P$_{var}$} & \texttt{GP\_DRW\_$\tau$} & \texttt{P$_{var}$} & \texttt{GP\_DRW\_$\tau$}\\
IV & \texttt{P$_{var}$} & \texttt{r-i} & \texttt{GP\_DRW\_$\sigma$} & \texttt{B-r} & \texttt{HST\_class\_star} & \texttt{Period\_fit} & \texttt{GP\_DRW\_$\tau$} & \texttt{P$_{var}$} & \texttt{GP\_DRW\_$\tau$} & \texttt{P$_{var}$}\\
V & \texttt{A$_{SF}$} & \texttt{i-z} & \texttt{GP\_DRW\_$\tau$} & \texttt{GP\_DRW\_$\tau$} & \texttt{P$_{var}$} & \texttt{GP\_DRW\_$\tau$} & \texttt{R$_{cs}$} & \texttt{A$_{SF}$} & \texttt{R$_{cs}$} & \texttt{A$_{SF}$}\\
\hline
\end{tabular}}
\end{table*}

\begin{figure*}[ht]
 \centering
\subfigure
            {\includegraphics[width=\columnwidth]{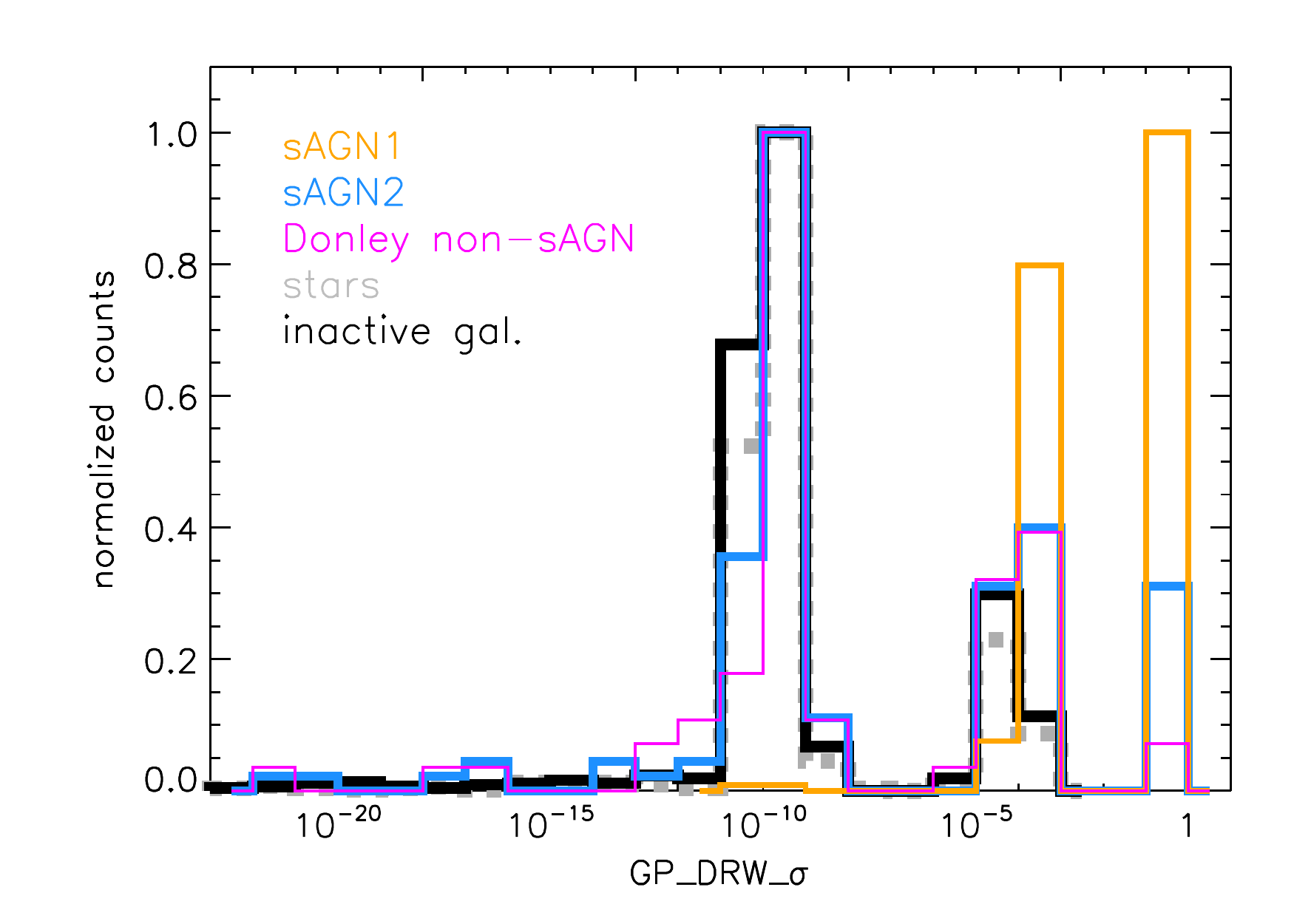}}
\subfigure
            {\includegraphics[width=\columnwidth]{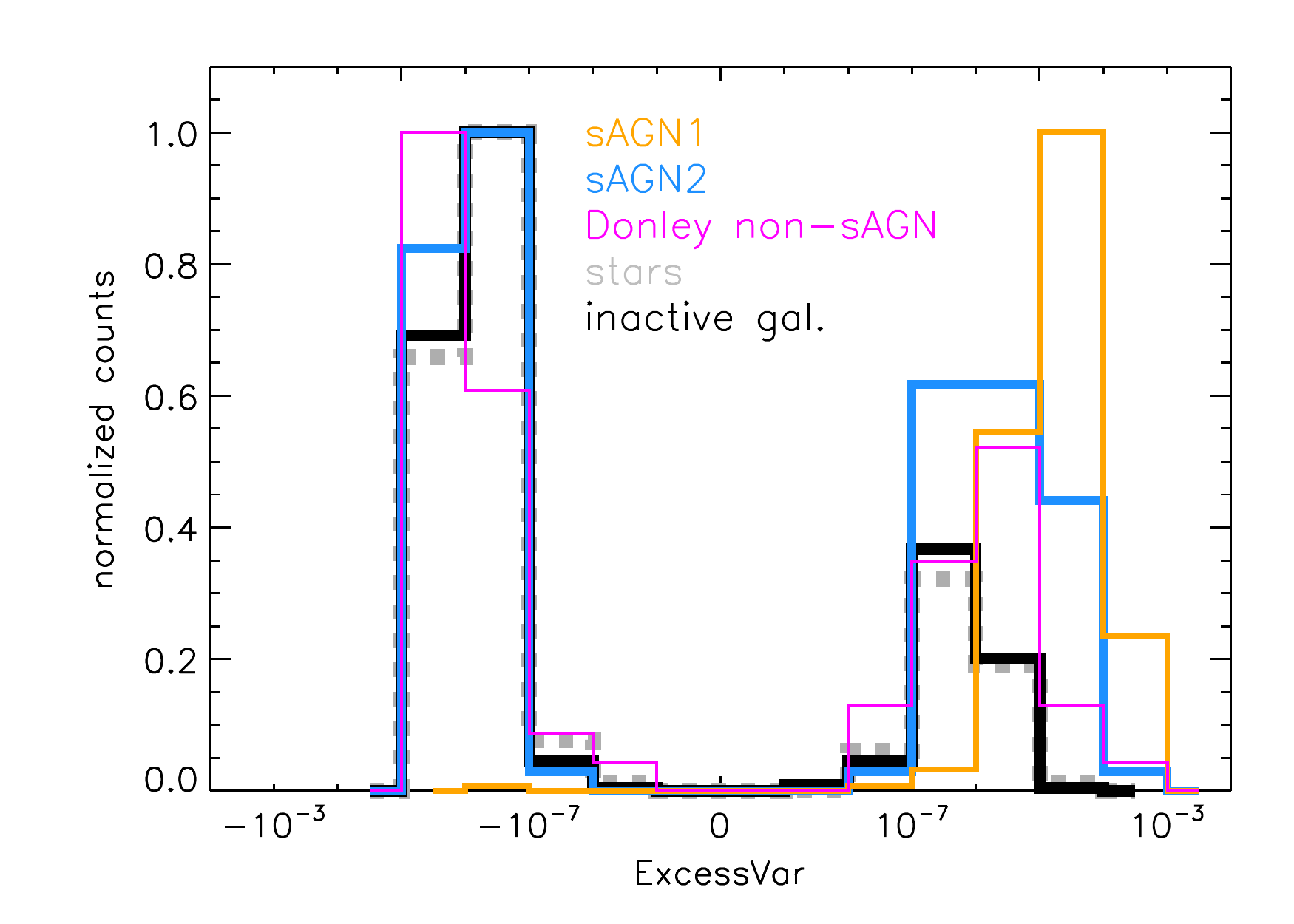}}
\subfigure
            {\includegraphics[width=\columnwidth]{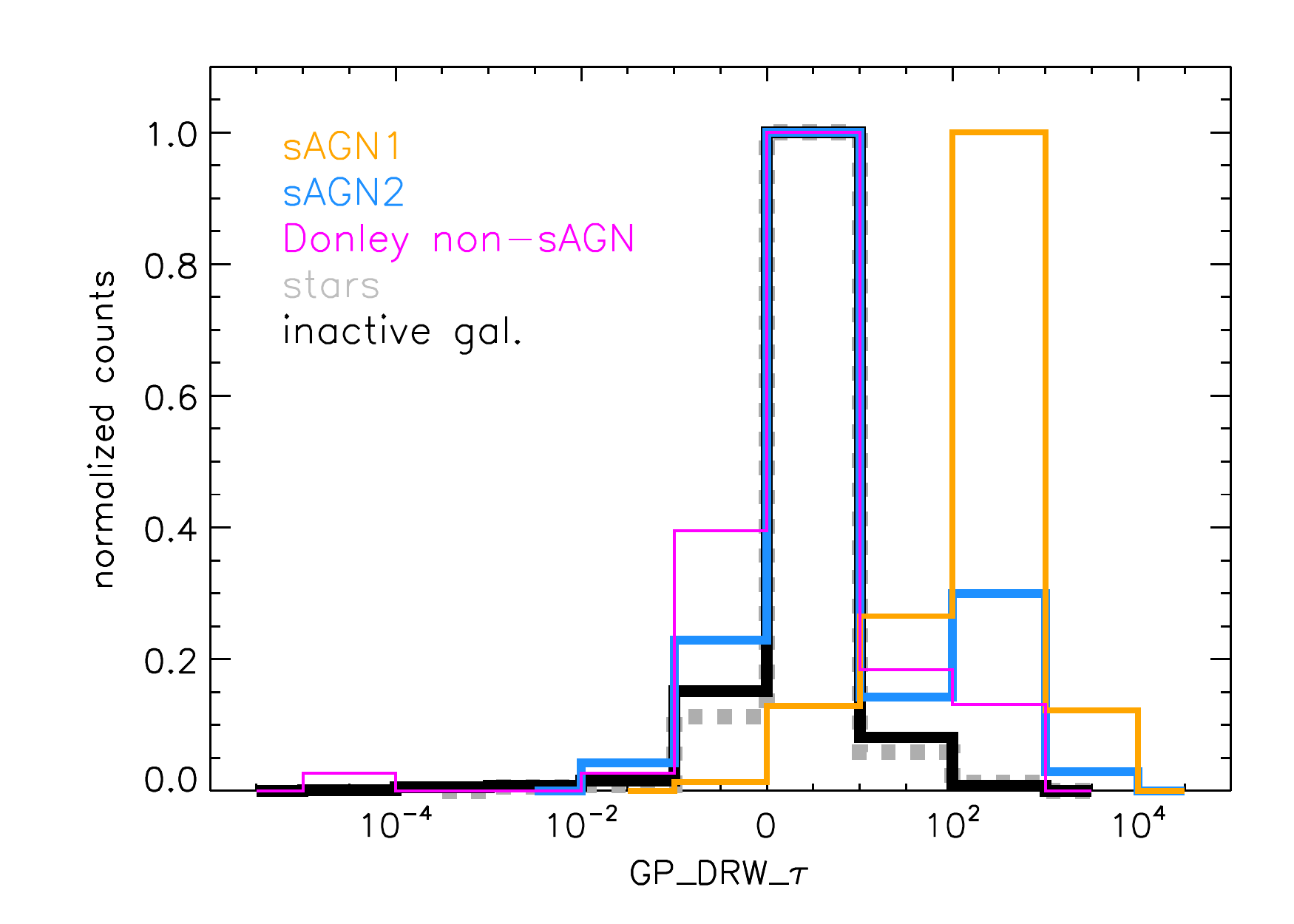}}\\
\subfigure
            {\includegraphics[width=\columnwidth]{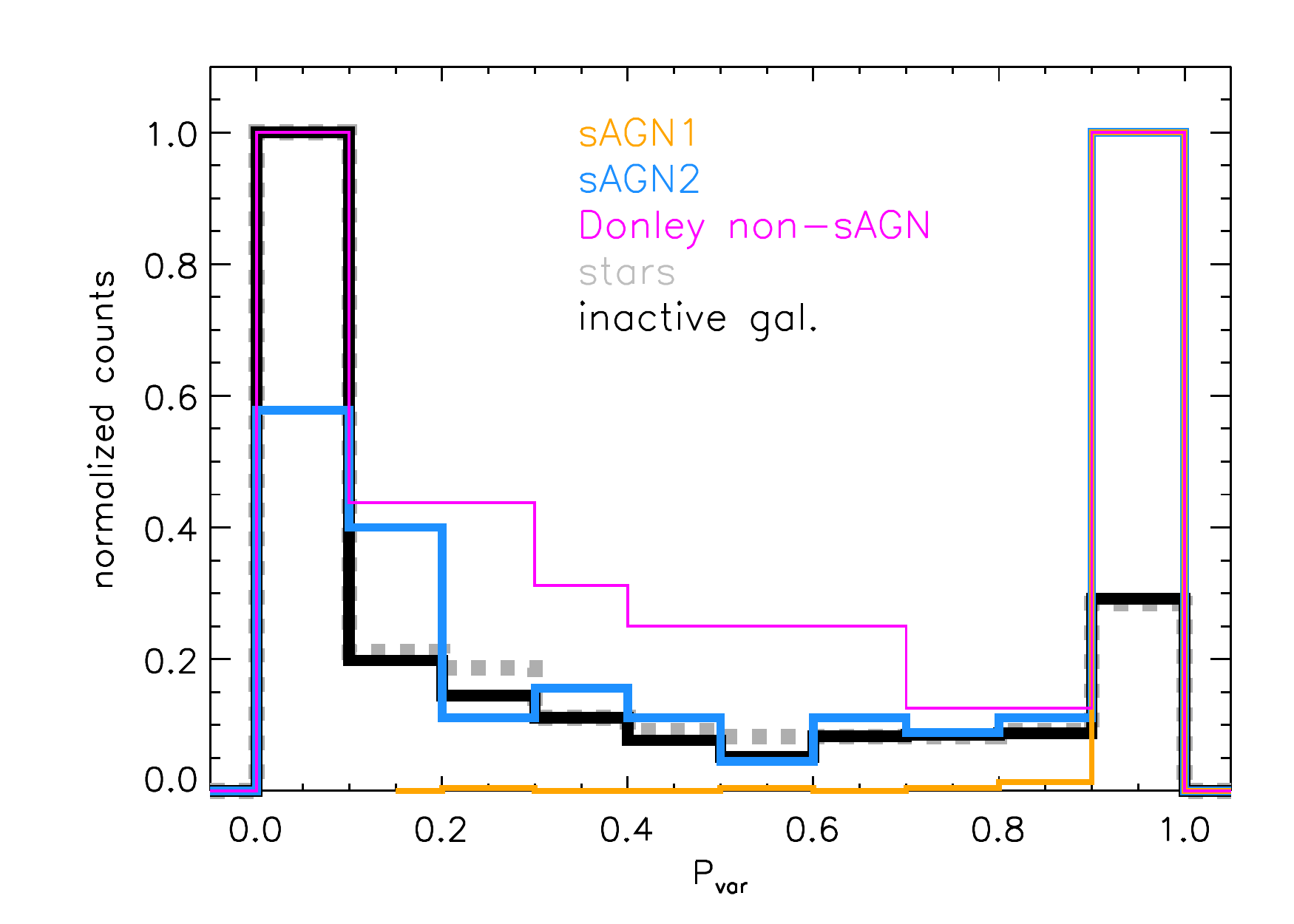}}
\subfigure
            {\includegraphics[width=\columnwidth]{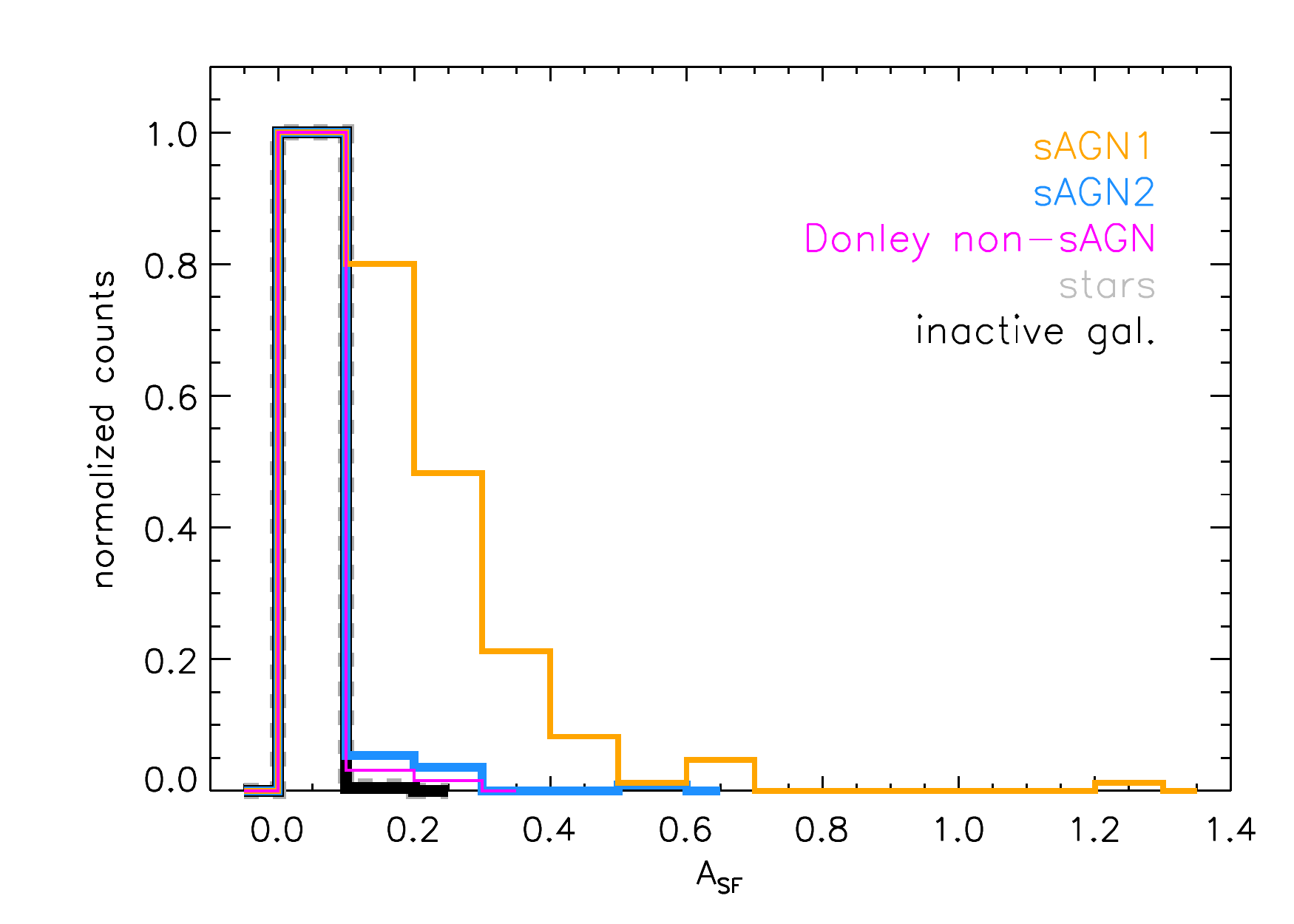}}
   \caption{\footnotesize{Distribution of the five most important features for the RF5 classifier (see Table \ref{tab:top5feat}) for known sAGN2 (light blue) and Donley non-sAGN (magenta). These are not included in the LS of our RF5 classifier, hence are part of the unlabeled set. We also show the corresponding distributions for sAGN1 (yellow), stars (dashed gray), and inactive galaxies (thick black) in the LS. The chosen bin size for each histogram is the one providing the best visualization and, when possible, separation of the distributions.}}\label{fig:fi_analysis}
   \end{figure*}

\begin{figure*}[ht]
 \centering
\subfigure
            {\includegraphics[width=\columnwidth]{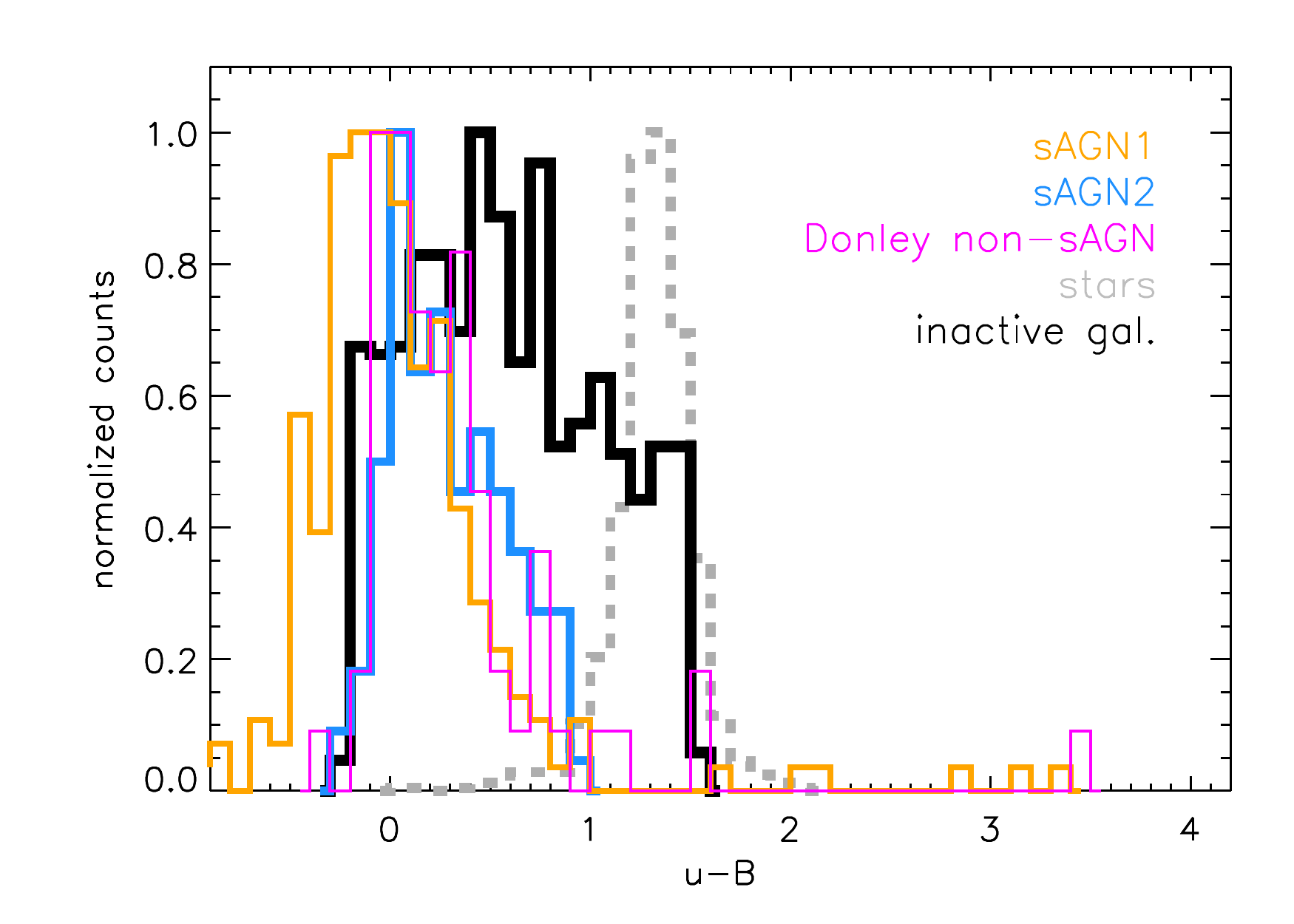}}
\subfigure
            {\includegraphics[width=\columnwidth]{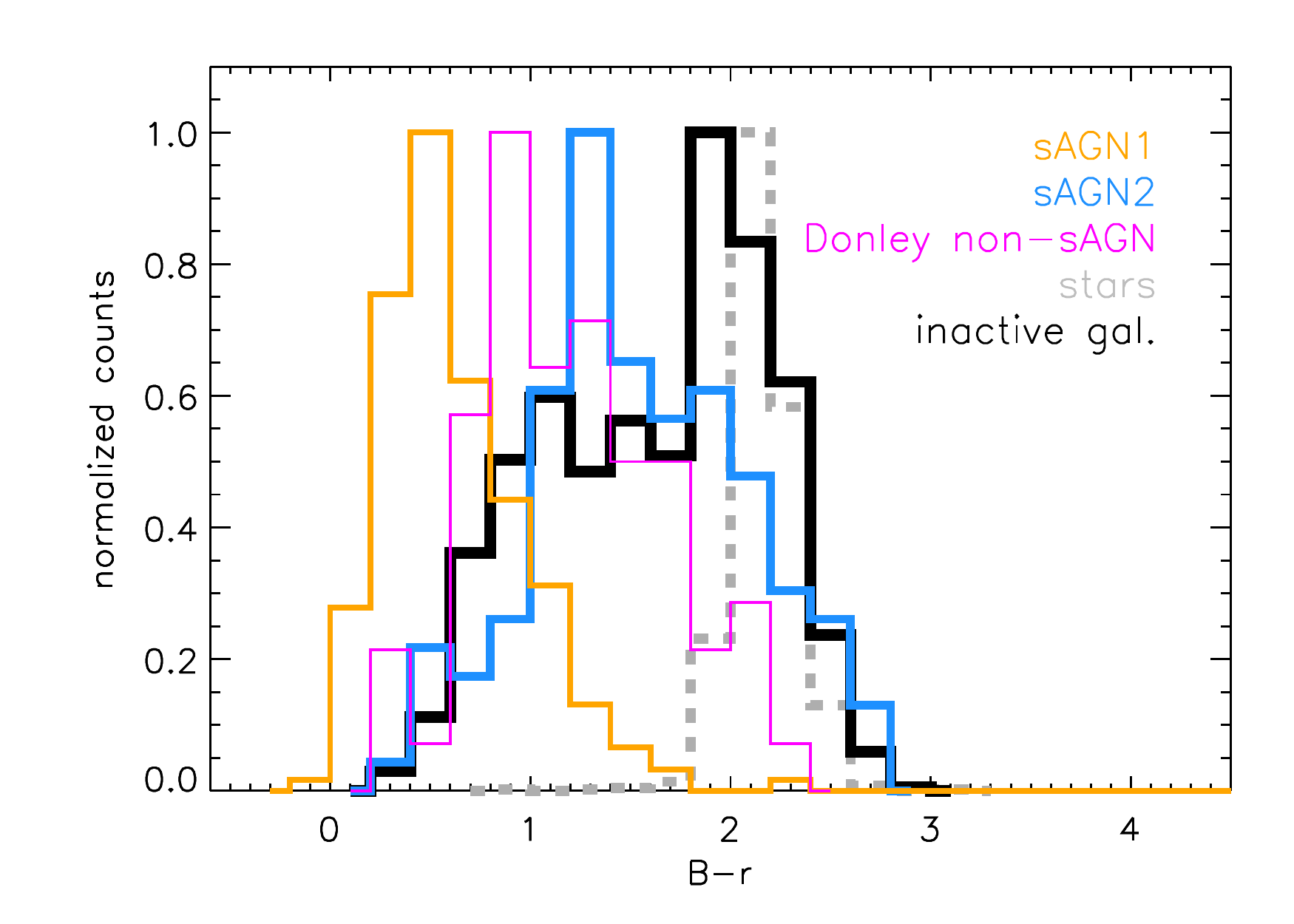}}
\subfigure
            {\includegraphics[width=\columnwidth]{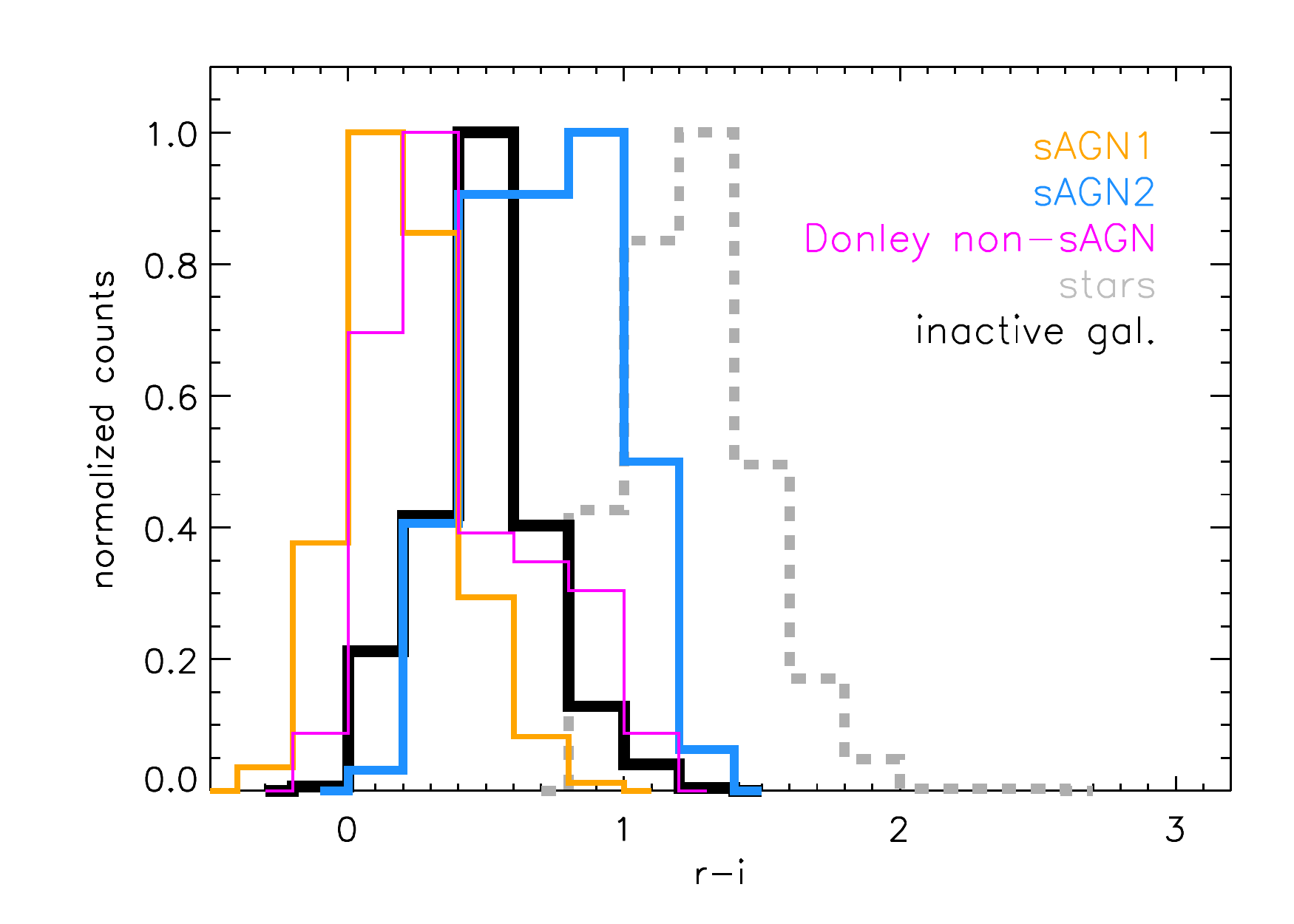}}\\
\subfigure
            {\includegraphics[width=\columnwidth]{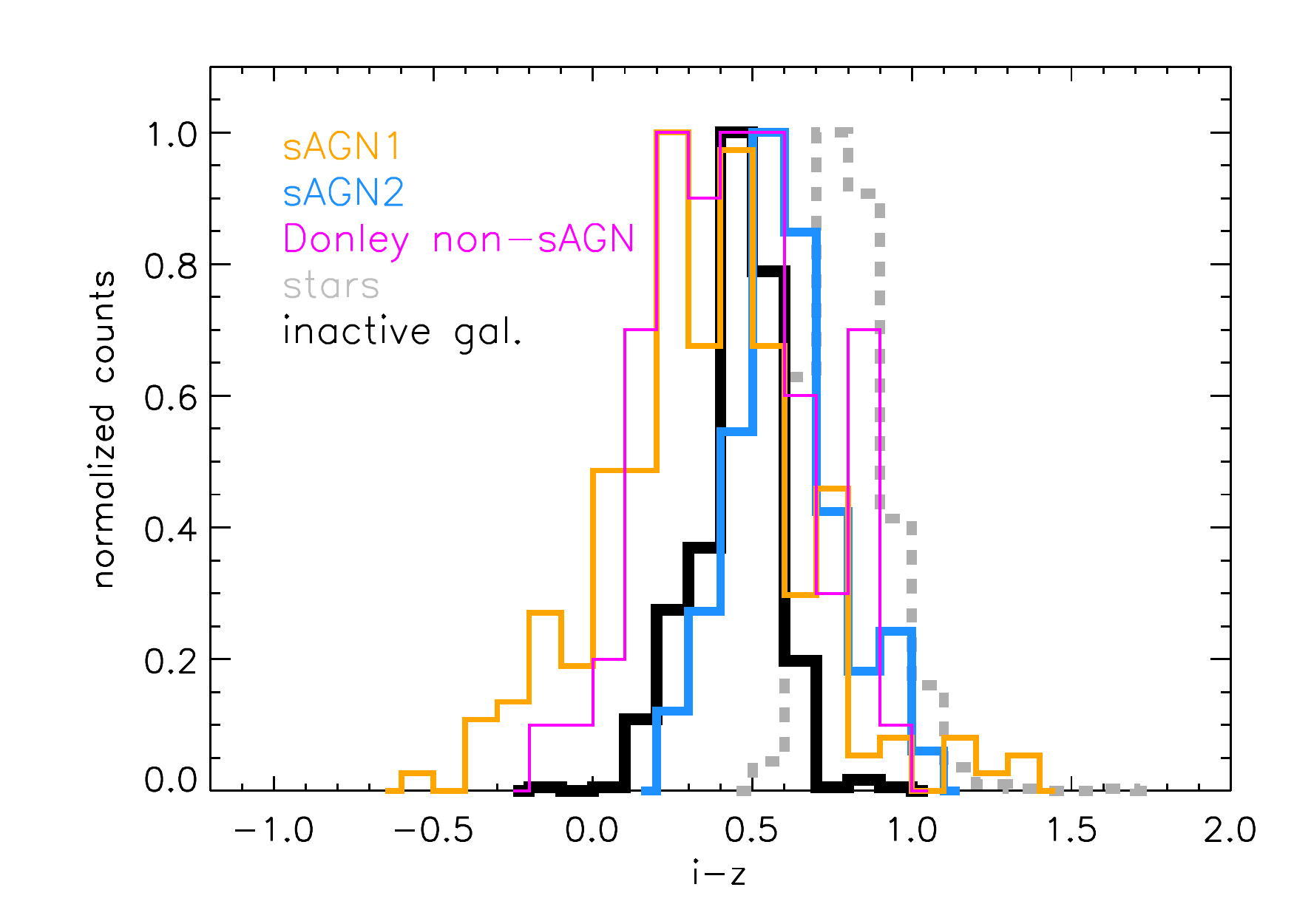}}
\subfigure
            {\includegraphics[width=\columnwidth]{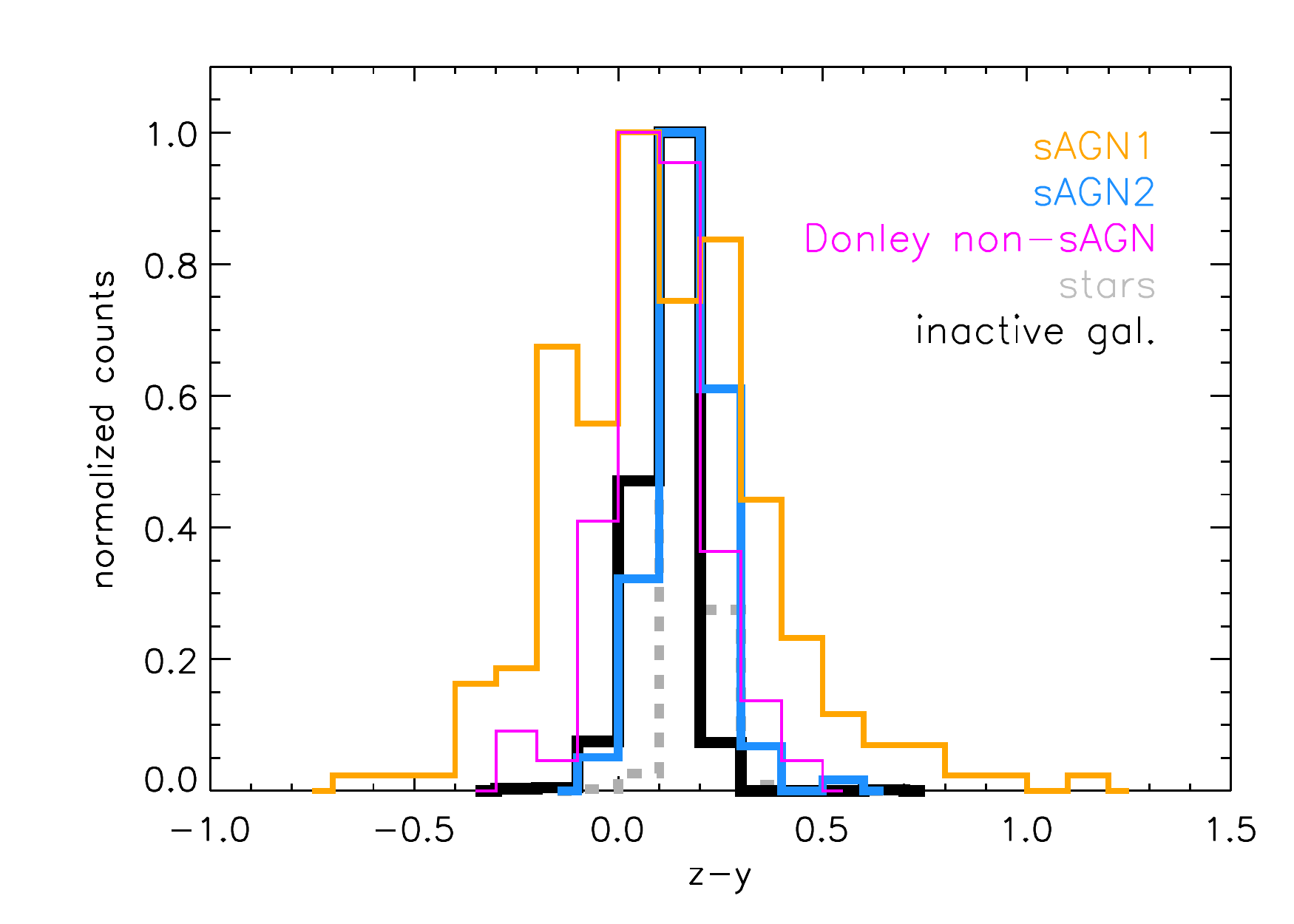}}
   \caption{\footnotesize{Distribution of the five optical and NIR color features used in this work for known sAGN2 (light blue)  and Donley non-sAGN (magenta). These are not included in the LS of our RF5 classifier, hence are part of the unlabeled set. We also show the corresponding distributions for sAGN1 (yellow), stars (dashed gray), and inactive galaxies (thick black) in the LS. The chosen bin size for each histogram is the one providing the best visualization and, when possible, separation of the distributions.}}\label{fig:fi_analysis_col}
   \end{figure*}

In order to highlight the difficulty of separating Type II and MIR AGN from galaxies among the various AGN LS combinations, we focus on the top five features obtained for the RF5 classifier and analyze the distributions of all known sAGN2 and Donley non-sAGN with respect to them, comparing these distributions to the ones corresponding to the other classes of known sources. Namely, these features are \texttt{GP\_DRW\_$\sigma$}, \texttt{ExcessVar}, \texttt{GP\_DRW\_$\tau$}, \texttt{P$_{var}$}, and \texttt{A$_{SF}$}; their sets of distributions are shown in Fig. \ref{fig:fi_analysis}. We also show the corresponding distributions for the five color features in Fig. \ref{fig:fi_analysis_col}. We include in the plots the sources in our LS, that is to say sAGN1 and non-AGN, separating stars from inactive galaxies, to ease the comparison among different classes of objects. From Fig. \ref{fig:fi_analysis} it is apparent that sAGN1 are a distinct population, with respect to the selected features, if compared to the other four classes, whose distributions have approximately coincident peaks. In the case of \texttt{A$_{SF}$}, the distribution of sAGN1 exhibits a peak in overlap with the distributions corresponding to all the other classes of objects, but the overall shape of the histogram for sAGN1 is different. The distributions of sAGN2, Donley non-sAGN, stars, and inactive galaxies corresponding to the five color features in Fig. \ref{fig:fi_analysis_col} have distinct peaks in some cases; but, overall, overlap widely and, except for stars, they partly overlap the distribution of sAGN1 as well. We notice that, for what concerns Donley non-sAGN, the overlap of their \texttt{r-i} and \texttt{i-z} distributions with non-AGN distributions is less broad. This class of AGN likely includes both Type I and Type II AGN, and therefore exhibits mixed properties on the histograms here shown. In summary, the various distributions in these two figures show how even the most relevant features used here optimize the selection of Type I AGN, but are not optimal for the disentangling of Type II/MIR AGN from non-AGN.

We compare the various pairs of distributions of interest, that is to say, each AGN class versus the others and versus non-AGN classes, via the Kolmogorov-Smirnov (K-S) test. We find that each distribution is distinct and well separated from the others: in terms of variability features, when comparing the sAGN1 distribution to the others we always obtain distances $D > 0.65$, and probabilities $p < 10^{30}$ to obtain larger distances assuming that the two distributions are drawn from the same distribution function; when comparing the color features distributions, we get relatively higher probabilities ($10^{-9}$ or lower). For what concerns sAGN2, we generally find probabilities $<10^{-5}$.

\subsection{Comparison with the findings of S\'{a}nchez-S\'{a}ez et al. (2019)}
\label{section:ss19}
\citet{sanchezsaez} tested different RF classifiers using part of the variability features used in this work, and colors obtained from the $griz$ bands. The utilized dataset comes from the QUEST-La Silla AGN Variability Survey \citep{cartier} and has a depth of $r$ ${\sim}21$ mag. In addition, the catalog is not free from misclassifications. The purest sample of AGN candidates is provided by the classifier making use of the two colors ($r-i$ and $i-z$). There are some differences in the approach followed in our work with respect to \citet{sanchezsaez}:
\begin{itemize}
    \item[--]We include the color features $z-y$ and $u-B$ and, in particular, the first is always used when color features are included.\\
    \item The number of variability features used here is approximately doubled. In particular, we now make use of DRW model-derived features, which turn out to be among the most relevant in the ranking.\\
    \item[--]The LS consists of Type I AGN and stars selected from the Sloan Digital Sky Survey (SDSS; \citealt{sdss}) spectroscopic catalog: the sources there included were mostly selected on the basis of optical colors. Our LS includes AGN selected through spectroscopy as well, but the selection was done on the basis of their X-ray properties. In addition, we include inactive galaxies in our LS. 
\end{itemize} 
In spite of these differences, we obtain similar results in our work:
\begin{itemize}
    \item[--]The variability features turn out to be more relevant than color features in each of the tested classifiers.\\
    \item[--]When all the three colors at issue are included in the classifier, \texttt{B-r} proves to be the most important among them.\\
    \item[--]The scores obtained for the three classifiers are not significantly different.
\end{itemize}

\section{Classification of the unlabeled set}
\label{section:results}
Once we selected the best set of features to build our RF classifier and the best sample of AGN (i.e., sAGN1) to include in our LS, we use the RF5 classifier to obtain a classification for our unlabeled set of sources. This consists of 18,445 sources (i.e., 20,670 sources in the main sample minus the 2,225 sources in the LS). Together with the classification, the classifier returns a probability for that prediction, computed as the mean of the predicted class probability for all the trees in the built forest. This probability is always $\geq0.5$; in what follows we will always require our AGN candidates to have a classification probability $\geq 0.6$, in order to stem the presence of contaminants. Based on this, the RF5 classifier returns a sample of 77 AGN candidates (hereafter, RF5 AGN candidates).

\begin{figure*}[tbh]
 \centering
\subfigure
            {\includegraphics[width=\columnwidth]{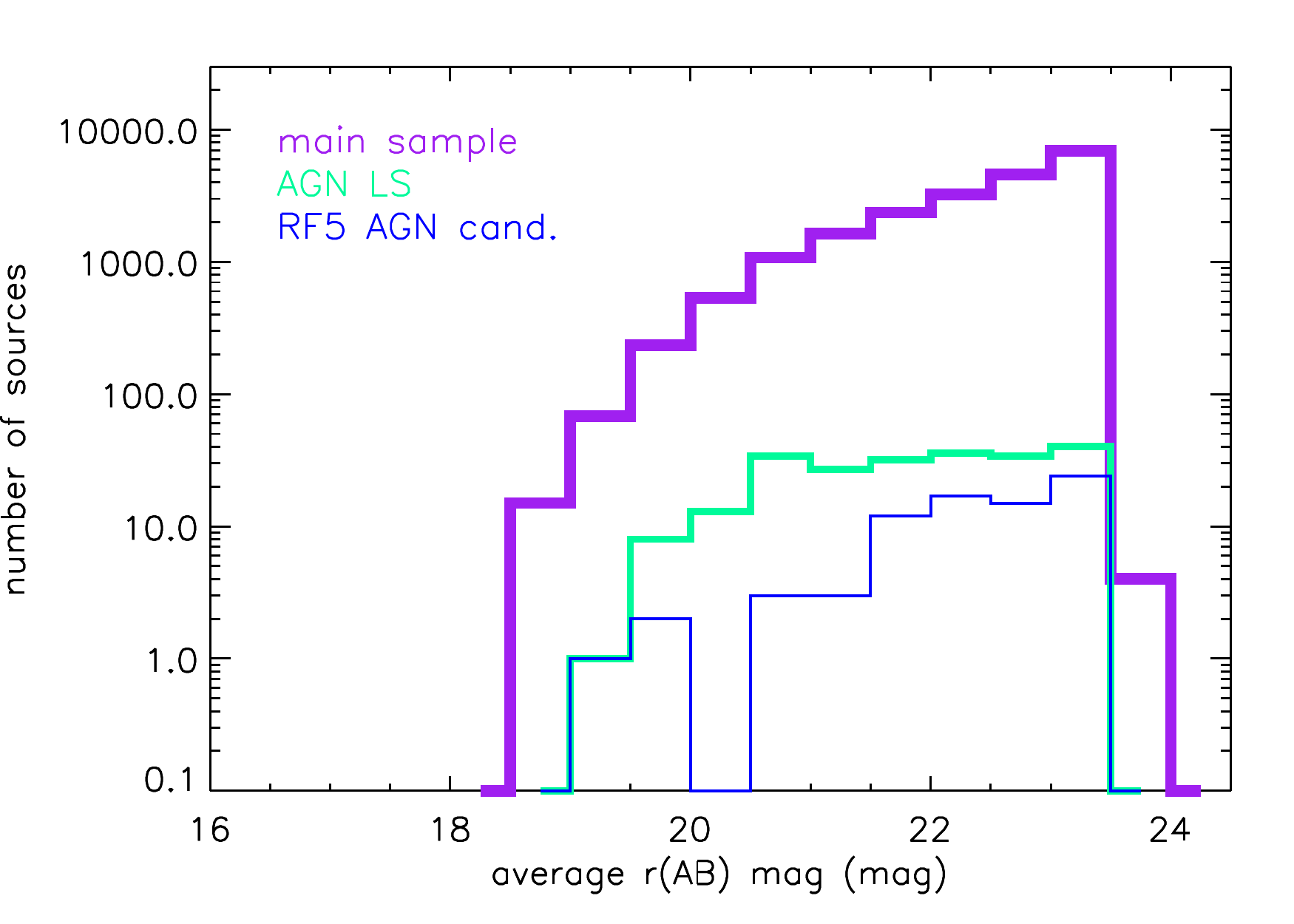}}
\subfigure
            {\includegraphics[width=\columnwidth]{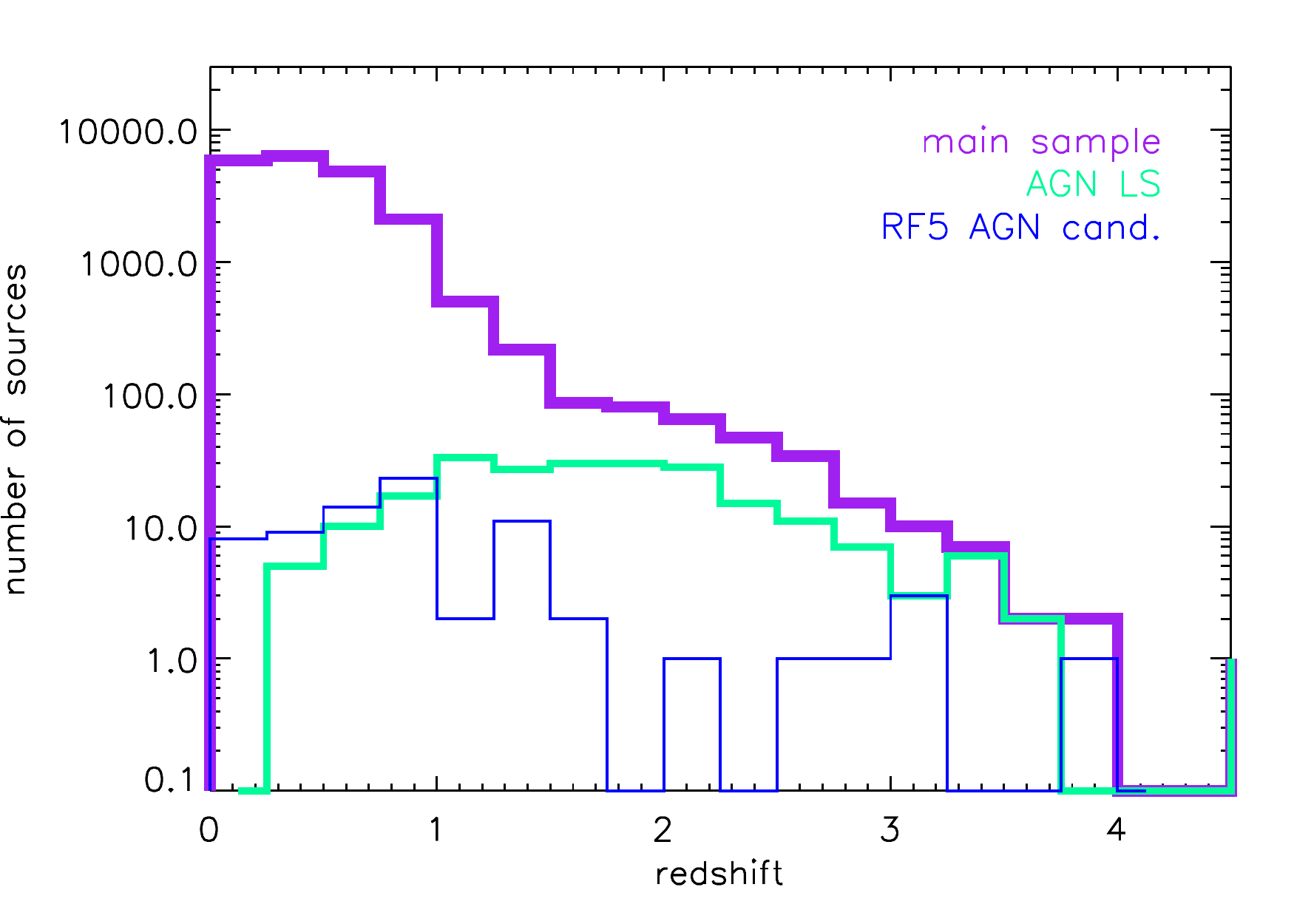}}
\subfigure
            {\includegraphics[width=\columnwidth]{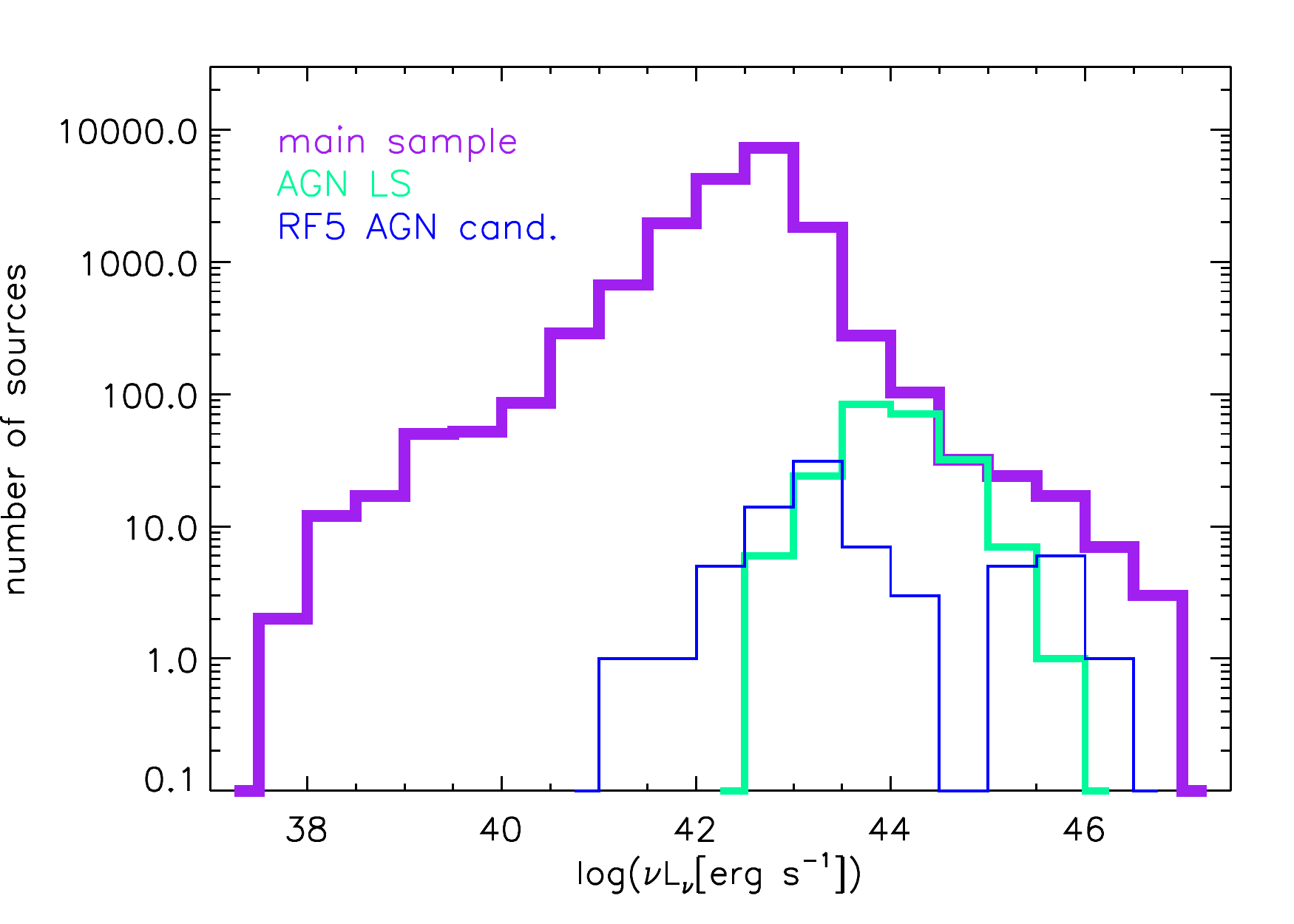}}\\
\subfigure
            {\includegraphics[width=\columnwidth]{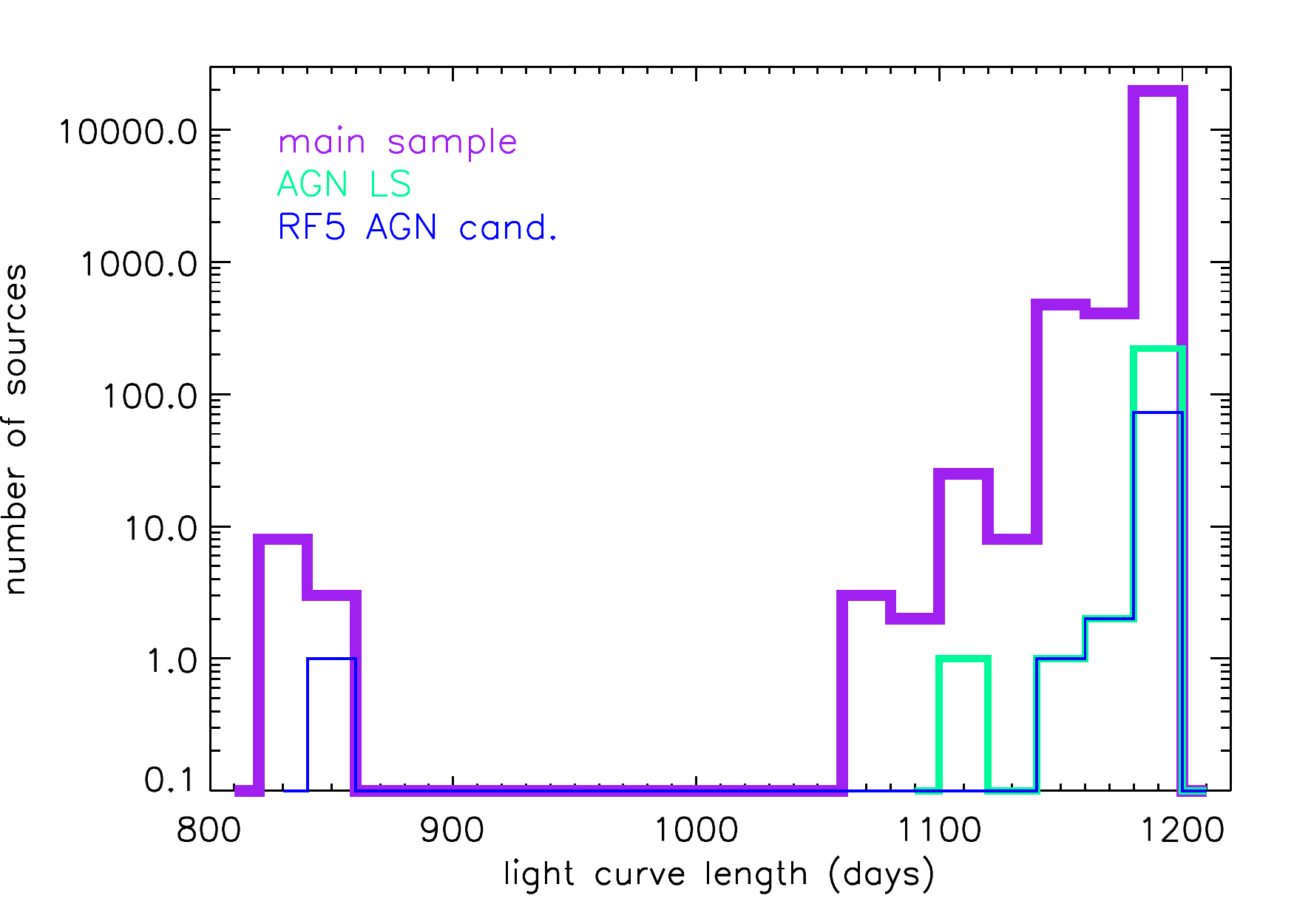}}
\subfigure
            {\includegraphics[width=\columnwidth]{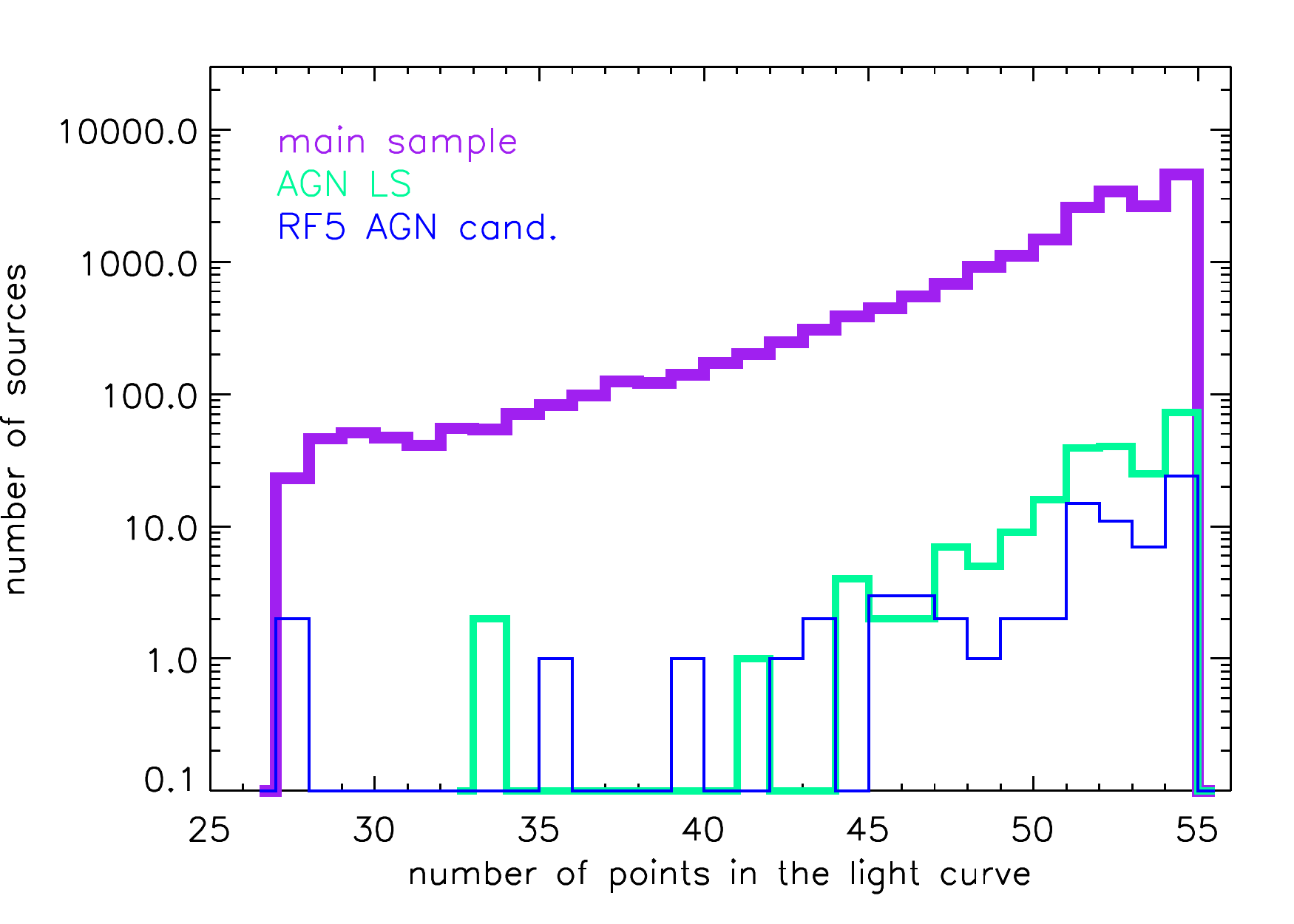}}
   \caption{\footnotesize{Distribution of magnitude (\emph{upper left panel}), redshift (\emph{upper right}), $r$-band luminosity (\emph{center}), length of the light curves (\emph{lower left}), and number of points in the light curves (\emph{lower right}) for the sources in the main sample (thick purple line), in the AGN LS (green line), and in the sample of RF5 AGN candidates (thin blue line).}}\label{fig:hists_with_agn}
   \end{figure*}

We characterize the sample of RF5 AGN candidates in Fig. \ref{fig:hists_with_agn}, where we show the same histograms as Fig. \ref{fig:hists_with_colors} for the main sample, the AGN LS (i.e., sAGN1), and the RF5 AGN candidates. A comparison of the AGN LS and RF5 AGN candidate histograms in each panel demonstrates that COSMOS has very complete redshift coverage of AGN over portions of relevant parameter spaces, but clearly suffers from some selection effects. Our classifier is able to increase by 34\% the number of classified AGN with respect to the LS used here, pushing in particular to lower bolometric luminosities at lower redshifts.

The histograms reveal that the RF5 AGN candidates, as well as the sAGN1 in the LS, have light curves spanning at least 1,150 days (except for two sources), and all but a dozen objects have at least 45 points in their light curves. This may support the thesis that a long and densely sampled light curve favors the identification of AGN, consistent with the results shown in \citetalias{decicco19}. 

The match of the RF5 AGN candidates with our sample of known AGN reveals that they include 26 of the 122 (21\%) Type II AGN confirmed by spectroscopy and eight of the 67 (12\%) Donley AGN. This is consistent with our discussion from Sect. \ref{section:misclass} about the low efficiency in retrieving Type II AGN (and hence many MIR AGN).

\begin{figure*}[tbh]
 \centering
\subfigure
            {\includegraphics[width=\columnwidth]{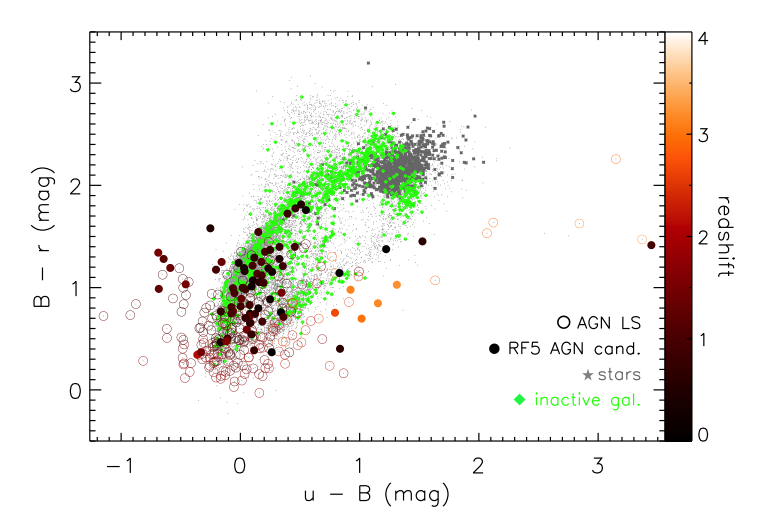}}
\subfigure
            {\includegraphics[width=\columnwidth]{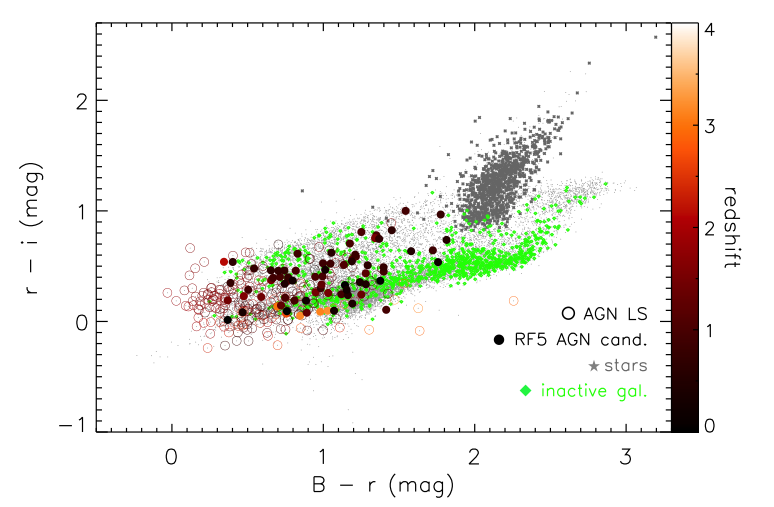}}
\subfigure
            {\includegraphics[width=\columnwidth]{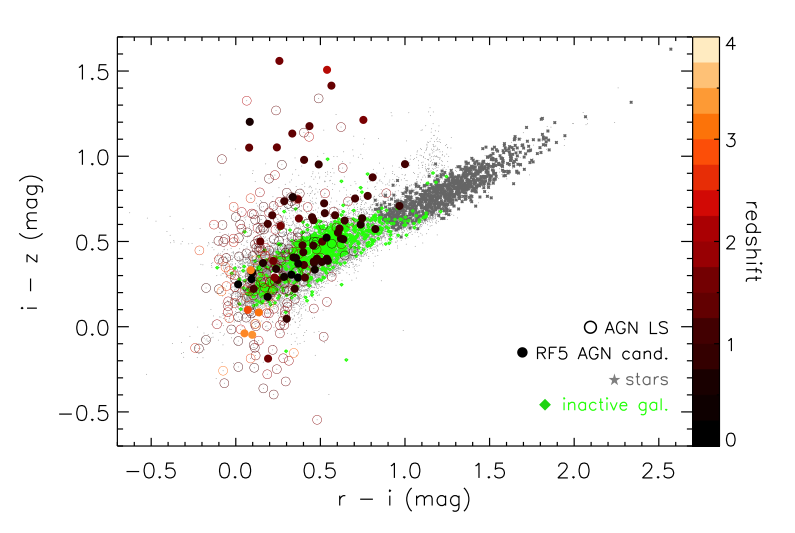}}
\subfigure
            {\includegraphics[width=\columnwidth]{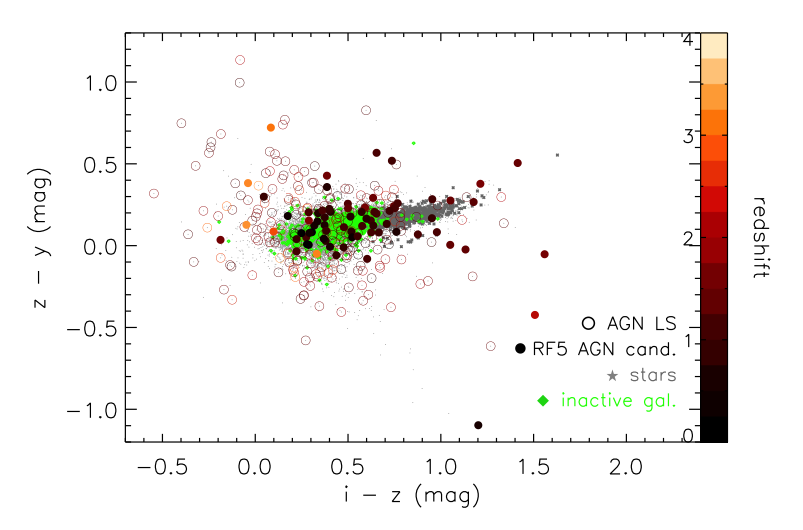}}
   \caption{\footnotesize{Color-color diagrams based on the use of the five optical and NIR colors adopted as features in this work: \texttt{u-B}, \texttt{B-r}, \texttt{r-i}, \texttt{i-z}, and \texttt{z-y}. The smaller gray dots represent the main sample. The sources in the  LS are shown as: gray stars (stars), green diamonds (inactive galaxies), and empty circles (AGN). The RF5 AGN candidates are shown as filled circles. All AGN are color-coded according to their redshift, based on the scale shown on the right vertical axis.
   }}\label{fig:color_diagrams}
   \end{figure*}

The RF5 AGN candidates, together with the AGN in the LS, are shown in Fig. \ref{fig:color_diagrams}, in four color-color diagrams obtained from the optical and NIR colors included as features in this work. While inactive galaxies can have the same colors as AGN, the stellar regions are fairly distinct from AGN in each diagram, which makes colors helpful in separating stars from AGN. The four diagrams show that our RF5 AGN candidates are generally characterized by redder colors than the AGN in our LS. This suggests that variability-based selection is quite effective in identifying host-dominated AGN, consistent with the findings of \citet{sanchezsaez,sanchezsaez20}, and highlights the strength of variability-based selection of AGN over color selection, which easily misses host-dominated AGN as their colors are similar to those of inactive galaxies. 

\subsection{Comparison with the sample of AGN candidates from De Cicco et al. (2019)}
\label{section:D19}
\citetalias{decicco19} used the same dataset as this work to identify optically variable AGN: they selected the 5\% most variable sources based on the rms distribution of their light curves, identified and removed spurious variable sources, and in the end obtained a sample of 299 optically variable AGN candidates. They confirmed 256 (86\%) of them as AGN using diagnostics obtained by means of ancillary multiwavelength catalogs available from the literature. In Sect. \ref{section:intro} we mentioned that they obtained a 59\% completeness with respect to spectroscopically confirmed AGN.

Here we compare our current results with those from \citetalias{decicco19}. In Sects. \ref{section:vst} and \ref{section:RF} we described how our initial sample was refined to the present main sample, removing sources with close neighbors and requiring color information and a measurement of the stellarity index to be available. Our present main sample includes 271 out of the 299 AGN candidates from \citetalias{decicco19}; the remaining 28 sources where excluded from the present analysis because they do not have all the $uBrizy$ magnitude measurements available in the COSMOS2015 catalog (21 sources), have no counterpart in the COSMOS-ACS catalog and hence no measurement of the stellarity index that we use as a feature (five sources), or have a close neighbor (two sources). Eighteen of these sources are confirmed AGN in \citetalias{decicco19}.
The 271 sources in common with this work consist of:
\begin{itemize}
    \item[--]186 sAGN1 included in our LS in the current work. The remaining sAGN1 in the LS do not belong to the AGN candidate sample in \citetalias{decicco19} because they were not classified as variable according to the selection criterion there adopted (most of them are just below the variability threshold).\\
    \item[--]One star in our LS, classified as stars in \citetalias{decicco19} as well.\\
    \item[--]84 sources in common with our unlabeled set of sources. The RF5 classifier in this work identifies 58 as AGN --56 of which with a classification probability $\geq0.6$-- and 26 as non-AGN, based on their features. Among the 58 RF5-identified AGN, \citetalias{decicco19} found 51 to be AGN on the basis of multiwavelength properties, while one was classified as a star and no diagnostic confirmed the nature of the remaining six sources. For the 26 RF5-identified non-AGN, only one was confirmed to be an AGN in \citetalias{decicco19}, while three turned out to be classified as stars, leaving 22 of them with no validation of their nature. This means, on one side, that we here classify as AGN 58 out of the 84 sources (69\%) in common with the sample of AGN candidates from \citetalias{decicco19} and, on the other hand, that 51 of our 77 RF5 AGN candidates with a classification probability $\geq 0.6$ (65\%) have a confirmation from \citetalias{decicco19}. If we restrict the comparison to the sample of confirmed AGN in \citetalias{decicco19}, the sources in common with our unlabeled sample are 52, and 50 of them (96\%) also belong to the sample of RF5 AGN candidates, thus showing an excellent agreement. After comparison with the results from \citetalias{decicco19}, the sources in the sample of RF5 AGN candidates that still require secondary confirmation are 27.
\end{itemize}

\subsection{Multiwavelength properties of the RF5 AGN candidates}
\label{section:mw_properties}
Given the major role of X-ray emission in the identification of AGN \citep[e.g.,][]{brandt&alexander}, we investigate the X-ray properties of our sample of RF5 AGN candidates. 
X-ray counterparts for VST-COSMOS sources were obtained making use of the already mentioned \emph{Chandra}-COSMOS Legacy Catalog as the main reference, and of the \emph{XMM}-COSMOS Point-like Source catalog \citep{brusa} as a back-up when information from the first catalog is not available. Details are provided in Sects. 4.1 and 4.2 of \citetalias{decicco19}. 

In our main sample of 20,670 sources there are 629 with an X-ray counterpart (hereafter, X-ray sample), while the RF5 AGN candidates with an X-ray counterpart are 55 out of 77 ($71\%$). Considering that all the 225 sAGN1 included in the LS come from X-ray catalogs, the total fraction of AGN (candidates) with an X-ray counterpart with respect to the X-ray sample is $280/629 = 45\%$. The corresponding fraction from \citetalias{decicco19} is $250/719 = 35\%$. We note once more that the main sample used in the present work only includes the VST-COSMOS sources with available colors from COSMOS2015 $uBrizy$ bands and stellarity index from the COSMOS-ACS catalog, which accounts for the differences in the X-ray samples used here and in \citetalias{decicco19}.

\begin{figure}[tb]
\centering
            {\includegraphics[width=\columnwidth]{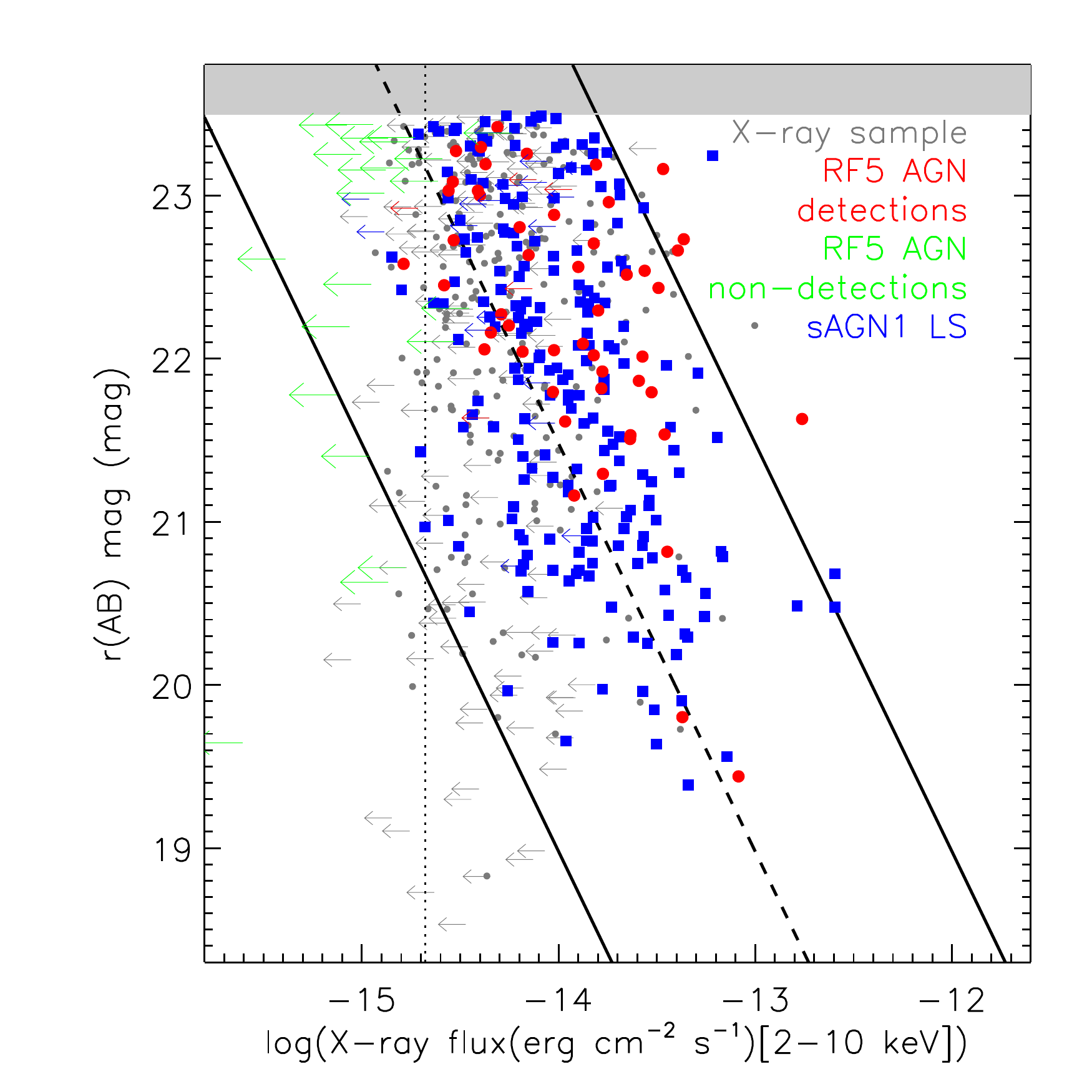}}
\caption{\footnotesize Comparison of X-ray to optical flux for several discussed samples: 55 X-ray detected (large red dots and arrows) and 22 X-ray undetected (large green arrows) sources in the RF5 AGN sample; 225 sAGN1 in the LS (blue squares and arrows); entire X-ray sample (i.e., all the VST sources with X-ray counterparts, gray dots and arrows), shown as a reference. Small leftward arrows denote VST-COSMOS sources which are detected in the \emph{Chandra}-COSMOS Legacy Catalog but only have 2--10 keV flux upper limits, while large green arrows denote the 22 RF5 AGN candidates which remain undetected in any X-ray band. For the 22 X-ray undetected sources, hard-band X-ray flux limits are estimated using CSTACK \citep{cstack}. The dashed line indicates $X/O =0$, while the lower and upper solid lines correspond to $X/O =-1$ and $X/O =1$, respectively, delimiting the traditional AGN locus. The vertical dotted line indicates the nominal hard-band flux limit of $2.1 \times 10^{-15}$ erg cm$^{-2}$ s$^{-1}$ of the \emph{Chandra}-COSMOS Legacy Catalog, estimated at 20\% of the area of the whole survey. The gray-shaded rectangle indicates that we do not find sources in that area as we set a magnitude limit of 23.5 mag for the VST-COSMOS dataset.}\label{fig:xo}
\end{figure}

We compare the X-ray and optical fluxes of the sources in the X-ray sample. The ratio of the X-ray-to-optical flux ($X/O$), first introduced by \citet{Maccacaro}, is
\begin{equation}
X/O=\log(f_{\scriptscriptstyle{X}}/f_{opt})=\log f_{\scriptscriptstyle{X}} + \frac{\mbox{mag}_{opt}}{2.5} + C \mbox{ ;}\label{eqn:XO}
\end{equation}
in this case, $f_{\scriptscriptstyle{X}}$ is the X-ray flux measured in the 2--10 \,keV band, $\mbox{mag}_{opt}$ is the VST \emph{r} magnitude, and the constant $C$, depending on the magnitude system adopted for the observations, equals 1.0752.
Traditionally, AGN are considered to lie in the region $-1\le X/O \le 1$, or even $X/O >1$ \citep[see, e.g.,][]{hornschemeier, Xue, civano}, while stars and inactive galaxies are generally characterized by lower $X/O$ levels. We show our $X/O$ diagram in Fig. \ref{fig:xo}: the plot includes the 629 sources in the X-ray sample, the 225 sAGN1 in the LS, and the RF5 AGN candidates. The sources in the X-ray sample that have $X/O >-1$ are 607 out of 629, and we therefore consider them to be AGN. All the sAGN1 belonging to our LS have $X/O >-1$ as well. For what concerns the 55 RF5 AGN candidates with an X-ray counterpart, all of them have $X/O >-1$, and most of this sample have higher $X/O$ values, with only seven sources having $X/O < 0$. The diagram hence confirms as AGN all the 55 X-ray emitters in our sample of RF5 AGN candidates. In particular, this sample includes 26 sAGN2 and confirms eight additional AGN not included in the sample of AGN candidates of \citetalias{decicco19} and with no further validation so far. For the 22 RF5 AGN candidates that do not have a counterpart in the \emph{Chandra}-COSMOS Legacy Catalog, we estimate upper limits for their hard-band X-ray fluxes using the CSTACK stacking analysis tool developed for \emph{Chandra} images \citep{cstack}, which also provides estimates for single sources. As expected, the upper limits generally lie below the nominal flux limit for the 2--10 keV band in the survey. It is also apparent that, based on these limits, all but three of these 22 sources are consistent with lying in the AGN locus on the $X/O$ diagram, which is consistent with their candidacy as AGN.

We also investigate the MIR properties of the RF5 AGN candidates and the sAGN1 LS. We summarize the results of the comparison of the RF selection and the various multiwavelength selection techniques in Table \ref{tab:mw}. The VST sources classified as AGN on the basis of the diagnostic proposed by \citet{donley} are 226. We find that 135 of these sources belong to the LS of sAGN1, while 16 additional sources belong to the RF5 AGN candidates. We also crossmatch our samples of sources with the R90 catalog of AGN candidates presented in \citet{assef18}: this consists of 4,543,530 AGN candidates with 90\% reliability, and were selected making use of data from the AllWISE Data Release of the Wide-field Infrared Survey Explorer (\emph{WISE}; \citealt{WISE}), according to the color-magnitude cut introduced in \citet{assef13}. This cut is based on the use of the two bands $W1$ (centered on 3.4 $\mu$m) and $W2$ (centered on 4.6 $\mu$m), and is defined for W2 $< 17.11$. Since our data are much deeper than this, we only find counterparts in the R90 catalog for 64 out of the 20,670 sources in the main sample: 40 of them are in the LS, and four of the remaining 24 are selected as AGN candidates by our RF5 classifier and are therefore included in the sample of RF5 AGN candidates. 

From Fig. \ref{fig:xo} we note that, among the 55 X-ray detected sources, five have only upper limits for the 2--10 keV flux; nevertheless, four of them are confirmed AGN according to their MIR properties (two are also in the sample of AGN candidates from \citetalias{decicco19}, so already confirmed), and the remaining is confirmed via its spectral energy distribution, after the information reported in the radio catalog described in \citet{smolcic}. Here we find radio counterparts for 20 of our RF5 AGN candidates; two of them are classified as AGN on the basis of their spectral energy distribution (both sources, one already confirmed in \citetalias{decicco19}) and of their X-ray properties (one of them, i.e., the source just mentioned). With regards to MIR properties, two sources are confirmed ex novo to be AGN following the Donley selection criterion; ten already confirmed sources are also validated by their MIR properties. To sum up, here we confirm eight new sources via X-ray properties, four via MIR properties, and one via SED properties. Since we have already confirmed 50 sources by comparison with the results from \citetalias{decicco19}, this leaves 14 sources in the RF5 AGN sample to confirm.

\begin{table*}[htb]
\renewcommand\arraystretch{1.4}
\caption{Comparison of the sample of AGN candidates selected via our RF5 classifier (1) and the samples of AGN candidates obtained via other selection techniques: the variability-based method from \citetalias{decicco19} (2); the $X/O$ diagram for the VST-COSMOS sources in the X-ray sample (3); the diagram introduced in \citet{donley} based on the use of MIR colors (4); and the color-magnitude cut used to obtain the catalog presented in \citet{assef18} (5). The first line in the table reports the number of AGN candidates from each selection. The second line reports the number of AGN candidates with a match in the sample of AGN candidates in this work; specifically, in the third line we split this number into the number of matched sources in the sample of RF5 AGN candidates and the number of matched sources in the LS.}\label{tab:mw}
\centering
\begin{tabular}{l c c c c c}
\hline  & RF5 & D19 & X & \citet{donley} & \citet{assef18}\\
 & (1) & (2) & (3) & (4) & (5)\\ 
\hline
\vspace{2mm}
AGN candidates & 302 (77+225) & 271 & 607 & 226 & 64\\
matched in this work & - & 240 & 280 & 151 & 44\\ 
(RF5 sample plus LS) & & (56+184) & (55+225) & (16+135) & (4+40)\\
\bottomrule
\end{tabular}
\end{table*}

An important consideration in all of the above is that the AGN LS, due to wealth of ancillary data available in COSMOS, appears to already contain a large majority of the total AGN found. As a consequence, to build an effective LS, we need to use most of the reliable AGN in this area that are known from the literature. As a consequence, limiting the analysis to the remaining, unclassified sources would result in a severe underestimate of the classifier performance, especially since most of the AGN-dominated sources are included in the LS, resulting in a biased unlabeled set. In fact the optimal use of such classifiers is based on the availability of  independent label and validation sets, as will be the case for large surveys. We ideally want to understand how our classifier will perform in surveys such as LSST, which will produce similar optical data in terms of depth and cadence, but largely lack the equivalent multiwavelength and spectroscopic information to develop a similar LS.
The use of the LOOCV (introduced in Sect. \ref{section:performance}) has the major advantage of considering each of the sources in the LS as if it were unlabeled: the source is excluded from the LS and a classification is provided for it using all the remaining sources in the LS. Essentially, this is equivalent to including each source, in turn, in the unlabeled set. In this way we can take the AGN in the LS into account when estimating the purity\footnote{We note that, when referring to the LS, we use the terms ``precision'' and ``recall''; we use ``purity'' and ``completeness'' when referring to the labeled plus unlabeled sets.} and completeness of our results for each classifier: we combine the obtained classifications for each labeled and unlabeled set, and hence the fact that our LS is internal to the area under investigation will not affect our results. In the case of the RF5 classifier, the known AGN classified correctly are 212/225 (TPs), leaving 13 misclassified AGN (FNs). Based on this, we can state that the purity of our total sample of AGN is 91\% ($212 + 63 = 275$ confirmed AGN in total, over 302 total candidates), which is higher than the result from \citetalias{decicco19}.



\citetalias{decicco19} estimate the completeness of the sample of AGN candidates with respect to the most robust sample of known AGN, that is, those with a spectroscopic confirmation and an X-ray counterpart. The obtained value is 59\%. If we estimate the completeness for each classifier in the same way we estimated the purity above, we obtain a 69\% completeness for the RF5 classifier, which represents a moderate improvement with respect to past results. This increase is mostly a reflection of the completeness increase for Type I AGN (+12\% with respect to \dd), while the completeness with respect to Type II AGN is raised by 3\%. Table \ref{tab:results} presents the main results obtained for the labeled plus unlabeled sets from various classifiers tested in this work. We note that, since the nature of the unconfirmed sources is unknown, purity estimates are always to be intended as lower limits.

\begin{table*}[htb]
\renewcommand\arraystretch{1.4}
\caption{Results from various classifiers tested in this work. 
The first division of six LSs focuses on the performance of the classifiers defined in Sects. \ref{section:ls_tests}. The middle division of three LSs (marked by stars) denotes the performance of certain classifiers when the \texttt{ch21} color is added to the other features used, as discussed in Sect. \ref{section:MIR_col}. In order to ease the comparison between the classifier with and without the \texttt{ch21} color, we indicate corresponding pairs with red, blue, or purple colors. In the third division, we include results from RF$_{col}$ to allow for comparison with a classifier where no variability features are used, as well as results from \citetalias{decicco19} in the fourth division to allow for comparison with our current findings. We report: {\it Col.1}: the AGN LS defining the classifier; {\it Col. 2}: the number of AGN in the LS; {\it Col.3}: the number of AGN candidates in the unlabeled set; {\it Col.4}: the number of multiwavelebngth confirmed AGN in the unlabeled set and the fraction with respect to the number of candidates; {\it Col.5}: for the three color-coded pairs of classifiers, the number of candidates in common and the percentage they represent among the total number of candidates; {\it Col.6}: purity (this is always a lower-limit estimate); {\it Col.7}: completeness with respect to spectroscopic AGN; {\it Col.8}: completeness with respect to spectroscopic Type I AGN; {\it Col.9}: completeness with respect to spectroscopic Type II AGN; {\it Col.10}: completeness with respect to MIR AGN. Purity and completeness are computed taking into account the AGN in each LS in addition to the confirmed AGN in the unlabeled set, as explained in Sect. \ref{section:mw_properties}.}\label{tab:results}
\centering
  \resizebox{\textwidth}{!}{  
\setlength{\tabcolsep}{3mm}
\begin{tabular}{l c c c c c c c c c}
\hline AGN LS & AGN in the LS & AGN candidates & Confirmed AGN & Matched & Purity & Completeness & Type I compl. & Type II compl. & MIR compl.\\
 & & in the unlabeled set & & & & & & \\
(1) & (2) & (3) & (4) & (5) & (6) & (7) & (8) & (9) & (10)\\ 
\hline
\red{sAGN1 (RF5)} & \red{225} & \red{77} & \red{63 (82\%)} & \red{65 (84\%)} & \red{91\%} & \red{69\%} & \red{94\%} & \red{21\%} & \red{64\%}\\
sAGN1+sAGN2, $r < 21$ mag & 69 & 182 & 173 (95\%) & - & 90\% & 59\% & 82\% & 18\% & 56\% \\
MIR-selected AGN & 226 & 136 & 116 (85\%) & - & 72\% & 66\% & 91\% & 19\% & 64\%\\
\blue{sAGN1+sAGN2} & \blue{347} & \blue{70} & \blue{38 (54\%)} & \blue{58 (83\%)} & \blue{68\%} & \blue{71\%} & \blue{95\%} & \blue{28\%} & \blue{65\%}\\
sAGN2 & 122 & 194 & 179 (92\%) & - & 65\% & 51\% & 67\% & 22\% & 46\%\\
\violet{all AGN types} & \violet{414} & \violet{99} & \violet{34 (34\%)} & \violet{60 (61\%)} & \violet{58\%} & \violet{72\%} & \violet{95\%} & \violet{30\%} & \violet{67\%}\\
\midrule
\red{sAGN1*} & \red{225} & \red{67} & \red{52 (78\%)} & \red{65 (97\%)} & \red{91\%} & \red{68\%} & \red{95\%} & \red{18\%} & \red{65\%}\\
\blue{sAGN1+sAGN2*} & \blue{347} & \blue{64} & \blue{38 (59\%)} & \blue{58 (91\%)} & \blue{69\%} & \blue{71\%} & \blue{96\%} & \blue{26\%} & \blue{66\%}\\
\violet{all AGN types*} & \violet{414} & \violet{81} & \violet{33 (41\%)} & \violet{60 (74\%)} & \violet{65\%} & \violet{76\%} & \violet{98\%} & \violet{36\%} & \violet{78\%}\\
\midrule
sAGN1, RF$_{col}$ & 225 & 499 & 46 (9\%) & - & 40\% & 61\% & 89\% & 9\% & 62\%\\ 
\midrule
\citetalias{decicco19} & - & 299 & 256 (86\%) & - & 86\% & 59\% & 82\% & 18\% & 55\%\\
\bottomrule
\end{tabular}}
\end{table*}

\begin{figure}[htb]
 \centering
\subfigure
            {\includegraphics[width=9cm]{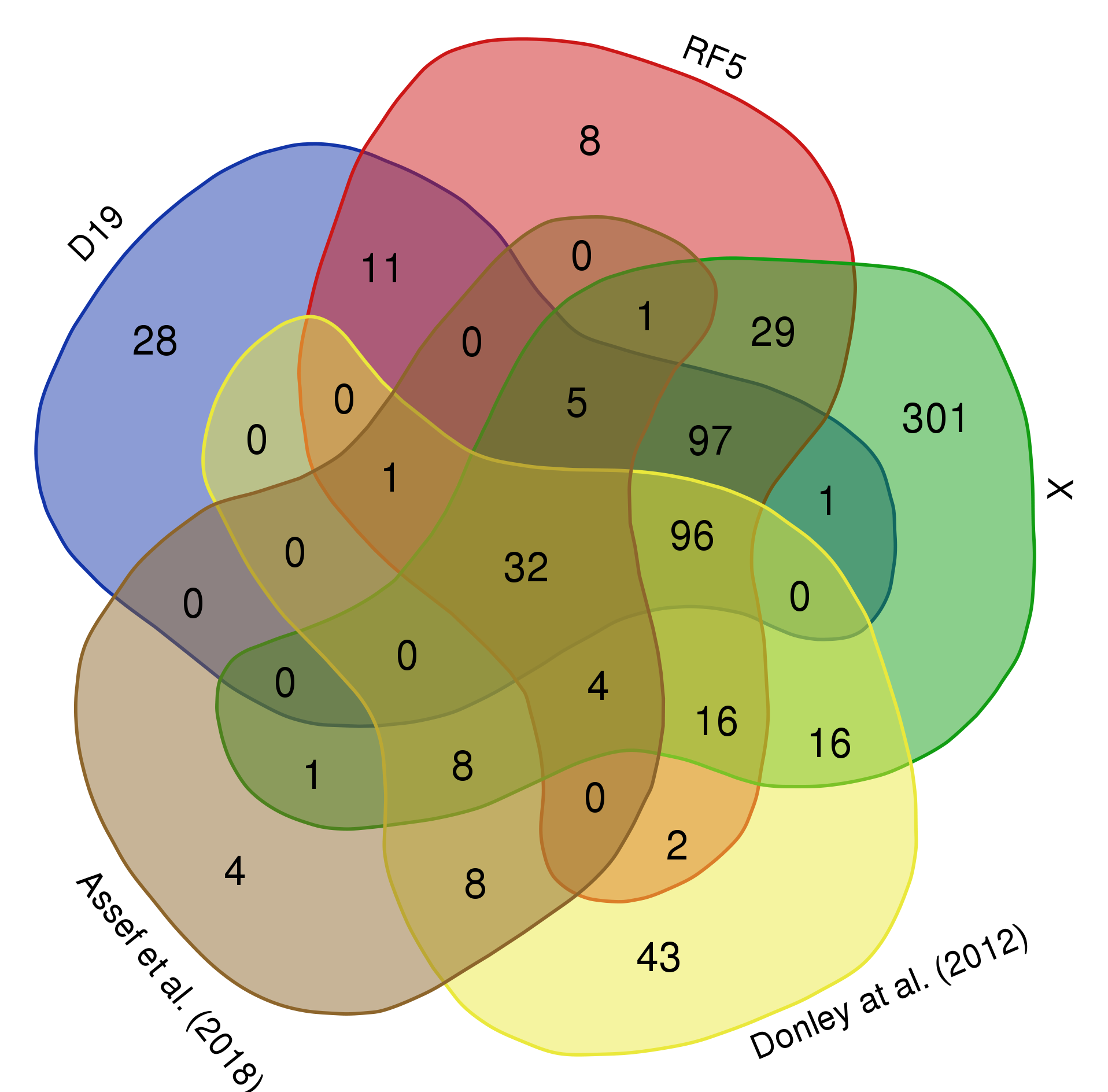}}
\caption{\footnotesize Venn diagram showing: the sample of AGN candidates returned by the RF5 classifier (RF); the sample of AGN candidates from D19; the VST sources in the X-ray sample classified as AGN on the basis of the $X/O$ diagram (X); the VST sources in the main sample classified as AGN on the basis of the MIR diagnostic from \citet{donley}; and the VST sources in the main sample classified as AGN on the basis of the criterion defined in \citet{assef18}. We caution that the area covered by each region is not related to the number of sources in that region. This diagram was rendered via \url{http://bioinformatics.psb.ugent.be/webtools/Venn/}.}\label{fig:venn}
\end{figure}

Figure \ref{fig:venn} shows a Venn diagram \citep{venn} comparing the selection of AGN candidates obtained in this work to: the sample selected in \citetalias{decicco19}; the sources in the X-ray sample classified as AGN based on the $X/O$ diagram; all the AGN in the main sample selected via the criterion defined in \citet{donley}; and all the sources in the main sample classified as AGN in the catalog presented in \citet{assef18}. Following the considerations above, the sample of RF5 AGN candidates shown here also includes the sAGN1 in the LS, in order to make a proper comparison with the other samples of sources. We also note that the sample from \citetalias{decicco19} is limited to the 271 sources in common with the main sample of sources used in the present work. 
The diagram shows how the selection of AGN candidates from this work largely overlaps the sample of AGN candidates from \citetalias{decicco19} but potentially expands the selection, and how a significant fraction of the X-ray and MIR samples do not overlap the other two: the bulk of these sources are heavily obscured at optical wavelengths, and hence missed by optical variability surveys.

\subsection{Results from the RF\texorpdfstring{$_{spec21}$}{} classifier}
\label{section:spec21}
In Sect. \ref{section:ls_tests}, among the various LS tests, we introduced the RF$_{spec21}$ classifier, where the sAGN1 and sAGN2 subsamples included in the LS are cut at $r \le 21$ mag. Such a selection for our LS is intriguing as it allows us to predict how well a classifier trained only with bright AGN can perform in the identification of fainter (down to 2.5 mag fainter, in this case) AGN. This is a likely scenario for the full-sky area of LSST WFD, as deep spectroscopic samples are only likely to exist over a few tens of sq. deg. given the current (e.g., COSMOS) and expected (e.g., LSST DDFs) instrumentation, meaning we will have relatively limited training sets with deep spectroscopic identification.

We shift all faint confirmed AGN in the parent LS --that is, 169 sAGN1 and 109 sAGN2-- to the unlabeled set for this test. The sample of optically variable AGN candidates returned by the RF$_{spec21}$ classifier consists of 182 sources (hereafter, RF$_{spec21}$ AGN candidates). These include 132 out of the 169 sAGN1 not included in the LS (78\%) and 21/109 (19\%) of the sAGN2 not included in the LS.

\begin{figure*}[tbh]
 \centering
\subfigure
            {\includegraphics[width=6cm]{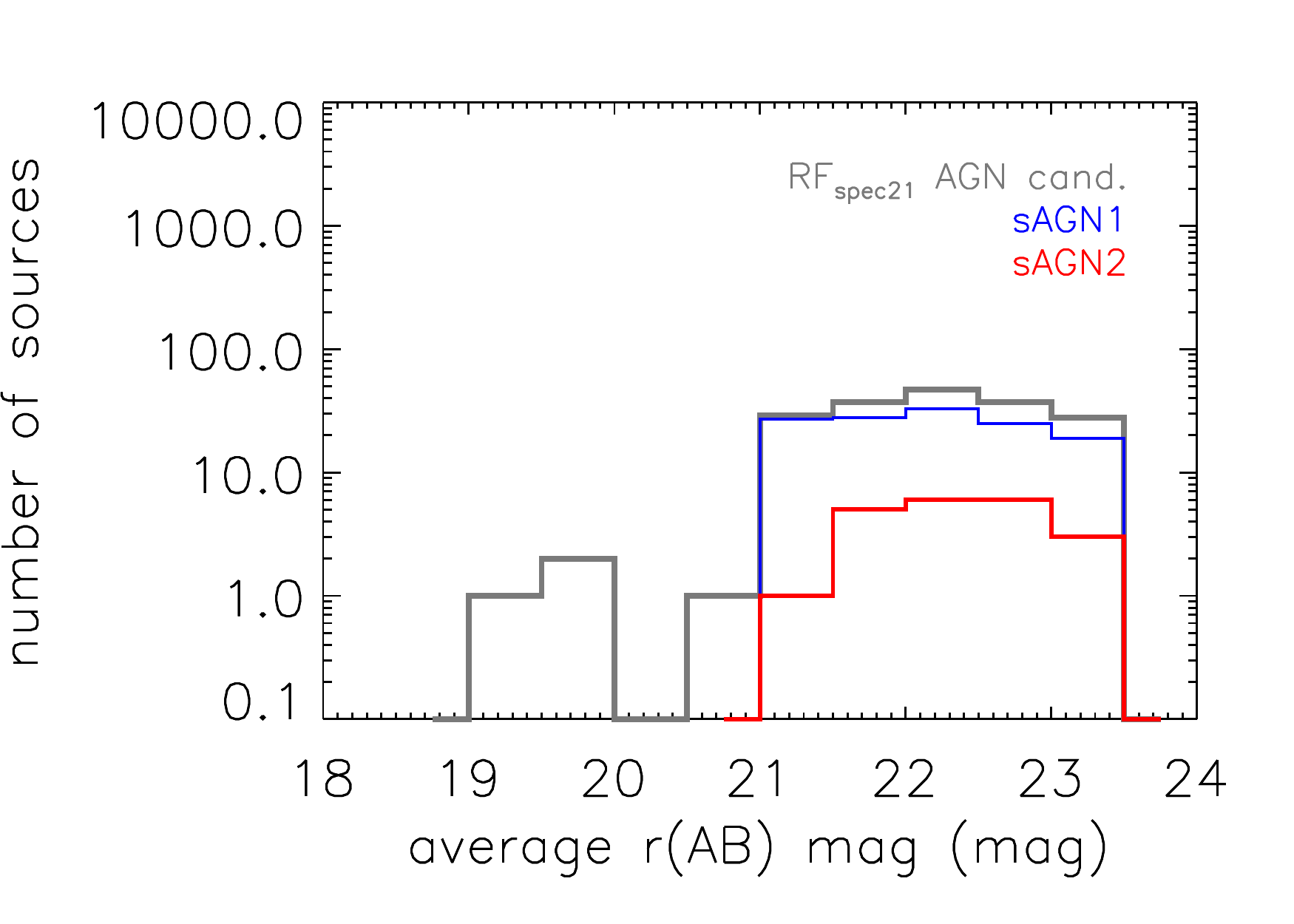}}
\subfigure
            {\includegraphics[width=6cm]{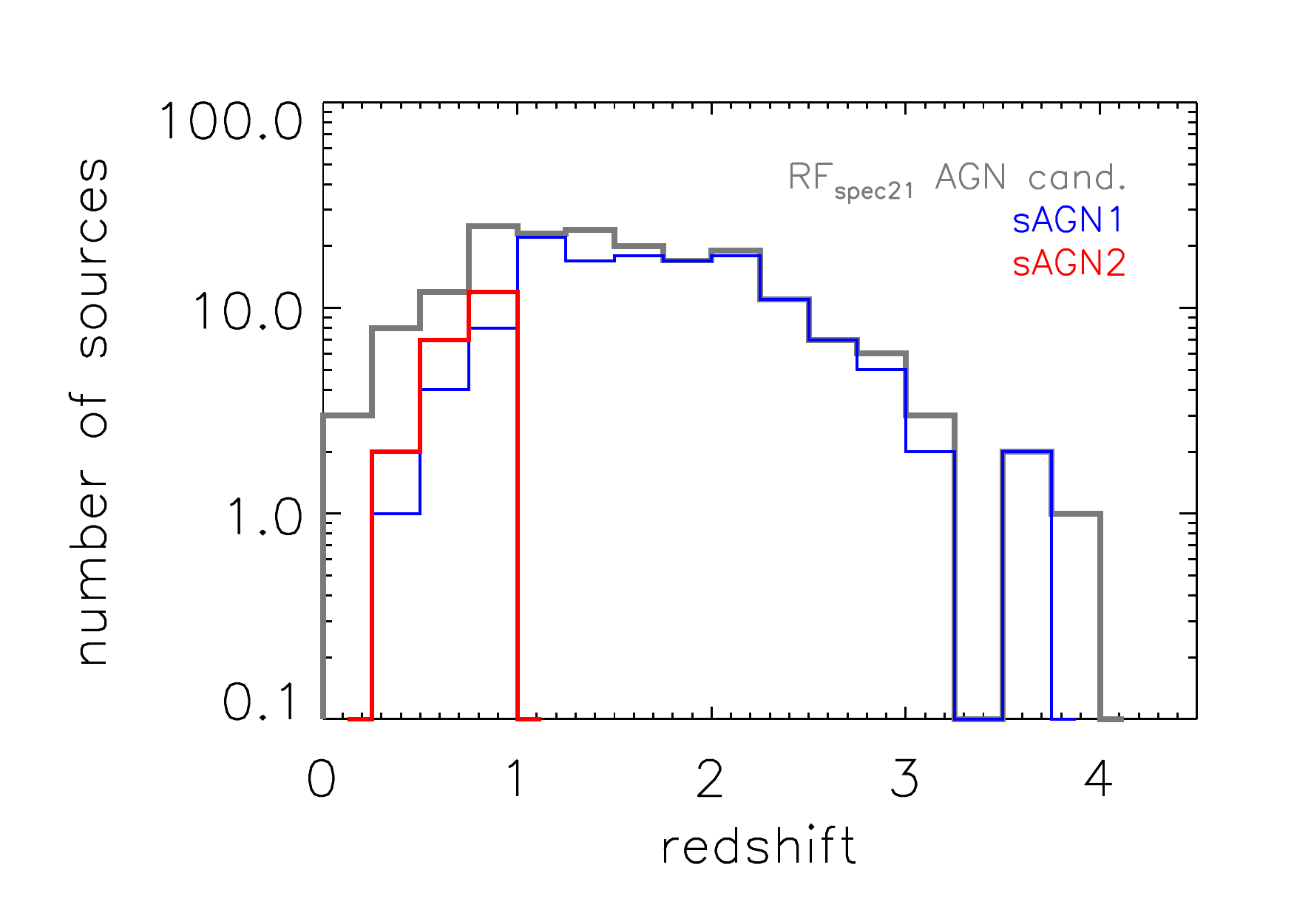}}
\subfigure
            {\includegraphics[width=6cm]{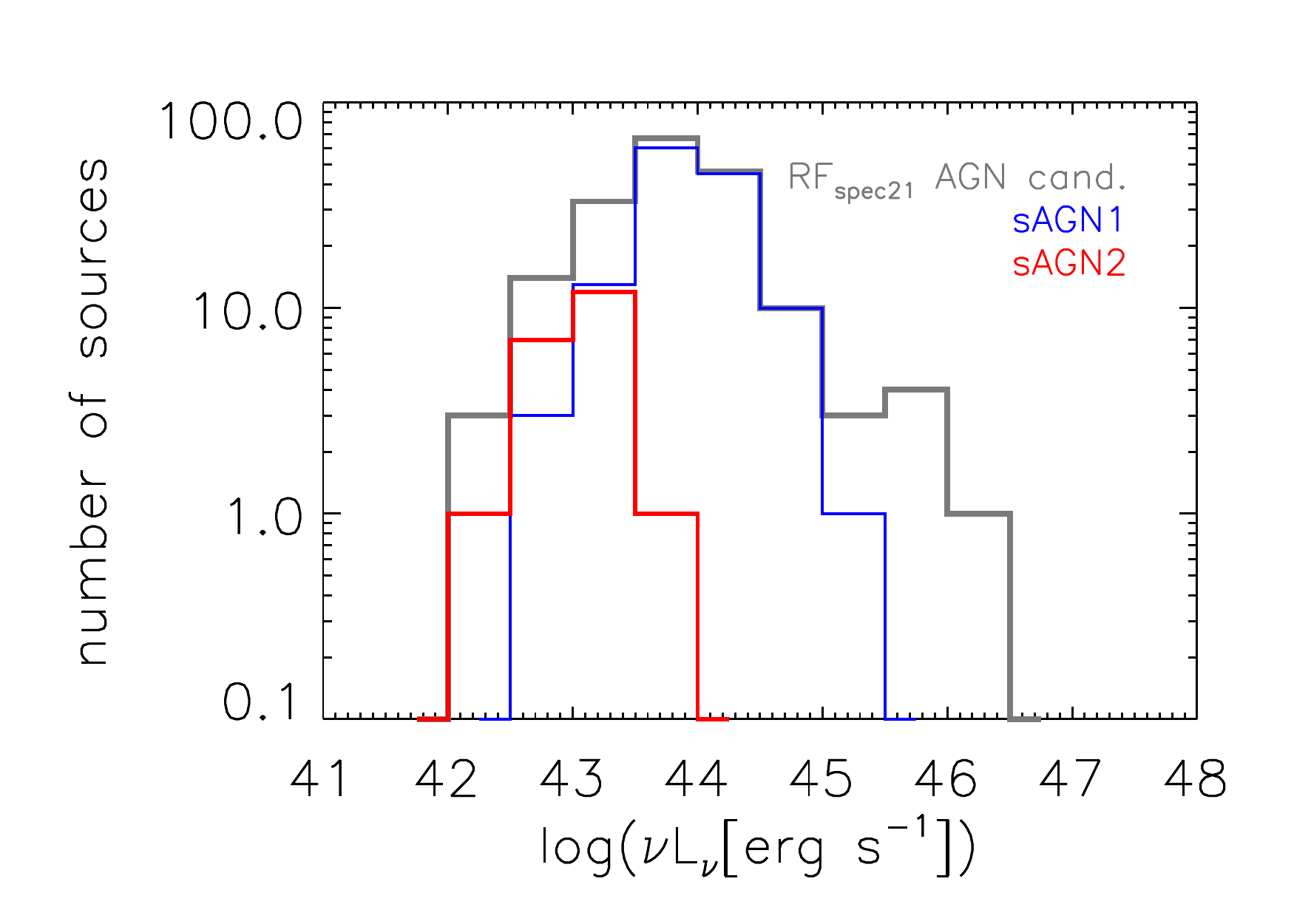}}\\
   \caption{\footnotesize{Distribution of magnitude (\emph{left}), redshift (\emph{center}), and $r$-band luminosity (\emph{right}) for the sample of 182 AGN candidates (gray line) obtained from the RF$_{spec21}$ classifier. sAGN1 and sAGN2 in the sample are also shown (blue and red lines, respectively).}}\label{fig:rf21cand}
   \end{figure*}

Fig. \ref{fig:rf21cand} shows the magnitude, redshift, and luminosity distributions for the RF$_{spec21}$ AGN candidates as well as the subsamples of sAGN1 and sAGN2 there included. The classifier is able to retrieve sAGN1 and sAGN2 with $r > 21$ mag: indeed, out of the 182 candidates, only four have a magnitude $r <$ 21 mag. We are able to recover sources across the whole redshift range of the main sample (see Fig. \ref{fig:hists_with_colors}). The $r$-band luminosity of this sample of AGN candidates is $> 10^{42}$ erg s$^{-1}$, hence it is ${\approx}4$ dex brighter than the faintest sources in the main sample (Fig. \ref{fig:hists_with_colors}). Consistent with the characteristics of the whole spectroscopic AGN sample, here we find sAGN1 at higher luminosities than sAGN2. We can conclude that, for what concerns magnitudes and redshifts, the sample of AGN candidates are quite representative of the sources in the main sample and, specifically, the subsamples of sAGN1 and sAGN2 included in the AGN candidates are representative of the whole samples of sAGN1 and sAGN2, respectively. For what concerns luminosities, the sample of AGN candidates is shifted toward the medium-high luminosity range if compared to the main sample, yet its faint end is faint enough to be of interest. All this shows that, even though our LS did not include AGN fainter than $r = 21$ mag, this does not prevent us from identifying fainter AGN.

It is worth mentioning that 172 out of the 182 RF$_{spec21}$ AGN candidates also belong to the sample of AGN candidates in \citetalias{decicco19} and 168 of them are confirmed to be AGN in that work; of the remaining 14 RF$_{spec21}$ AGN candidates, nine are confirmed by their X-ray and/or MIR properties. This leaves five out of 182 sources that do not have any confirmation, meaning that this sample of AGN candidates has a purity $\ge 97\%$. As reported in Table \ref{tab:results}, this classifier returns one of the lowest completeness values (59\%) if compared to the others.

When we compare the RF$_{spec21}$ AGN candidates to the RF5 AGN candidates, we find that:
\begin{itemize}
    \item[--] 50 sources are in common between RF$_{spec21}$ and RF5: 41 of them are confirmed to be AGN based on their multiwavelength properties, while the remaining nine sources have no secondary confirmation.\\
    \item[--] 132 RF$_{spec21}$ AGN candidates belong to the 
    RF5 LS of AGN and hence have no match among the RF5 AGN candidates; they have not been included in the RF$_{spec21}$ LS of AGN as they have $r > 21$ mag.\\
    \item[--]27 RF5 AGN candidates have no match among the RF$_{spec21}$ AGN candidates. One of them belongs to the RF$_{spec21}$ LS of AGN and is an sAGN2, five are sAGN2, and 12 are multiwavelength confirmed AGN. For the remaining nine candidates there is no secondary confirmation.
\end{itemize}

\begin{figure}[tbh]
 \centering
\subfigure
            {\includegraphics[width=\columnwidth]{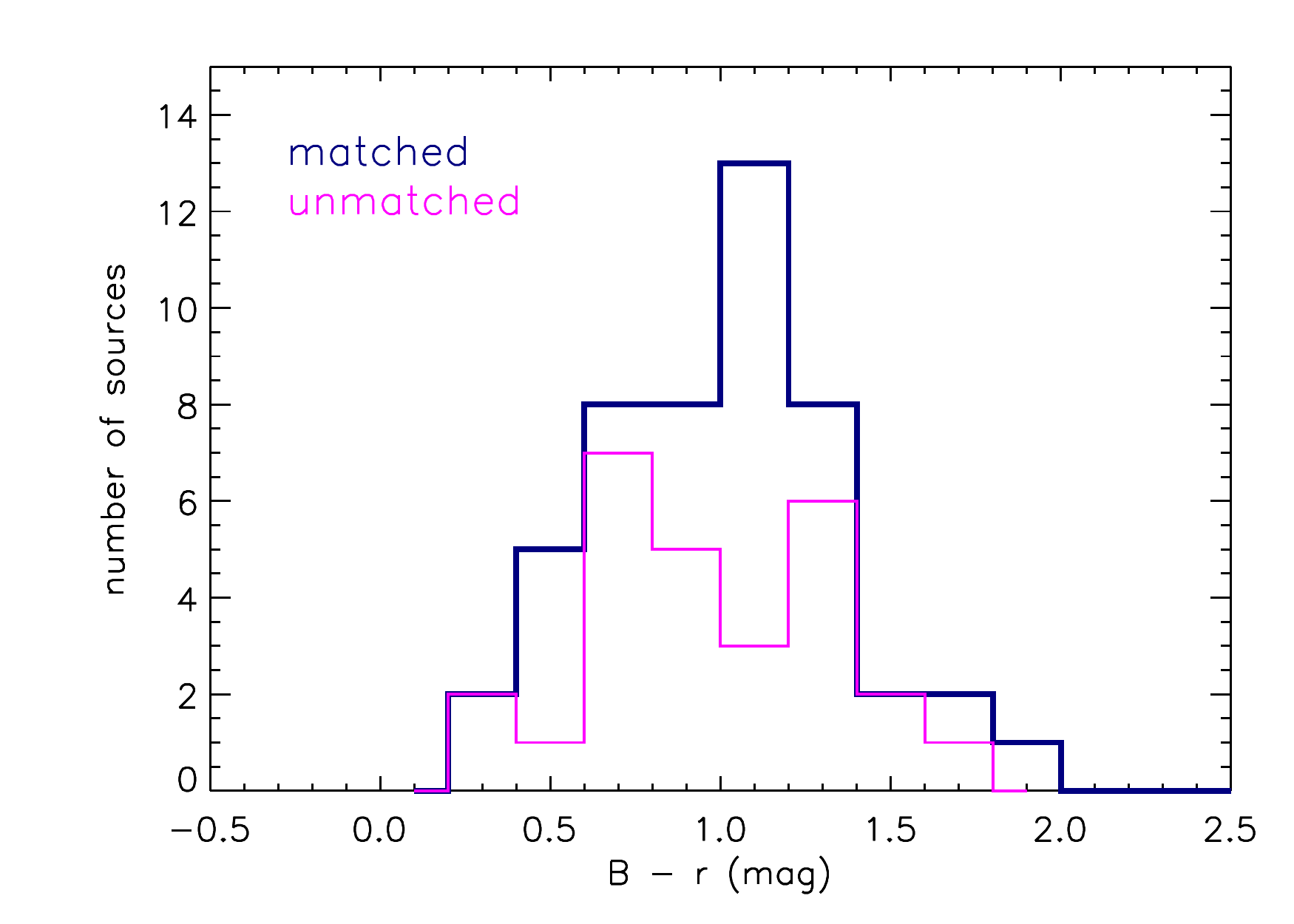}}
\subfigure
            {\includegraphics[width=\columnwidth]{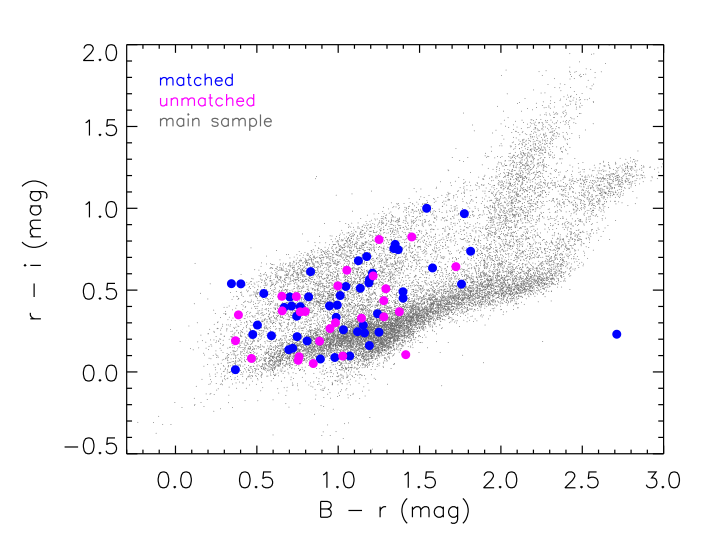}}
   \caption{\footnotesize{Distribution of the color \texttt{B-r} (\emph{upper panel}) and color-color diagram comparing \texttt{B-r} and \texttt{r-i} (\emph{lower panel}) for the sample of matched (blue) and unmatched (magenta) sources obtained comparing the RF5 and RF$_{spec21}$ samples of AGN candidates. In the color-color diagram, the main sample sources are also shown as a reference (gray).}}\label{fig:rf5_rfspec21}
   \end{figure}

We inspect the distributions of each color feature for the 50 matched sources as well as the 27 unmatched RF5 AGN candidates, and find that both are well distributed in terms of colors and extend to the same red end. This stands for each color, meaning that the unmatched RF5 AGN candidates do not have peculiar color properties. This suggests that, even when limiting the AGN LS to a bright ($r < 21$ mag) sample, we are able to identify faint red AGN, likely hosted by a dominant galaxy or obscured by dust. This is a considerable result in view of future surveys, such as LSST. In Fig. \ref{fig:rf5_rfspec21}, as an example, we show the distribution for the color \texttt{B-r} and the color-color diagram comparing \texttt{B-r} and and \texttt{r-i} for these two subsamples of sources. A key point to thoroughly characterize and evaluate the performance of the RF$_{spec21}$ classifier is whether the nine unmatched RF5 AGN candidates with no secondary confirmation can be confirmed as AGN or not. With \emph{r} magnitudes of 22--23.5 mag, these will require spectroscopic confirmation with 4--8m telescopes to unveil their nature.

\subsection{Results from the RF\texorpdfstring{$_{col}$}{} classifier}
\label{section:col_only}
In Sect. \ref{section:color_test} we introduced the $RF_{col}$ classifier, where the only features we use are the five colors \texttt{u-B}, \texttt{B-r}, \texttt{r-i}, \texttt{i-z}, \text{z-y}, and the stellarity index. Based on the obtained CM (Fig. \ref{fig:cm_12345_col}), the performance of this classifier is comparable to that of the other classifiers that add colors to variability features and stellarity, suggesting that the variability-based approach does not bring any substantial improvement to a ``color only''-based classification.

Here we describe the results obtained using RF$_{col}$ to classify our unlabeled set of sources. Including only sAGN1 in the AGN LS, we obtain a sample of 499 AGN candidates: this is much larger than the sample of 77 RF5 AGN candidates or the 182 RF$_{spec21}$ AGN candidates introduced above and, more importantly, only 46 out of 499 candidates are confirmed AGN based on their multiwavelength or spectroscopic properties. A cross-match with the COSMOS2015 catalog (see Sect. \ref{section:labeled_set}) reveals an 8\% contamination by inactive galaxies based on the best-fit templates reported in the catalog; of course in principle there could be AGN hidden among apparently inactive galaxies. We also test the performance of RF$_{col}$ adopting the other AGN LSs defined in Sect. \ref{section:labeled_set}. As we include different AGN types in the LS, we see that the size of the sample grows up to almost 1,000 sources while the purity and completeness tend to decrease and increase, respectively. Depending on the LS used, only ${\sim}8$--9\% of these candidates are confirmed by other methods, while 5 to 23\% of the candidates are likely inactive galaxies based on the information provided in the COSMOS2015 catalog.

We crossmatch the above-mentioned sample of 499 RF$_{col}$ AGN candidates and the sample of 77 RF5 AGN candidates, and find 36 sources in common, including 30 AGN confirmed by their multiwavelength properties. The RF$_{col}$ sample includes 16 multiwavelength confirmed AGN (3\% of 499) not found in the RF5 sample, while the RF5 includes 29 multiwavelength confirmed AGN (38\% of 77) not found in the RF$_{col}$ sample.

\begin{figure}[tbh]
 \centering
\subfigure
            {\includegraphics[width=\columnwidth]{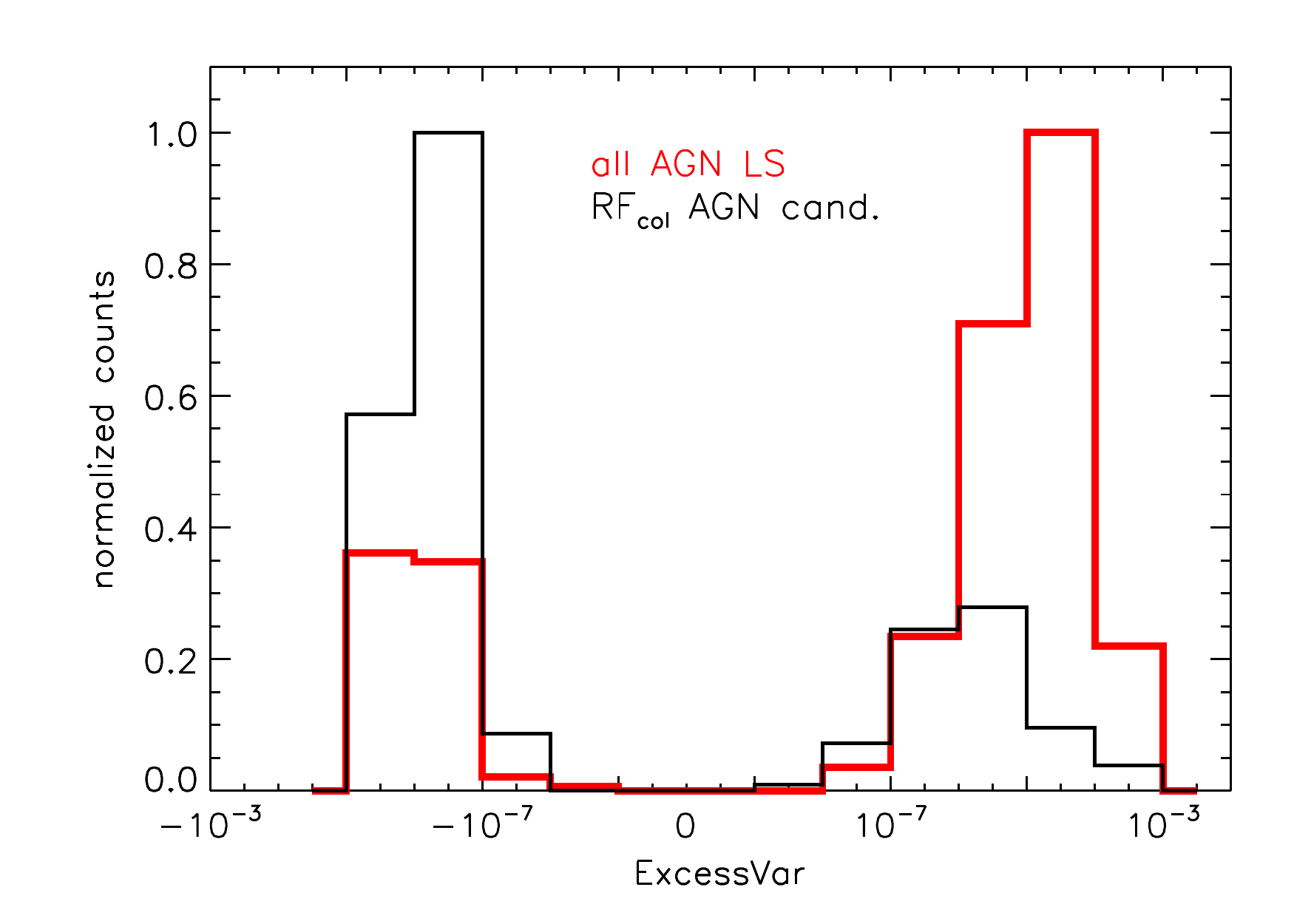}}
\subfigure
            {\includegraphics[width=\columnwidth]{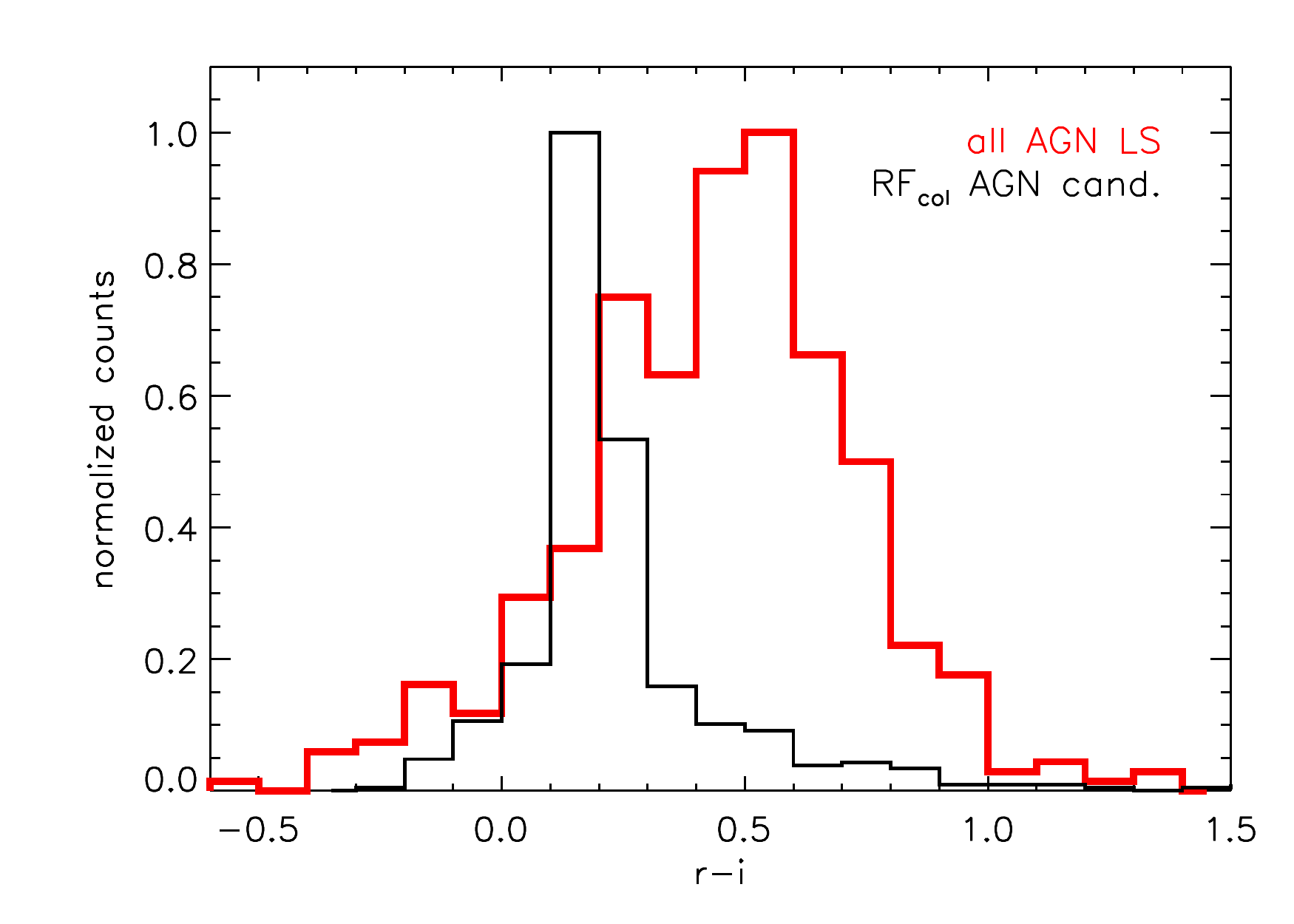}}
\subfigure
            {\includegraphics[width=\columnwidth]{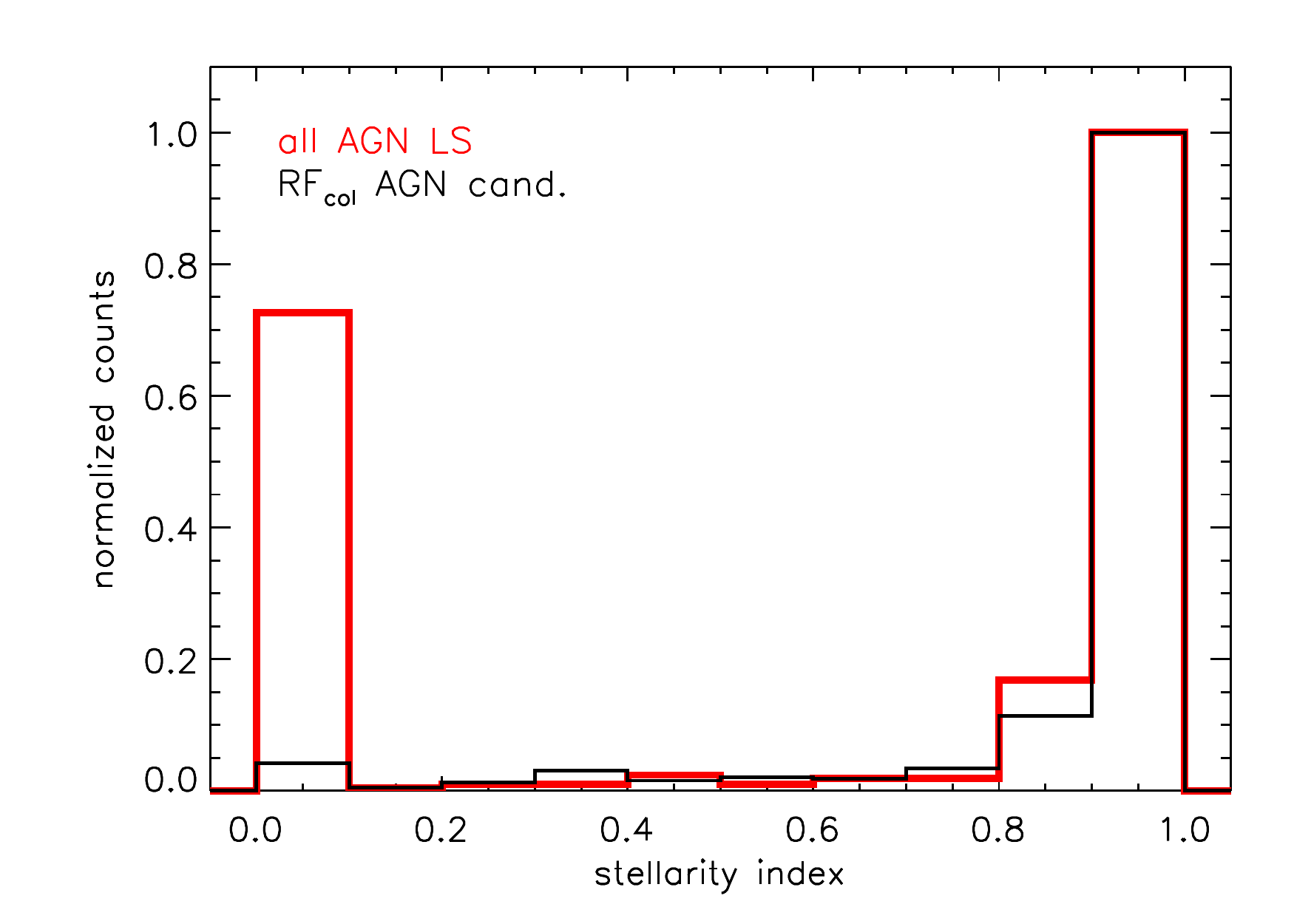}}
   \caption{\footnotesize{\texttt{ExcessVar} (\emph{top}), \texttt{r-i} (\emph{center}), and stellarity index (\emph{bottom}) distributions for the sample of 414 known AGN used to build the various LSs in this work (red) and the sample of AGN candidates obtained from RF$_{col}$ when only sAGN1 are included in the LS (black).}}\label{fig:distrib_comp_rfcol}
   \end{figure}

We inspect the distributions of each feature for the sample of 414 known AGN used to build the various LSs in this work (see Sect. \ref{section:labeled_set}), and compare them to the corresponding sample of 499 RF$_{col}$ AGN candidates. Three examples of these distribution pairs (one for each type of feature: variability, stellarity, and color) are shown in Fig. \ref{fig:distrib_comp_rfcol}. We point out that the sample of 414 AGN is not homogeneous, as it was selected based on different characteristics; nonetheless, the AGN distributions are always very different from the ones obtained for the RF$_{col}$ AGN candidates at issue. This is also supported by the results of the K-S test, which returns a maximum distance $D \geq 0.36$ between the distributions in each pair, while the probability to obtain a larger distance $D$ assuming that the distributions are drawn from the same distribution function is $p < 10^{-25}$. This indicates that consistency in each pair is very unlikely, suggesting that the sample of AGN candidates probably includes a large number of contaminants. From Fig. \ref{fig:distrib_comp_rfcol} we also note that the \texttt{r-i} distribution of the sample of AGN candidates is bluer than the distribution of the 414 AGN: this also holds for all the other colors used here, and confirms that colors alone are not effective in identifying AGN characterized by redder colors, that is to say, with a strong host-galaxy component in their emission.

As seen in Fig. \ref{fig:fi_analysis_col}, the colors of Type II AGN, inactive galaxies, and stars show substantial overlap; to a lesser extent, the colors of Type I AGN also overlap, but the combination of colors and stellarity is sufficient to identify most of them in the LS, and this explains the good results from the confusion matrix. When dealing with the unlabeled set we need to take into account that, very likely, it will include a large fraction of Type II AGN, which are harder to separate from non-AGN, and hence we obtain a larger sample of AGN candidates but with substantial contamination. We should also keep in mind that, by excluding variability features, we are leaving out some of the most important features in the ranking, independent of the specific AGN LS used, and this inevitably leads to increasing contamination as it worsens the ability of the classifier in breaking degeneracy among different classes of sources. 

From Table \ref{tab:results} we can see that, while this classifier is associated with the lowest purity (40\%) if compared to the others, its completeness is 61\%: in particular, it is of 89\% for Type I AGN and of 62\% for MIR AGN, consistent with other classifiers in the table, while it drops to 9\% for Type II AGN.

\section{Test including the MIR color \texorpdfstring{$4.5-3.6$}{} \texorpdfstring{$\mu m$}{}}
\label{section:MIR_col}
In Sect. \ref{section:tests} we discussed how optical variability features and colors are not ideal for identifying obscured AGN. Since this work is LSST-oriented, so far we have made use of features that, in the future, we will be able to derive from LSST data only. Nonetheless, we already know that LSST data can benefit from complementary MIR information coming from various surveys. With this in mind, here we test the inclusion of a MIR color in our set of features, as it proves effective in identifying Type II AGN given their properties, as discussed in Sect. \ref{section:labeled_set}; the color is obtained as the difference of the magnitudes in the 4.5 and 3.6 $\mu$m MIR bands. Specifically, these two bands will be available for the DDFs via, for example, the SWIRE/SERVS/DeepDrill surveys from the \emph{Spitzer Space Telescope} \citep[e.g.,][]{swire,servs} down to ${\approx}23$ AB magnitudes in both bands, while for the WFD survey they will be provided by the \emph{WISE} and \emph{NEOWISE} \citep{NEOWISE} missions, where the two bands correspond to the two bands $W2$ and $W1$ mentioned in Sect. \ref{section:mw_properties}. We should point out that, based on the CATWISE 2020 release \citep{eisenhardt}, the $5\sigma$ AB magnitude limits of $W1$ and $W2$ are 20.37 and 19.81, respectively, which means that the chosen color feature will prove useful only for the brighter AGN population observed by LSST.

The data we use here are from the \emph{Spitzer} Large Area Survey with Hyper-Suprime-Cam (SPLASH) program, where the two bands at 3.6 and 4.5 $\mu$m are known as \emph{channel1} and \emph{channel2}, respectively. We find that 23 out of the 20,670 sources in our main sample do not have a detection in at least one of these two bands, hence we exclude them from this test. Since the new feature we are using is defined as the difference between the \emph{channel2} and \emph{channel1} magnitudes, we name it \texttt{ch21}. Consistent with the rest of this work, we test different AGN LSs, consisting of: 
\begin{itemize}
    \item[--]only the 225 sAGN1;
    \item[--]only the 347 sAGN1 + sAGN (i.e., the AGN LS used to build the RF$_{spec}$ classifier);
    \item[--]the full LS of 414 AGN.
\end{itemize}

\begin{figure*}[tbh]
 \centering
\subfigure
            {\includegraphics[width=7cm]{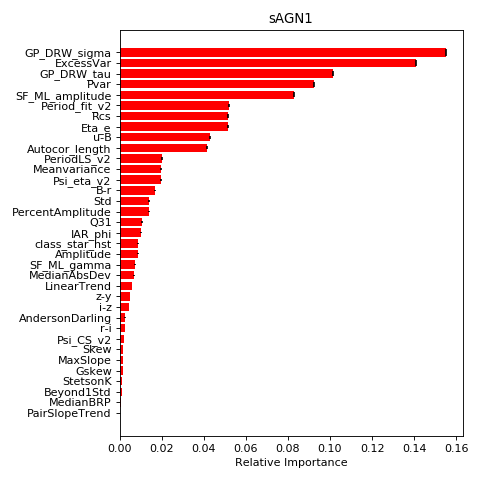}}
\subfigure
            {\includegraphics[width=7cm]{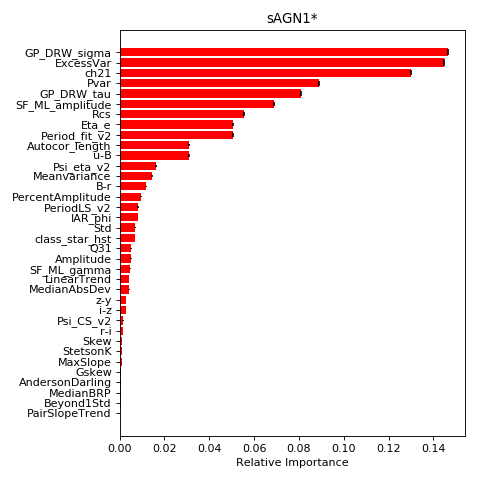}}
\subfigure
            {\includegraphics[width=7cm]{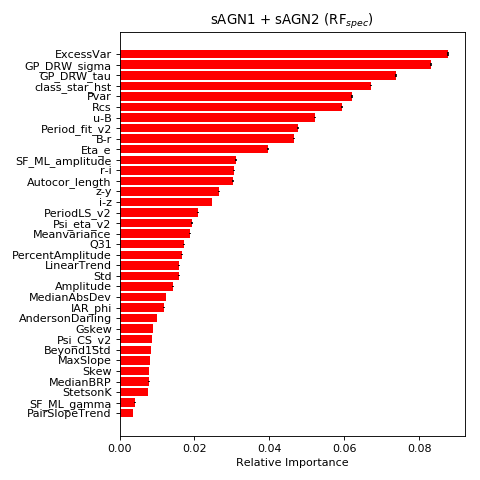}}
\subfigure
            {\includegraphics[width=7cm]{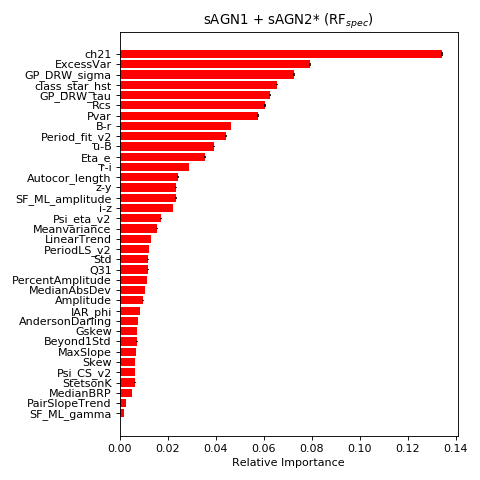}}
\subfigure
            {\includegraphics[width=7cm]{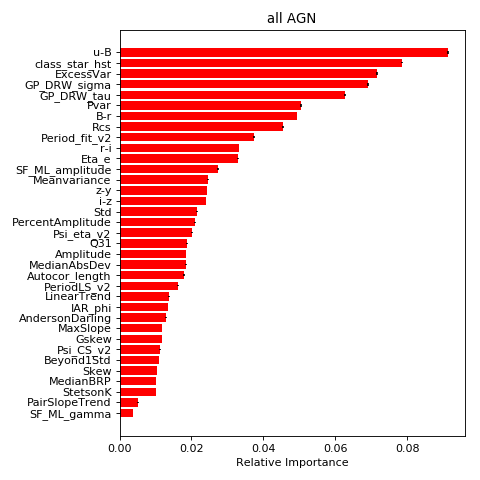}}
\subfigure
            {\includegraphics[width=7cm]{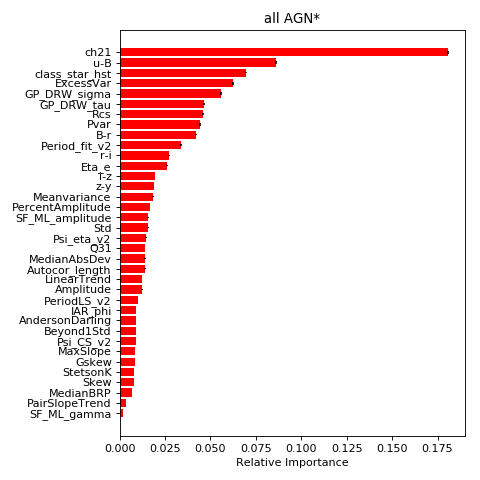}}
            
   \caption{\footnotesize{Feature importance ranking corresponding to the LS where: only sAGN1 are included (\emph{top panels}); spectroscopic AGN (i.e., sAGN1 plus sAGN2) are included (\emph{center}); all AGN types are included (\emph{bottom}). Left panels correspond to the feature set consisting of variability features plus optical and NIR colors plus stellarity index; right panels correspond to the same set plus the MIR color $4.5 - 3.6$ $\mu m$ (labeled as \texttt{ch21}), and the names of the AGN LS identifying these three classifiers are marked by a star. Each plot includes error bars for each feature (small vertical black bars at the right end of each feature bar): these are defined as the standard deviation of the importance of each specific feature over the total number of trees in the forest. The standard deviation depends on how important a feature is for the various trees, therefore larger error bars indicate that the feature importance varies considerably depending on the tree.
   }}\label{fig:f_i_comparison}
   \end{figure*}

Fig. \ref{fig:f_i_comparison} compares the feature importance ranking for each pair of classifiers tested with these LS, with and without the \texttt{ch21} feature. It is apparent that this feature is the third in the ranking for the classifier with only sAGN1 in the LS, while it is the most important feature for the other two classifiers. Specifically, it is dramatically more important than all the other features when all AGN types are included in the LS. 

\begin{figure}[htb]
 \centering
\subfigure
            {\includegraphics[width=\columnwidth]{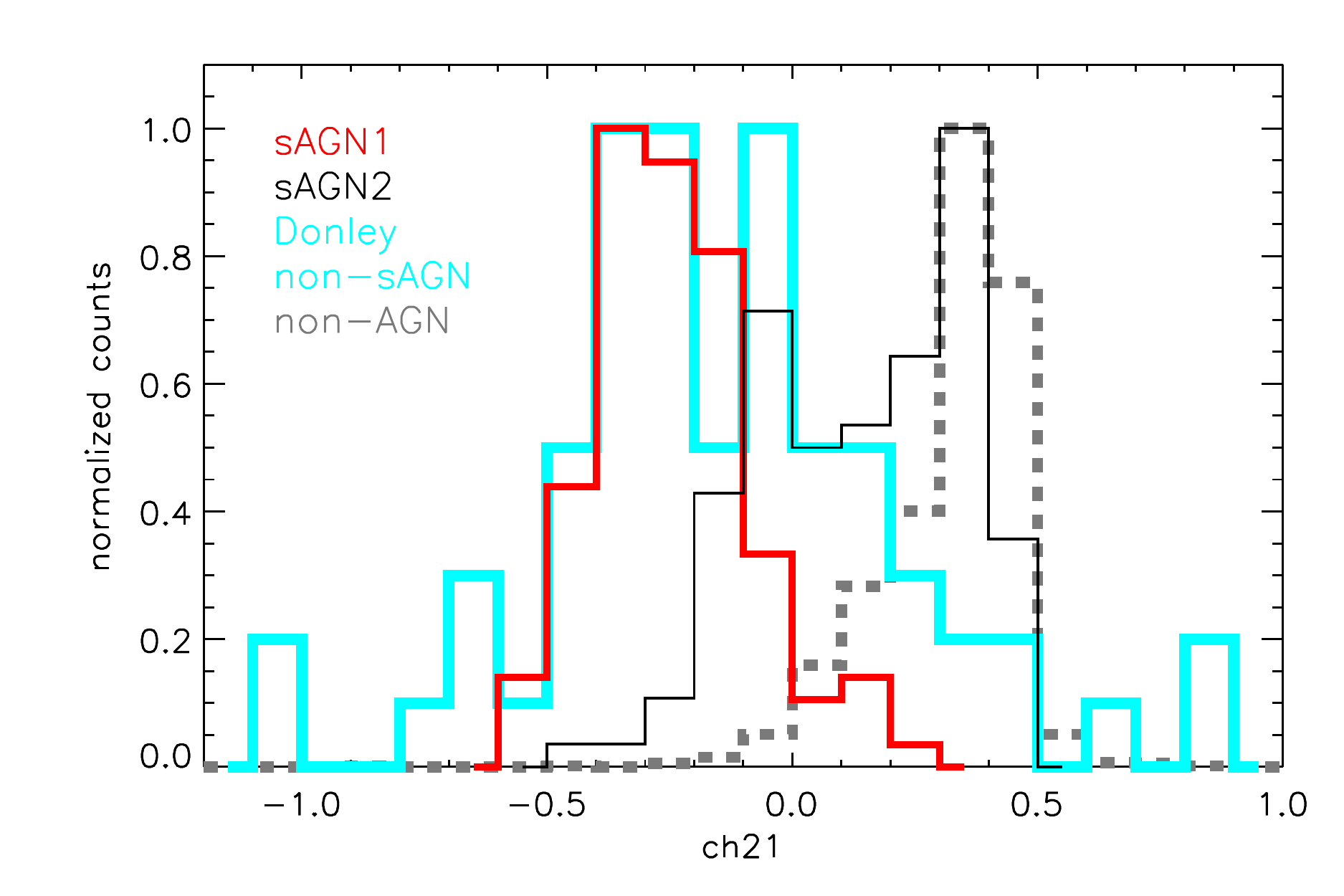}}
\caption{\footnotesize Distribution of the \texttt{ch21} feature for the various classes of sources constituting the full LS used in this work: non-AGN (dashed gray line), sAGN1 (red line), sAGN2 (thin black line), and Donley non-sAGN (thick cyan line).}\label{fig:ch21}
\end{figure}

In Sect. \ref{section:mw_properties} we introduced Table \ref{tab:results}, which presents the results obtained for the labeled plus unlabeled sets from various classifiers tested in this work. This table also allows for comparison of the results obtained with and without the \texttt{ch21} color in the above-mentioned pairs of classifiers. It is apparent that the results obtained for the two classifiers in the first pair (corresponding to the sAGN1 LS) are very similar, and the same holds for the second pair (sAGN1 plus sAGN2 LS): differences of purity and completeness are of 1--3\% among all the indicators in the table. A spectroscopic follow-up of the unmatched and unconfirmed AGN candidates in each pair would help understand whether the inclusion of the \texttt{ch21} feature actually helps retrieve a purer sample of AGN or instead fails in the identification of some AGN.
The last pair of classifiers compared (all AGN types LS) exhibit more remarkable variations, the largest of which concern purity (+7\%), completeness with respect to Type II AGN (+6\%), and completeness with respect to MIR AGN (+11\%). This supports the thesis that the introduction of MIR colors significantly helps disentangle obscured AGN from non-AGN and, based on this, the effect of adding the \texttt{ch21} color to our set of features becomes relevant when a large sample of obscured AGN is included in the LS. This is also supported by Fig. \ref{fig:ch21}, where we show the distribution of the \texttt{ch21} color for the various classes of sources that constitute the LS adopted in this work. It is apparent that sAGN1 are well separated from non-AGN; conversely, sAGN2 partly overlap non-AGN. For both classes of AGN, the situation is the same as it is for most of the other features used in this work. Donley non-sAGN, on the other hand, exhibit a distribution separated from non-AGN, and mostly overlapping sAGN1; this means that, when we include Donley non-sAGN in our LS (all AGN* in Table \ref{tab:results} and Fig. \ref{fig:f_i_comparison}), the \texttt{ch21} feature significantly helps disentangle them from non-AGN, thus explaining the improvement in the performance of the classifier. It is worth mentioning that 47 out of the 81 AGN candidates identified in this last case have \emph{channel1} or \emph{channel2} magnitude values below the \emph{WISE} all-sky limits.

\section{Summary and conclusions}
\label{section:discussion}
In this work we performed an optical variability study aimed at the identification of AGN through an RF classifier, in view of future LSST studies. In order to build our classifier, we chose a set of features that will be available from LSST data: 29 variability features computed from the light curve of each source, five optical and NIR colors obtained from the $uBrizy$ bands, and a morphology feature, the stellarity index (see Table \ref{tab:features}). We also analyzed the impact of an additional feature, the MIR color $4.5 - 3.6$ $\mu m$ (\texttt{ch21} feature), which will be available for LSST-related studies. In the following we summarize our main findings.
\begin{enumerate}
\item[--]The best results, based on the obtained confusion matrices and the computed scores (see Fig. \ref{fig:cm_ls_tests} and Table \ref{tab:scores}, respectively), are obtained when the full set of features is used.\\
However, the addition of colors to the feature list does not lead to substantial improvement in the results of the classification. We investigate five different RF classifiers: one where only variability features are used (RF1), and four more where two to five colors are added to the list of variability features progressively (RF2 to RF5). We chose to work with the RF5 classifier based on the slightly better results obtained for its confusion matrix (Fig. \ref{fig:cm_12345_col}) and the corresponding scores (Table \ref{tab:scores}). 
If we consider the samples of AGN candidates returned by each of the five classifiers, we obtain 81, 82, 79, 79, and 77 sources, respectively, with a very strong overlap: indeed, 71 sources are common to all the samples (87\% to 92\% of the various samples of AGN candidates). Fifty-seven out of these 71 sources (80\%) are confirmed to be AGN by their multiwavelength properties. On the other hand, in total, there are 16 sources missed by at least one of the samples, and it is worth mentioning that only three of them can be confirmed as AGN on the basis of their multiwavelength properties. We estimate the purity obtained from each classifier in the same way described in Sect. \ref{section:mw_properties}, and obtain the following values for the first four RF classifiers: 87\% (RF1, RF2, RF3), and 89\% (RF4). We note that each of these values is a bit higher than the 86\% value obtained in \citetalias{decicco19} and lower than the 91\% obtained when the full set of optical and NIR color features is used. All this suggests that, although variability features alone are already effective in providing high-purity samples of AGN, even higher purity values can be obtained if colors are included in the feature list.
\\
\item[--]Consistent with the fact that optical variability favors the identification of Type I AGN, a LS consisting of Type I AGN only is the first we tested. In addition we explored the possibility to include various types of AGN in our LS, selected by means of different techniques: spectroscopically confirmed Type II AGN, and also MIR AGN, based on the diagram proposed by \citet{donley}. Type II AGN, as well as MIR AGN, do not provide much benefit in the context of the present study, as the distributions of their variability features, in general, cannot be disentangled from the corresponding distributions for non-AGN (see Figs. \ref{fig:fi_analysis} and \ref{fig:fi_analysis_col}). \\
Table \ref{tab:top5feat} shows how the most important features for the classification process vary depending on the selected LS of AGN, which will prove useful for future studies. The ranking observed for the various classifiers suggests that colors become important when we include Type II/MIR AGN in the LS; on the contrary, they are not dominant features when we resort to a pure LS including only sAGN1.
We made a test including only sAGN1 in the LS and using only the five most important features in the RF5 ranking, and obtained a purity of 78\% and a completeness of 73\%, versus 91\% and 69\% obtained respectively with the full set of features. This test shows how the contribution of the other features, though not dominant, is not negligible.
\\
\item[--]Table \ref{tab:results} is a summary of the results obtained from various classifiers tested here and clearly shows how, in general, a purer selection of AGN in the LS corresponds to a higher purity. Specifically, the highest-purity value obtained is 91\% and is associated with the RF5 classifier; the corresponding completeness values with respect to spectroscopically confirmed AGN and MIR-confirmed AGN are 69\% and 64\% , respectively. These values are higher than the corresponding values obtained in \citetalias{decicco19} (86\%, 59\%, and 55\%, respectively). With the RF5 classifier we are able to identify +12\% Type I AGN and +3\% Type II AGN compared to \citetalias{decicco19}.\\ 
On the other hand, when all classes of AGN are included in the LS (all AGN*), we obtain the highest values for each estimate of the completeness: in particular, we double the Type II AGN completeness, identifying 36\% of known AGN in this class. This comes at the cost of much lower purity (65\% vs. 86\% in \citetalias{decicco19}). Based on our results, one can select the most suitable LS depending on whether their study aims at favoring purity over completeness or vice versa. 
\\
\item[--]We find that an AGN LS consisting only of a bright selection ($r \leq$ 21 mag) of spectroscopically confirmed AGN performs quite well in the identification of fainter AGN (Sect. \ref{section:spec21}). The RF$_{spec21}$ classifier returns a sample of AGN candidates with 90\% purity and 59\% completeness, largely overlapping the RF5 sample of AGN candidates, and is able to identify AGN candidates at fainter magnitudes, covering the same color ranges as the sample of candidates selected through the RF5 classifier (Fig. \ref{fig:rf5_rfspec21}). This results are encouraging in the context of future AGN studies based on LSST data, as LSs of AGN reaching approximately the image depth will generally not be available, except for deeply surveyed areas like the DDFs. 
\\
\item[--]We find that an RF classifier making use only of the five selected color features plus stellarity returns samples of AGN candidates that are at least one order of magnitude larger than the ones obtained when variability features are included. These samples very likely include a large number of contaminants, and are characterized by very low fractions of confirmed AGN, as detailed in Sect. \ref{section:col_only}. This, together with the inadequacy of color selection alone in identifying AGN characterized by redder colors, highlights the importance of variability selection for AGN.
\\
\item[--]We test the inclusion of an additional feature, the MIR color defined as $4.5-3.6$ $\mu m$, to the set of optical and NIR colors plus stellarity and variability features (see Sect. \ref{section:MIR_col} and Table \ref{tab:results}). We test this on different AGN LSs and register mild variations (${\sim}1$--3\%) in the purity and completeness of the selected samples of AGN when only sAGN1 or sAGN1 plus sAGN2 are included in the AGN LS. When we include MIR-selected AGN, purity and completeness values increase by ${\sim}3$--11\%; Type II AGN and MIR AGN are the ones that benefit the most from the inclusion of the new feature, their completeness rising from 30\% to 36\% and from 67\% to 78\%, respectively. 
\\
\item[--]We inspected the properties of the 14 RF5 AGN candidates with no confirmation of their nature based on the diagnostics used here. Five out of 14 (43\%) have an average magnitude $r < 23$ mag. Moreover, five out of 14 are close to the edges of the imaged field or to areas masked because of the presence of a halo due to a close saturated star; in both cases the area is affected by a higher noise compared to the rest of the image. In addition, four of these five AGN candidates have a magnitude $r > 23$ mag. These two effects together could favor the detection of spurious variability. The brightest of the remaining nine sources not too close to the edges of the imaged field is classified as a star in various COSMOS catalogs, while the others are classified as galaxies. In Sect. \ref{section:mw_properties} we provided an estimate of the upper limits for the hard-band X-ray fluxes of these sources, which do not have a counterpart in the \emph{Chandra}-COSMOS Legacy Catalog. These estimates and the positions these sources consequently occupy in the $X/O$ diagram shown in Fig. \ref{fig:xo} support their candidacy as AGN. Spectroscopic confirmation, at least for the sources in this sample that are not located in questionable regions, will help refine the purity of our selection of AGN candidates. 
\\
\item[--]We estimate the number of AGN found in the VST-COSMOS area. Taking into account the number of confirmed AGN in the RF5 sample of candidates (63), and also the known AGN included in the LS and correctly classified (212), we have 275 AGN in an area of 0.8 sq. deg. (the area is not the full square degree imaged by the VST as ${\approx}20$\% of it was masked due to artifacts and defects; see \citetalias{decicco15} and \citetalias{decicco19} for details).
This means that we obtain a sky density of 344 AGN/sq. deg. and, based on this estimate, we expect to find ${\approx}6.2$ million AGN in the 18,000 sq. deg. area of the LSST main survey down to our current magnitude limit. This is likely only a lower limit, given the higher-resolution imaging of LSST and the availability of a multi-visit dataset in a larger number of bands compared to the VST-COSMOS dataset.
\end{enumerate}

\begin{acknowledgements}
We acknowledge support from: ANID grants FONDECYT Postdoctorado Nº 3200222 (D.D.) and 3200250 (P.S.S.); FONDECYT Regular Nº 1190818 and 1200495 (F.E.B.); Millennium Science Initiative ICN12\_009 (D.D., F.E.B., P.S.S., G.P.); Basal-CATA AFB-170002 (D.D., F.E.B., P.S.S.); ASI-INAF agreement n.2017-14-H.O. (M.P.); Fondo di Finanziamento per le Attività Base di Ricerca (FFABR 2017, S.C.); NASA grant 80NSSC19K0961 (W.N.B.); Inter-University Institute for Data Intensive Astronomy (IDIA, M.V.); Italian Ministry of Foreign Affairs and International Cooperation (MAECI Grant Number ZA18GR02, M.V.); South African Department of Science and Technology's National Research Foundation (DST-NRF Grant Number 113121, M.V.).
\end{acknowledgements}

\bibliographystyle{aa}
\bibliography{cosmos_rf}

\end{document}